\documentclass{article}

\usepackage{amsthm}
\newtheorem{assumption}{Assumption}
\newtheorem{theorem}{Theorem}
\usepackage{arxiv}
\usepackage{amsmath}
\usepackage[utf8]{inputenc} 
\usepackage[T1]{fontenc}    
\usepackage{hyperref}       
\usepackage{url}            
\usepackage{booktabs}       
\usepackage{amsfonts}       
\usepackage{nicefrac}       
\usepackage{microtype}      
\usepackage{lipsum}
\usepackage{multirow}
\usepackage{footnote}
\usepackage{threeparttable}

\usepackage{tikz}
\usepackage{tikz-cd}
\usepackage{graphicx}
\graphicspath{ {./images/} }
\usepackage{natbib}
\usepackage{doi}

\title{Two-Stage Estimation for Causal Inference Involving Semi-Continuous Exposures}

\author{ \href{https://orcid.org/0000-0000-0000-0000}{Xiaoya Wang}\thanks{Use footnote for providing further
		information about author (webpage, alternative
		address)---\emph{not} for acknowledging funding agencies.} \\
	Department of Statistics and Actuarial Science\\
	University of Waterloo\\
	Waterloo, ON N2L 3G1, Canada\\
	\texttt{x932wang@uwaterloo.ca} \\
	\And
   {Richard J. Cook} \\
	Department of Statistics and Actuarial Science\\
	University of Waterloo\\
	Waterloo, ON N2L 3G1, Canada\\
	\texttt{rjcook@uwaterloo.ca} \\
	\And
   {Yeying Zhu} \\
	Department of Statistics and Actuarial Science\\
	University of Waterloo\\
	Waterloo, ON N2L 3G1, Canada\\
	\texttt{yeying.zhu@uwaterloo.ca} \\
    \And
   {Tugba Akkaya-Hocagil} \\
	Department of Biostatistics, School of Medicine\\
	Ankara University\\
	Ankara, Ankara 06230, Turkey\\
	\texttt{tahocagil@ankara.edu.tr} \\
    \And
   {R. Colin Carter} \\
	Departments of Emergency Medicine and Pediatrics\\
	Columbia University\\
	New York, NY 10032, USA\\
	\texttt{rcc2142@cumc.columbia.edu} \\
    \And
   {Sandra W. Jacobson} \\
	Department of Psychiatry and Behavioral Neurosciences\\
	Wayne State University\\
	Detroit, MI 48202, USA\\
	\texttt{sandra.jacobson@wayne.edu} \\
    \And
   {Joseph L. Jacobson} \\
	Department of Psychiatry and Behavioral Neurosciences\\
	Wayne State University\\
	Detroit, MI 48202, USA\\
	\texttt{joseph.jacobson@wayne.edu} \\
    \And
   {Louise M. Ryan} \\
	School of Mathematical and Physical Sciences\\
	University of Technology Sydney\\
	Sydney, NSW 2007, Australia\\
	\texttt{louise.m.ryan@uts.edu.au} \\
}

\begin{document}
\maketitle
\begin{abstract}
Methods for causal inference are well developed for binary and continuous exposures, but in many settings, the exposure has a substantial mass at zero---such exposures are called semi-continuous. We propose a general causal framework for such semi-continuous exposures, together with a two-stage estimation strategy. A two-part propensity structure is introduced for the semi-continuous exposure, with one component for exposure status (exposed vs unexposed) and another for the exposure level among those exposed, and incorporates both into a marginal structural model that disentangles the effects of exposure status and dose. The two-stage procedure sequentially targets the causal dose--response among exposed individuals and the causal effect of exposure status at a reference dose, allowing flexibility in the choice of propensity score methods in the second stage. We establish consistency and asymptotic normality for the resulting estimators, and characterize their limiting values under misspecification of the propensity score models. Simulation studies evaluate finite sample performance and robustness, and an application to a study of prenatal alcohol exposure and child cognition demonstrates how the proposed methods can be used to address a range of scientific questions about both exposure status and exposure intensity.
\end{abstract}



\section{Introduction} 
\subsection{Background}
In many areas of public health research the goal is to assess whether an exposure has a causal effect on an outcome and to quantify that effect. In observational studies, propensity score (PS) methods are now widely used to mitigate bias due to measured confounding variables \citep{rubin1974random,rosenbaum1983central,rosenbaum1984reducing}. For binary exposures, propensity scores are used for matching, stratification, and inverse probability weighting \citep{heckman1998matching,robins2000marginal}. For continuous exposures, the generalized PS is based on the conditional of the exposure evaluated at the observed exposure level \citep{hirano2004propensity,imai2004causal}. Although inverse density weighting can be used to estimate causal dose-response relations in principle, in practice it can lead to highly variable estimators when extreme weights arise (e.g., due to outliers or limited overlap) \citep{naimi2014constructing}. Moreover, modeling the exposure distribution can be challenging in practice---discretizing the exposure or the generalized PS has been advocated by some \citep{wei2021emulating}.

In many applications, samples comprise both unexposed individuals with variation in the exposure level among exposed individuals. This results in an exposure distribution with a point mass at zero and a continuous component among the exposed. Semi-continuous exposures are also common in behavioral settings---for example, smoking, cannabis use, or alcohol consumption---where many individuals abstain and the level of exposure varies among users. This also arises in environmental research where toxicants may be absent in some locations, yet vary in concentration among locations where it is present \citep{begu2016method}. There has been relatively limited methodological attention to causal analysis involving such exposure distributions. 

A motivating research question involves the study of prenatal alcohol exposure (PAE) and its relation to child cognition. A substantial proportion of mothers report no drinking during pregnancy, while consumption intensity varies widely among drinkers, yielding a semi-continuous exposure with a point mass at zero and a continuous right-skewed distribution among the exposed. Although epidemiological studies provide evidence of an association between PAE and cognitive outcomes \citep{jacobson2004maternal,jacobson2011number,lewis2015verbal,lewis2016prospective}, the causal dose--response relation remains unclear. This motivates two causal targets: the dose--response effect among exposed individuals and the contrast between exposure at a clinically meaningful reference dose and no exposure.

Existing work has proposed two-part PS strategies for semi-continuous exposures by modeling exposure status and dose among exposed using separate regression models \citep{akkaya2021propensity,li2023use}; however, these approaches are often presented primarily as modeling and adjustment tools and do not explicitly define and target both causal estimates within a unified potential outcomes framework, together with corresponding identification assumptions and large sample properties. In this article, we develop and evaluate a two-stage framework using propensity scores for causal inference with a semi-continuous exposure and a continuous outcome. The approach targets two causal estimands using separate PS models: in Stage~I the dose-response effect is estimated among exposed individuals using PS regression adjustment, and in Stage~II the effect
of exposure at a specified reference dose versus no exposure is estimated using a second PS for exposure status. The two-stage structure enables a use of an augmented inverse probability weighted (AIPW) estimating equation \citep{bang2005doubly,robins2007comment} in Stage II. To reflect realistic data-generating mechanisms, we allow distinct (but possibly overlapping) sets of confounders affecting exposure status and dose among the exposed. We formalize the approach in the potential outcomes framework, define interpretable causal estimands aligned with the semi-continuous exposure distribution, and state identifying assumptions. We establish large sample properties for the resulting estimators, evaluate finite sample performance through extensive simulation studies including settings involving PS misspecification, and illustrate the methods in an analysis of the effect of prenatal alcohol exposure and child cognition \citep{jacobson2002validity}.

The remainder of this article is organized as follows. In Section~\ref{sec-1.2}, we introduce the PAE study as a motivating example. In Section \ref{sec2}, we introduce notation, models, and identification assumptions for causal analysis involving a semi-continuous exposure. Regression adjustment based on a two-part PS is introduced to estimate the two causal estimands of interest. In Section~\ref{sec3}, we present the two-stage estimators for the binary status and continuous dose effects, and derive their asymptotic properties. In Section~\ref{sec4}, we study the impact of propensity score misspecification by evaluating the limiting values of the estimators. Simulation studies are reported in Section~\ref{sec5}, and in Section~\ref{sec6} we apply the methods to data from the Detroit study on the effect of PAE on child cognition. Conclusions and future directions are discussed in Section~\ref{sec7}.
\begin{figure}[t]
    \centering
    \includegraphics[width=0.7\textwidth]{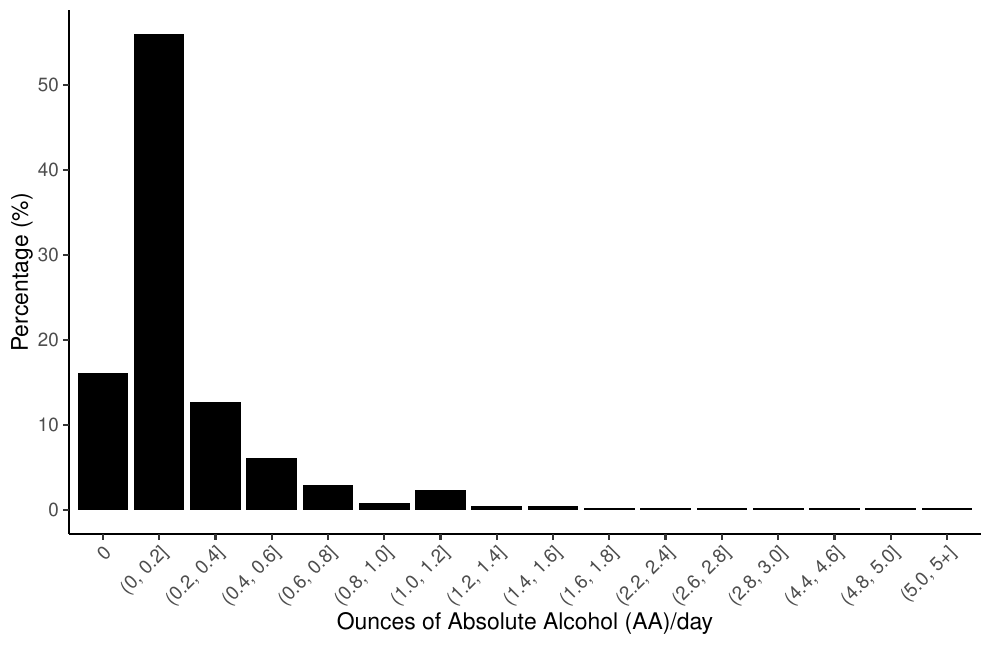}
    \caption{Frequency distribution of ounces of absolute alcohol (AA)/day in the Detroit longitudinal cohort study}
            \label{fig-1}
\end{figure}
\subsection{A Study of Prenatal Alcohol Exposure and Child Cognition}
\label{sec-1.2}
The Detroit longitudinal cohort study followed children born between 1986 and 1989 from birth through adolescence and early adulthood and has been used to examine the effects of PAE on neurocognitive outcomes \citep{jacobson2004maternal,jacobson2002validity}. A total of 480 pregnant African-American women were recruited from inner-city areas and interviewed during pregnancy about alcohol use using a timeline follow-back method. At each interview, average daily alcohol consumption during pregnancy was recorded and expressed as ounces of absolute alcohol (AA) per day.

We focus on the 377 children who completed the 7-year follow-up and use the Full Scale IQ (FSIQ) score from the Wechsler Intelligence Scale for Children, Third Edition (WISC-III), as the outcome. The WISC-III is a standardized measure of cognitive functioning for children aged 6--16 years that provides norm-referenced IQ scores, including the Full Scale IQ summary measure \citep{wechsler1991wisc}. The FSIQ score is norm-referenced (mean = 100) and typically ranges from 40 to 160, with higher values indicating better overall cognitive performance relative to age-matched peers \citep{wechsler1991wisc}. Figure~\ref{fig-1} displays the distribution of AA/day; 16.2\% of mothers reported no drinking during pregnancy, yielding a point mass at zero together with a continuous distribution among drinkers. This distribution motivates two causal targets: the dose–response effect among mothers who reported drinking and the effect of a clinically meaningful non-zero exposure level relative to no exposure, which are addressed by the two-stage PS framework developed in the following sections.

\section{Causal analysis under a semi-continuous exposure}\label{sec2}
\subsection{Notation, Potential Outcomes, and Causal Estimands}\label{sec2.1}

Let $Y$ be a continuous response variable, $T$ be a non-negative random variable representing the exposure dose (e.g. the ounces of absolute alcohol per day), and $A=I(T>0)$ indicate the exposure status. Let $\mathbf{X}=(\mathbf{X}_1',\mathbf{X}_2',\mathbf{X}_3')'$ be a vector of three sets of confounders with $\mathbf{X}_{1}=(X_{11},..., X_{1k_1})'$, $\mathbf{X}_{2}=(X_{21},..., X_{2k_2})'$, and $\mathbf{X}_{3}=(X_{31},..., X_{3k_3})'$. $\mathbf{X}_1$ and $\mathbf{X}_3$ confound the $T-Y$ association given $A=1$, and $\mathbf{X}_2$ and $\mathbf{X}_3$ confound the $A-Y$ association; see the directed acyclic graph in Figure \ref{fig-DAG}. In practice, $\mathbf{X}_1$, $\mathbf{X}_2$ and $\mathbf{X}_3$ could be the same set of covariates, but we use separate notations here to allow them to vary and represent their different roles with respect to confounding. For a sample of $n$ independent mother-child pairs, the observed data is denoted as $\{(Y_i, T_{i}, A_i, \mathbf{X}_{i}'),i=1,\ldots,n\}$. 

\begin{figure}[t]
    \centering
        \begin{tikzcd}[row sep=1em, column sep=4em]
            \mathbf{X}_1 & T\\
            \mathbf{X}_3&& Y \\
           \mathbf{X}_2 & A
            \arrow[from=1-1, to=1-2]
            \arrow[from=1-2, to=2-3]
            \arrow[from=3-1, to=3-2]
            \arrow[from=3-2, to=2-3]
            \arrow[from=2-1, to=3-2]
             \arrow[from=2-1, to=1-2]
             \arrow[from=3-2, to=1-2]
             \arrow[bend left=60, from=1-1, to=2-3]
            \arrow[bend right=60, from=3-1, to=2-3]
             \arrow[out=125, in=70, looseness=2, from=2-1, to=2-3]
        \end{tikzcd}
      \caption{A directed acyclic graph for a two-part exposure model.}
       \label{fig-DAG}
\end{figure}

As is common in dose-response modelling, we consider a log transformation of the continuous exposure $T$ if $T>0$ and refer to $D=\log T$ as the dose, with $d$ its realized value. We let $Y_i(1,d)$ represent the potential outcome for the $i$-th individual exposed at dose $d$ \citep{rubin1974random,splawa1990application}; if the $i$-th individual is unexposed, then the dose is undefined, but we denote the potential outcome as $Y_i(0,0)$ for consistency of notation, taking it as understood that $d$ is undefined. The semi-continuous nature of the exposure motivates two causal estimands. First, among exposed individuals we consider the causal dose-response effect
\begin{equation}
\label{causal-dose-response-effect}
\Delta_D=E\{Y(1,d+1)-Y(1,d)\mid A=1\}.
\end{equation}Conditioning on $A=1$ defines the target population as individuals who are exposed in the observed population. Second, to compare exposed versus unexposed individuals, we must specify a reference dose $c$ and consider this reference dose as applying to the exposed when assessing the effect of exposure (exposed versus unexposed). Thus, we define  
\begin{equation}
\label{reference-dose-effect}
\Delta_0(c)=E\{Y(1,c)-Y(0,0)\}
\end{equation}
as one causal effect of interest. In practice, $c$ may be chosen as a clinically meaningful value or a central value of the dose distribution among the exposed.


To provide a summary of the two causal targets in \eqref{causal-dose-response-effect} and \eqref{reference-dose-effect}, we consider a MSM involving a semi-continuous exposure as
\begin{equation}
\label{msm}
E\{Y(a,d)\}=\psi_0+\psi_{11}a(d-c)+\psi_{12}a.
\end{equation}
The terms $a(d-c)$ and $a$ ensure that the dose component contributes only for exposed individuals; when $a=0$ both terms drop out and the mean reduces to $E\{Y(0,0)\}=\psi_0$. Under \eqref{msm}, $\Delta_0(c)=\psi_{12}$ and, for the exposed target population, $\Delta_D(d_1,d_0)=\psi_{11}(d_1-d_0)$, so $\psi_{12}$ and $\psi_{11}$ directly parameterize the reference-dose effect and the dose--response effect, respectively. The assumptions in Section~\ref{sec2.2} ensure that these parameters (and hence $\Delta_0(c)$ and $\Delta_D$) are causally interpretable and identifiable from the observed data.

\subsection{Identification Assumptions}\label{sec2.2}
We next state assumptions for identifiability the semi-continuous exposure setting \citep{rubin1980randomization,rubin1990application,rosenbaum1983central,cole2008constructing}.

\begin{assumption}[Stable Unit Treatment Value Assumption]\label{ass:sutva}
There is no interference between individuals and consistency holds:
$Y_i=Y_i(0,0)$ if $A_i=0$, and $Y_i=Y_i(1,d)$ if $A_i=1$ and $D_i=d$.
\end{assumption}

\begin{assumption}[Exchangeability]\label{ass:exchange}
For all $d$, $Y_i(1,d)\perp D_i\mid (A_i=1,\mathbf{Z}_{i1})$ and
$\{Y_i(0,0),Y_i(1,d)\}\perp A_i\mid \mathbf{Z}_{i2}$.
\end{assumption}

\begin{assumption}[Positivity]\label{ass:positivity}
For covariate values with positive density, $0<P(A_i=1\mid \mathbf{Z}_{i2})<1$ and
$f(D_i\mid A_i=1,\mathbf{Z}_{i1})>0$ for all $d$ in the dose range of interest.
\end{assumption}

Assumption~\ref{ass:sutva} links the observed outcome to the corresponding potential outcome under the observed exposure and rules out interference between individuals. Assumption~\ref{ass:exchange} states that, conditional on measured covariates, exposure status is independent of the potential outcomes and, among exposed individuals, dose is independent of the potential outcomes. Assumption~\ref{ass:positivity} requires overlap in exposure status across covariate values and adequate support for the dose levels needed to identify the dose-response effect among exposed and the reference-dose effect. Under Assumptions~\ref{ass:sutva}--\ref{ass:positivity}, $\psi_{11}$ identifies $\Delta_D$ and $\psi_{12}$ identifies $\Delta_0(c)$.

\subsection{Model Settings and a Two-Part Propensity Score}\label{sec2.3}

For the data generating model, we adopt a linear regression model relating the semi-continuous exposure and baseline covariates to the conditional mean of the outcome,
\begin{equation}\label{eq:outcome_mean}
E\!\left(Y_i \mid A_i,D_i,\mathbf X_i;\boldsymbol{\theta}\right)
=\theta_0+\theta_{11}A_i(D_i-c)+\theta_{12}A_i+\mathbf X_i'\boldsymbol\theta_{2},
\end{equation}
where $c$ is a pre-specified reference dose, $\boldsymbol{\theta}=(\theta_0,\theta_{11},\theta_{12},\boldsymbol{\theta}_2')'$ with
$\boldsymbol{\theta}_2=(\boldsymbol{\theta}_{21}',\boldsymbol{\theta}_{22}',\boldsymbol{\theta}_{23}')'$ and $\boldsymbol{\theta}_{21}=(\theta_{211},\ldots,\theta_{21k_1})'$, $\boldsymbol{\theta}_{22}=(\theta_{221},\ldots,\theta_{22k_2})'$ and $\boldsymbol{\theta}_{23}=(\theta_{231},\ldots,\theta_{23k_3})'$.
To represent the semi-continuous exposure distribution, we adopt a two-part exposure model with one component for the continuous dose among the exposed and one for exposure status \citep{akkaya2021propensity,li2023use}.
Among exposed individuals, we specify a linear model for the continuous dose conditional on covariates,
\begin{equation}\label{eq:dose_model}
E(D_i \mid A_i=1,\mathbf Z_{i1};\boldsymbol\alpha_1)=\bar{\mathbf Z}_{i1}'\boldsymbol\alpha_1,
\end{equation}
where $\bar{\mathbf{Z}}_1=(1, \mathbf{Z}_1')'$ and $\boldsymbol{\alpha}_1=(\alpha_{10}, \boldsymbol{\alpha}_{11}',\boldsymbol{\alpha}_{12}')'$ where $\boldsymbol{\alpha}_{11}=(\alpha_{111},...,\alpha_{11k_1})'$ and $\boldsymbol{\alpha}_{12}=(\alpha_{121},...,\alpha_{12k_3})'$.
For the exposure status, we define a logistic regression model
\begin{equation}\label{eq:status_model}
\log\left\{\frac{P(A_i=1\mid \mathbf Z_{i2})}{1-P(A_i=1\mid \mathbf Z_{i2})}\right\}
=\bar{\mathbf Z}_{i2}'\boldsymbol\alpha_2,
\end{equation}
with $\bar{\mathbf{Z}}_2=(1, \mathbf{Z}_2')'$ and $\boldsymbol{\alpha}_2=(\alpha_{20}, \boldsymbol{\alpha}_{21}',\boldsymbol{\alpha}_{22}')'$ where $\boldsymbol{\alpha}_{21}=(\alpha_{211},...,\alpha_{21k_2})'$ and $\boldsymbol{\alpha}_{22}=(\alpha_{221},...,\alpha_{22k_3})'$. 

For a binary exposure the PS is the conditional probability of exposure given covariates \citep{rosenbaum1983central} while for continuous exposures the generalized PS is the conditional exposure density given covariates; more generally any function of the covariates that balances the exposure distribution can serve as a balancing score \citep{imai2004causal}. In our setting with a binary indicator $A$ and a continuous dose $D$ among the exposed, we construct a two-part PS
$\mathbf S(\mathbf X;\boldsymbol\alpha)=(S_1(\mathbf Z_1;\boldsymbol\alpha_1),\,S_2(\mathbf Z_2;\boldsymbol\alpha_2))'$,
where $S_1(\mathbf Z_1;\boldsymbol\alpha_1)=E(D\mid A=1,\mathbf Z_1;\boldsymbol\alpha_1)$ and
$S_2(\mathbf Z_2;\boldsymbol\alpha_2)=P(A=1\mid \mathbf Z_2;\boldsymbol\alpha_2)$.
Under the dose model in \eqref{eq:dose_model} and assuming normal errors with constant variance,
$D\mid(A=1,\mathbf Z_1)\sim N(S_1(\mathbf Z_1;\boldsymbol\alpha_1),\sigma_D^2)$, the conditional distribution of $D$
depends on $\mathbf Z_1$ only through $S_1(\mathbf Z_1;\boldsymbol\alpha_1)$; hence among exposed individuals,
$D\perp \mathbf Z_1 \mid \{A=1,S_1(\mathbf Z_1;\boldsymbol\alpha_1)\}$.
Accordingly, $\mathbf S(\mathbf X;\boldsymbol\alpha)$ can be treated as a two-part PS for adjusting confounding in the semi-continuous exposure
setting under the exposure models \eqref{eq:dose_model} and \eqref{eq:status_model}.

\subsection{Regression Adjustment with a Two-Part Propensity Score}\label{sec2.4}

Using the two-part propensity score $\mathbf S(\mathbf X;\boldsymbol\alpha)$ defined in Section~\ref{sec2.3}, we consider regression adjustment based on $(A,D,\mathbf S)$:
\begin{equation}
\label{equa-2.5}
E\{Y|A,D,\mathbf{S}(\mathbf{X};\boldsymbol{\alpha})\}
= \eta_{0} + \eta_{11}A(D-c) + \eta_{12}A
+ \mathbf{S}(\mathbf{X};\boldsymbol{\alpha})'\boldsymbol{\eta}_{2}
\end{equation}
where $\boldsymbol{\eta}=(\eta_{0},\boldsymbol{\eta}_{1}',\boldsymbol{\eta}_{2}')'$ with
$\boldsymbol{\eta}_{1}=(\eta_{11},\eta_{12})'$ and $\boldsymbol{\eta}_{2}=(\eta_{21},\eta_{22})'$. Under Assumptions~\ref{ass:sutva}--\ref{ass:positivity} and correct specification of the exposure models used to construct $\mathbf S(\mathbf X;\boldsymbol\alpha)$, we have the balancing properties
$A\perp \mathbf Z_2\mid S_2(\mathbf Z_2;\boldsymbol\alpha_2)$ and, among exposed individuals,
$D\perp \mathbf Z_1\mid\{A=1,S_1(\mathbf Z_1;\boldsymbol\alpha_1)\}$; see Appendix \ref{sup-A} for detailed derivations establishing the balancing properties of $\mathbf{S}(\mathbf{X};\boldsymbol{\alpha})$. Consequently, $\eta_{11}$ identifies the causal dose-response effect among exposed and $\eta_{12}$ identifies the causal effect of the exposure at the reference dose compared to no exposure.

\section{A Framework for Two-stage Estimation} \label{sec3}

Regression adjustment hinges on the correct specification of both PS models, motivating us to develop a two-stage formulation that separates estimation of the dose–response from the contrast between exposure and no exposure at a reference dose. In Stage~I we estimate the causal dose–response effect among exposed individuals and in Stage~II we estimate the causal effect of exposure at the reference dose compared to no exposure. We use PS regression adjustment in Stage~I, and in Stage II use PS regression adjustment, inverse probability weighted estimating equations, or augmented inverse probability weighted estimating equations, and derive the corresponding estimating equations, and present asymptotic properties.
\subsection{Stage I: The Causal Dose-Response Effect}
\label{sec3.1}
Let $n$ denote the sample size and index individuals by $i=1,\ldots,n$. The observed data are ${\mathcal{D}}_i=(Y_i,\mathbf{X}_i,A_i,D_i)$, $i=1,\ldots,n$. Among exposed individuals ($A_i=1$), we consider a PS regression adjustment to assess the causal dose–response effect by fitting the model 
\begin{equation}
\label{new.estimating.equation3.1}
E\left\{Y_i | A_i=1,D_i,S_{1}(\mathbf{Z}_{i1};\boldsymbol{\alpha}_1);\boldsymbol{\gamma}_1\right\}
   = \gamma_{10}+\gamma_{11}D_i+\gamma_{12}S_1(\mathbf{Z}_{i1};\boldsymbol{\alpha}_1)=\mu_{i11}(\boldsymbol{\phi}_1),
\end{equation}
by least squares where $S_{1}(\mathbf{Z}_{i1};\boldsymbol{\alpha}_1)
   = E(D_i | A_i=1,\mathbf{Z}_{i1};\boldsymbol{\alpha}_1)$, $\boldsymbol{\gamma}_1=(\gamma_{10},\gamma_{11},\gamma_{12})'$, and $\boldsymbol{\phi}_1=(\boldsymbol{\gamma}_1',\boldsymbol{\alpha}_1')'$.
 The estimating equation for $\boldsymbol{\gamma}_1$ is
\begin{equation}
\label{equa-3.1}
    \mathbf{U}_{11}(\mathcal{D};\boldsymbol{\phi}_1)
    = \sum_{i=1}^n \mathbf{U}_{i11}(\mathcal{D}_i;\boldsymbol{\phi}_1) 
    = \mathbf{0},
\end{equation}
where 
$$\mathbf{U}_{i11}(\mathcal{D}_i;\boldsymbol{\phi}_1)=A_i
      \frac{\partial \mu_{i11}(\boldsymbol{\phi}_1)}{\partial \boldsymbol{\gamma}_1}
      \{Y_i-\mu_{i11}(\boldsymbol{\phi}_1)\}.$$ 
The parameter $\boldsymbol{\alpha}_1$ can be estimated by solving 
\begin{equation}
\label{new.estimating.equation3.2}
\mathbf{U}_{12}(\mathcal{D};\boldsymbol{\alpha}_1)=\sum^n_{i=1}\mathbf{U}_{i12}(\mathcal{D}_i;\boldsymbol{\alpha}_1) =\mathbf{0}
\end{equation}
where 
$$\mathbf{U}_{i12}(\mathcal{D}_i;\boldsymbol{\alpha}_1)=A_i\frac{\partial \mu_{i12}(\boldsymbol{\alpha}_1)}{\partial\boldsymbol{\alpha}_1}\left\{D_i-\mu_{i12}(\boldsymbol{\alpha}_1)\right\}$$ with $\mu_{i12}(\boldsymbol{\alpha}_1)=\bar{\mathbf{Z}}_{i1}'\boldsymbol{\alpha}_1$. Considering (\ref{equa-3.1}) and (\ref{new.estimating.equation3.2}) jointly, the Stage I estimating equation is 
\begin{equation}
\label{new.estimating.equation-3.4}
    \mathbf{U}_{1}(\mathcal{D}_i;\boldsymbol{\phi}_1)=\sum_{i=1}^{n}\mathbf{U}_{i1}(\mathcal{D}_i;\boldsymbol{\phi}_1)=\mathbf{0}
\end{equation}
with
$\mathbf{U}_{i1}(\mathcal{D}_i;\boldsymbol{\phi}_1)=\left(\mathbf{U}_{i11}'(\mathcal{D}_i;\boldsymbol{\phi}_1),\mathbf{U}_{i12}'(\mathcal{D}_i;\boldsymbol{\alpha}_1)\right)'.$
Under Assumptions \ref{ass:sutva}-\ref{ass:positivity} and  the correct specification of the PS model for the continuous exposure to obtain $S_1(\mathbf{Z}_{i1};\boldsymbol{\alpha}_1)$, the causal dose–response effect among exposed is represented by $\gamma_{11}$, which corresponds to $\psi_{11}$ in (\ref{msm}). 

\subsection{Stage II: The Effect of Exposure at the Reference Dose}

In Stage II, we assess the causal effect of exposure at a pre-specified reference dose $c$ compared to no exposure. This is achieved by conceptualizing ${\gamma}_{11}A_i(D_i-c)$ as an ``offset''---in practice $\gamma_{11}$ will be estimated in Stage I and $\hat{\gamma}_{11}A_i(D_i-c)$ will be treated as a fixed quantity in Stage II analysis. Intuitively, the offset removes the portion of the outcome variation attributable to departures of the observed dose from the reference level among exposed individuals, thereby isolating the causal effect of exposure at dose $c$ versus no exposure. The uncertainty in $\hat{\gamma}_{11}$ from Stage I is addressed in the large sample variance estimation. We consider three approaches to estimation at Stage II: (i) PS regression adjustment \citep{vansteelandt2014regression}, (ii) inverse probability weighting \citep{robins2000marginal}, and (iii) augmented inverse probability weighting \citep{bang2005doubly,robins2007comment}.

\subsubsection{Propensity Score Regression Adjustment} 
To assess the effect of $A=1$ at dose $D=c$ via a PS regression adjustment, we consider a Stage II linear predictor of the form
\begin{equation} 
    E\left\{Y_i| D_i,A_i,S_{2}(\mathbf{Z}_{i2};\boldsymbol{\alpha}_2);\boldsymbol{\gamma}_2\right\}=\mu_{i21}(\boldsymbol{\phi}_2)+ \text{offset}\left\{{\gamma}_{11}A_i(D_i-c)\right\}\nonumber
\end{equation}
where $\mu_{i21}(\boldsymbol{\phi}_2)=\gamma_{20}+\gamma_{21}A_i+\gamma_{22}S_2(\mathbf{Z}_{i2};\boldsymbol{\alpha}_2)$ with the Stage II PS $S_2(\mathbf{Z}_{i2};\boldsymbol{\alpha}_2)=E(A_i| \mathbf{Z}_{i2};\boldsymbol{\alpha}_2)$, and $\boldsymbol{\gamma}_{2}=(\gamma_{20},\gamma_{21},\gamma_{22})'$ with $\boldsymbol{\phi}_2=(\boldsymbol{\gamma}_2',\boldsymbol{\alpha}_2')'$. The regression parameter $\boldsymbol{\gamma}_2 $ can be estimated for specified $(\boldsymbol{\gamma}_{1}',\boldsymbol{\alpha}_2')'$ by the estimating equation
\begin{equation}
\label{new.estimating.equation-3.5}
    \tilde{\mathbf{U}}_{21}(\mathcal{D};\boldsymbol{\gamma}_1,\boldsymbol{\phi}_2)=\sum^n_{i=1} \tilde{\mathbf{U}}_{i21}(\mathcal{D}_i;\boldsymbol{\gamma}_1,\boldsymbol{\phi}_2)=\boldsymbol{0} 
\end{equation}
where $$\tilde{\mathbf{U}}_{i21}(\mathcal{D}_i;\boldsymbol{\gamma}_1,\boldsymbol{\phi}_2)=\frac{\partial{\mu}_{i21}(\boldsymbol{\phi}_2)}{\partial \boldsymbol{\gamma}_2}\big[Y_i-\left[{\mu}_{i21}(\boldsymbol{\phi}_2)+\text{offset}\left\{{\gamma}_{11}A_i(D_i-c)\right\}\right]\big]. $$
The parameter $\boldsymbol{\alpha}_2$ can be estimated by solving
\begin{equation}
\label{new.estimating.equation3.3}
\mathbf{U}_{22}\left(\mathcal{D} ; \boldsymbol{\alpha}_2\right)=\sum^n_{i=1}\mathbf{U}_{i22}\left(\mathcal{D}_i; \boldsymbol{\alpha}_2\right)=\mathbf{0}
\end{equation}
where
$$\mathbf{U}_{i22}\left(\mathcal{D}_i; \boldsymbol{\alpha}_2\right)=\frac{\partial S_2(\mathbf{Z}_{i2};\boldsymbol{\alpha}_2)}{\partial\boldsymbol{\alpha}_2}\frac{1}{S_2(\mathbf{Z}_{i2};\boldsymbol{\alpha}_2)\left\{1-S_2(\mathbf{Z}_{i2};\boldsymbol{\alpha}_2)\right\}}\left\{A_i-S_2(\mathbf{Z}_{i2};\boldsymbol{\alpha}_2)\right\}.$$ Any binary regression model can be used but we employ a logistic link. To combine \eqref{new.estimating.equation-3.5} and \eqref{new.estimating.equation3.3} we let $\tilde{\mathbf{U}}_{i2}(\mathcal{D}_i;\boldsymbol{\gamma}_1,\boldsymbol{\phi}_2)=\big\{\tilde{\mathbf{U}}_{i21}'(\mathcal{D}_i;\boldsymbol{\gamma}_{1},\boldsymbol{\phi}_2),\mathbf{U}_{i22}'(\mathcal{D}_i;\boldsymbol{\alpha}_2)\big\}'$ and define $$\tilde{\mathbf{U}}_{2}(\mathcal{D}_i;\boldsymbol{\gamma}_1,\boldsymbol{\phi}_2)=\sum_{i=1}^{n}\tilde{\mathbf{U}}_{i2}(\mathcal{D}_i;\boldsymbol{\gamma}_1,\boldsymbol{\phi}_2)=\mathbf{0}$$ as the joint Stage II estimating equations under PS regression adjustment. Taken together with the Stage I estimating equation, the full set of estimating equations for the two-stage analysis under PS regression adjustment in Stage I and II is
\begin{equation}
\label{equa-3.3}
\tilde{\mathbf{U}}(\mathcal{D};\boldsymbol{\Omega} )=\left(\begin{array}{l}\mathbf{U}_{1}(\mathcal{D};\boldsymbol{\phi}_1 )\\
   \tilde{\mathbf{U}}_{2}(\mathcal{D};\boldsymbol{\gamma}_1,\boldsymbol{\phi}_2 )
\end{array}\right) = \boldsymbol{0}
\end{equation}
where $\boldsymbol{\Omega}=(\boldsymbol{\phi}_1',\boldsymbol{\phi}_2' )'$. Under Assumptions \ref{ass:sutva}-\ref{ass:positivity} and the correct specification of the PS model for the binary exposure status to obtain $S_2(\mathbf{Z}_{i2};\boldsymbol{\alpha}_2)$, 
$\gamma_{21}$ is the causal effect of the exposure at reference dose compared to no exposure. and this holds likewise for the IPW and AIPW approaches described in Sections \ref{subsection3.2.2} and \ref{subsection3.2.3}. 

\subsubsection{Inverse Probability Weighting}
\label{subsection3.2.2}
We construct inverse probability weights based on $S_2(\mathbf{Z}_{i2};\boldsymbol{\alpha}_2)$ to fit the Stage~II MSM
$$E\left\{Y_i( a, d);\boldsymbol{\gamma}_1,\boldsymbol{\gamma}_2\right\}=\gamma_{20}+\gamma_{21}a+\text{offset}\left\{{\gamma}_{11}a(d-c)\right\}.$$
An IPW estimating equation \citep{hernan2000marginal,robins2000marginal} is  \begin{equation}
\label{equa-3.4}
    \bar{\mathbf{U}}_{21}(\mathcal{D};\boldsymbol{\gamma}_1,\boldsymbol{\phi}_2 )=\sum^n_{i=1} \bar{\mathbf{U}}_{i21}(\mathcal{D}_i;\boldsymbol{\gamma}_1,\boldsymbol{\phi}_2)=\mathbf{0}
\end{equation} where 
$$\bar{\mathbf{U}}_{i21}(\mathcal{D};\boldsymbol{\gamma}_1,\boldsymbol{\phi}_2 )=\sum_{a=0}^{1}w_i(a;\boldsymbol{\alpha}_2)
\frac{\partial{\mu}_{21}(a;\boldsymbol{\gamma}_2)}{\partial\boldsymbol{\gamma}_2}\big[Y_i-\left[{\mu}_{21}(a;\boldsymbol{\gamma}_2)+\text{offset}\left\{{\gamma}_{11}A_i(D_i-c)\right\}\right]\big]$$
where ${\mu}_{21}(a;\boldsymbol{\gamma}_2)=\gamma_{20}+\gamma_{21}a$,
and  
\begin{equation}
\label{equa-3.6}
    w_i(a;\boldsymbol{\alpha}_2)=\frac{I(A_i=a)}{S_2(\mathbf{Z}_{i2};\boldsymbol{\alpha}_2)^{a}\left\{1-S_2(\mathbf{Z}_{i2};\boldsymbol{\alpha}_2)\right\}^{1-a}}
\end{equation} 
for $a=0,1$. The joint Stage~II estimating equation under IPW is$$\bar{\mathbf{U}}_{2}(\mathcal{D}_i;\boldsymbol{\gamma}_1,\boldsymbol{\phi}_2)=\sum_{i=1}^{n}\bar{\mathbf{U}}_{i2}(\mathcal{D}_i;\boldsymbol{\gamma}_1,\boldsymbol{\phi}_2)=\mathbf{0}$$ where $\bar{\mathbf{U}}_{i2}(\mathcal{D}_i;\boldsymbol{\gamma}_1,\boldsymbol{\phi}_2)=\left(\bar{\mathbf{U}}_{i21}'(\mathcal{D}_i;\boldsymbol{\gamma}_{1},\boldsymbol{\phi}_2),\mathbf{U}_{i22}'(\mathcal{D}_i;\boldsymbol{\alpha}_2)\right)'$. The two-stage estimating equations with IPW are then
\begin{equation}
\label{equa-3.7}
\bar{\mathbf{U}}(\boldsymbol{\Omega})=\left(\begin{array}{l}\mathbf{U}_{1}(\mathcal{D};\boldsymbol{\phi}_1 )\\
    \bar{\mathbf{U}}_{2}(\mathcal{D};\boldsymbol{\gamma}_1,\boldsymbol{\phi}_2)
\end{array}\right) = \mathbf{0}.
\end{equation}

\subsubsection{Augmented Inverse Probability Weighting}
\label{subsection3.2.3}
An AIPW estimating function is introduced by \cite{bang2005doubly} to obtain consistent estimators of causal effects when at least one of the PS model and the imputation model is correctly specified \citep{funk2011doubly}; the augmentation term we describe ensures this double robustness property. We again estimate ${\boldsymbol{\gamma}}_2$ via an augmented estimating equation 
\begin{equation}
\label{new.estimating.equation-3.12}
    \bar{\bar{\mathbf{U}}}_{21}(\mathcal{D};\boldsymbol{\gamma}_{1},\boldsymbol{\phi}_2)=
    \sum^n_{i=1}\bar{\bar{\mathbf{U}}}_{i21}(\mathcal{D}_i;\boldsymbol{\gamma}_{1},\boldsymbol{\phi}_2)=\boldsymbol{0}, 
\end{equation}
and $\bar{\bar{\mathbf{U}}}_{i21}(\mathcal{D}_i;\boldsymbol{\gamma}_{1},\boldsymbol{\phi}_2)$ is given by 
$$
\begin{aligned}
\sum_{a=0}^1w_i(a;\boldsymbol{\alpha}_2) \frac{\partial\mu_{21}(a;\boldsymbol{\gamma}_2)}{\partial \boldsymbol{\gamma}_2}\big[Y_i-\left[\mu_{21}(a;\boldsymbol{\gamma}_2)+\text{offset}\left\{{\gamma}_{11}A_i(D_i-c)\right\}\right]\big]-\{w_i(a;\boldsymbol{\alpha}_2)-1\}\boldsymbol{g}_i(a;\boldsymbol{\theta},\boldsymbol{\gamma}_2)
\end{aligned}
$$ where $\boldsymbol{\phi}_2 = (\boldsymbol{\gamma}_2',\boldsymbol{\alpha}_2',\boldsymbol{\theta}')'$ with $\boldsymbol{\theta}=(\boldsymbol{\theta}_1',\boldsymbol{\theta}_0')'$, $w_i(a;\boldsymbol{\alpha}_2)$ is given in (\ref{equa-3.6}), and  
$$\boldsymbol{g}_i(a;\boldsymbol{\theta},\boldsymbol{\gamma}_2)=\frac{\partial\mu_{12}(a;\boldsymbol{\gamma}_2)}{\partial \boldsymbol{\gamma}_2}\left\{m_{ia}(\boldsymbol{\theta})-\mu_{12}(a;\boldsymbol{\gamma}_2)\right\}$$
where $m_{ia}(\boldsymbol{\theta})$ denotes the imputed model based expression for the expected response under treatment $a \in {0,1}$; from (\ref{eq:outcome_mean}). The imputed values are specified as
$$
m_{ia}(\boldsymbol{\theta}_a) 
= E\left\{Y_i - \gamma_{11}a(D_i-c)\,| A_i=a, \mathbf{X}_i;\boldsymbol{\theta}_a\right\} 
= \theta_{a0}+\boldsymbol{\theta}_{a1}'\mathbf{X}_{i1}+\boldsymbol{\theta}_{a2}'\mathbf{X}_{i2}+\boldsymbol{\theta}_{a3}'\mathbf{X}_{i3},
$$ 
where $\boldsymbol{\theta}_{ap}$ is $k_p$-dimensional, $p=1,2,3$ for $a=0,1$.
The estimating equation for $\boldsymbol{\theta}$ with specified $\boldsymbol{\gamma}_{1}$ is
\begin{equation}
\label{new.estimating.equation-3.13}
    \mathbf{U}_{23}(\mathcal{D};\boldsymbol{\gamma}_1,\boldsymbol{\theta})=\sum^{n}_{i=1}\mathbf{U}_{i23}(\mathcal{D}_i;\boldsymbol{\gamma}_1,\boldsymbol{\theta})=\mathbf{0} 
\end{equation}
where $\mathbf{U}_{i23}(\mathcal{D}_i;\boldsymbol{\gamma}_1,\boldsymbol{\theta})=\left\{\mathbf{U}_{i231}'(\mathcal{D}_i;\boldsymbol{\gamma}_1,\boldsymbol{\theta}_1),\mathbf{U}_{i230}'(\mathcal{D}_i; \boldsymbol{\theta}_0)\right\}'$ with 
$$
\mathbf U_{i23a}(\mathcal D_i;\boldsymbol{\gamma}_1,\boldsymbol{\theta}_a)
= I(A_i=a)\,
\frac{\partial m_{ia}(\boldsymbol{\theta}_a)}{\partial \boldsymbol{\theta}_a}
\left\{Y_i-\gamma_{11}a(D_i-c)-m_{ia}(\boldsymbol{\theta}_a)\right\}$$ for $a=0,1$.
By combining (\ref{new.estimating.equation-3.12}), (\ref{new.estimating.equation3.3}), and (\ref{new.estimating.equation-3.13}), let $\bar{\bar{\mathbf{U}}}_{i2}(\mathcal{D}_i;\boldsymbol{\phi}_2)=\big\{\bar{\bar{\mathbf{U}}}_{i21}'(\mathcal{D}_i;\boldsymbol{\gamma}_1,\boldsymbol{\phi}_2), { {\mathbf{U}}}_{i22}'(\mathcal{D}_i;\boldsymbol{\alpha}_2), {{\mathbf{U}}}_{i23}'(\mathcal{D}_i;\boldsymbol{\gamma}_1,\boldsymbol{\theta} )\big\}'$ and consider the joint Stage II AIPW estimating equation
\begin{equation}
\label{new.estimating.equation-3.14}
    \bar{\bar{\mathbf{U}}}_{2}(\mathcal{D};\boldsymbol{\gamma}_1,\boldsymbol{\phi}_2)=\sum_{i=1}^n\bar{\bar{\mathbf{U}}}_{i2}(\mathcal{D}_i;\boldsymbol{\gamma}_1,\boldsymbol{\phi}_2)=\mathbf{0}.
\end{equation}
By combining (\ref{new.estimating.equation-3.4}) and (\ref{new.estimating.equation-3.14}), the joint estimating equation is 
\begin{equation}
\label{equa-3.8}
    \bar{\bar{\mathbf{U}}}(\boldsymbol{\Omega})=\left(\begin{array}{l}\mathbf{U}_{1}(\mathcal{D};\boldsymbol{\phi}_1)\\
   \bar{\bar{\mathbf{U}}}_{2}(\mathcal{D};\boldsymbol{\gamma}_1,\boldsymbol{\phi}_2)
     \end{array}\right) = \mathbf{0}.
\end{equation}

\subsection{Estimation and Statistical Inference}
\label{sec-variance}
The two-stage analysis can be characterized by 
\begin{equation}
\label{equa-3.joint}
\mathbf{U}(\mathcal{D};\boldsymbol{\Omega}) = \left(
\begin{array}{l}
\mathbf{U}_{1}(\mathcal{D};\boldsymbol{\phi}_1) \\
\mathbf{U}_{2}(\mathcal{D};\boldsymbol{\gamma}_1,\boldsymbol{\phi}_2)
\end{array}\right) = \boldsymbol{0},
\end{equation}
where $\mathbf{U}_{1}(\mathcal{D};\boldsymbol{\phi}_1)$ is defined in (\ref{new.estimating.equation-3.4}) and $\mathbf{U}_{2}(\mathcal{D};\boldsymbol{\gamma}_1,\boldsymbol{\phi}_2)$ denotes the Stage II estimating equations under regression adjustment, IPW, or AIPW, as given in (\ref{equa-3.3}), (\ref{equa-3.7}), and (\ref{equa-3.8}), respectively. From a theoretical perspective, treating (\ref{equa-3.joint}) as a unified system is convenient for deriving the large sample properties of the resulting estimators, and the estimates $\hat{\boldsymbol{\Omega}}=(\hat{\boldsymbol{\phi}}_{1}',\hat{\boldsymbol{\phi}}_{2}')'$ can be obtained by solving (\ref{equa-3.joint}) directly.  
In practice, however, estimation proceeds sequentially: the Stage I estimating equation (\ref{new.estimating.equation-3.4}) is solved to obtain $\hat{\boldsymbol{\phi}}_{1}$ and hence $\hat{\gamma}_{11}$, which is then substituted into the Stage II estimating equation. Solving the latter yields $\hat{\boldsymbol{\phi}}_{2}$ which includes $\hat{\gamma}_{21}$, the estimator of the causal effect of exposure at the reference dose.

We now establish the large sample properties of the two-stage estimator with AIPW in Stage II; results for regression adjustment and IPW follow analogously. Let $\hat{\boldsymbol{\Omega}}$ denote the solution to the joint estimating equation (\ref{equa-3.joint}), where $\mathbf{U}_i(\cdot)$ is the stacked vector of estimating functions contributions from individual $i$ from Stages I and II, $i=1,\ldots,n$. 

\begin{theorem}
\label{thm-1}
If Assumptions \ref{ass:sutva}–\ref{ass:positivity} hold and the PS model for $S_1(\mathbf Z_1;\boldsymbol\alpha_1)$ is correctly specified, then the solution $\hat{\boldsymbol\phi}_1=(\hat{\boldsymbol\gamma}_1',\hat{\boldsymbol\alpha}_1')'$ to (\ref{new.estimating.equation-3.4}) is consistent for $\boldsymbol\phi_1$. In particular, $\hat{\gamma}_{11}$ is a consistent estimator of the causal dose–response effect among exposed $\psi_{11}$ in the MSM (\ref{msm}).
\end{theorem}

\begin{theorem}
\label{thm-2}
If Assumptions \ref{ass:sutva}–\ref{ass:positivity} hold and at least one of the PS model for $S_2(\mathbf Z_2;\boldsymbol\alpha_2)$ or the imputation model $m_a(\boldsymbol{\theta})$ is correctly specified, if $\hat{\gamma}_{11}$ is a consistent estimator of $\psi_{11}$, then the solution $\hat{\boldsymbol\gamma}_2=(\hat{\gamma}_{20},\hat{\gamma}_{21},\hat{\gamma}_{22})'$ to the estimating equation (\ref{new.estimating.equation-3.12}) is consistent for $\boldsymbol{\gamma}_2$. In particular, $\hat{\gamma}_{21}$ is a consistent estimator of the causal effect of the exposure at the reference dose compared to no exposure $\psi_{12}$ in the MSM (\ref{msm}).
\end{theorem}
\begin{theorem}
\label{thm-3}
If Assumptions \ref{ass:sutva}–\ref{ass:positivity} hold and standard regularity conditions \citep{van2000asymptotic} are satisfied, then the joint estimator $\hat{\boldsymbol{\Omega}}$ is consistent for $\boldsymbol{\Omega}$, and  
$\sqrt{n}\,(\hat{\boldsymbol{\Omega}}-\boldsymbol{\Omega})$
converges in distribution to a mean-zero multivariate normal vector with covariance matrix  
\[
\boldsymbol{\Sigma}(\boldsymbol{\Omega}) = \mathcal{A}^{-1}(\boldsymbol{\Omega})\,\mathcal{B}(\boldsymbol{\Omega})\,\{\mathcal{A}^{-1}(\boldsymbol{\Omega})\}' ,
\]
where $\mathcal{A}(\boldsymbol{\Omega}) = E\left(-{\partial \mathbf{U}(\boldsymbol{\Omega})}/{\partial \boldsymbol{\Omega}'}\right)$ and $\mathcal{B}(\boldsymbol{\Omega}) = E\left\{\mathbf{U}_{i}(\boldsymbol{\Omega})\mathbf{U}_{i}'(\boldsymbol{\Omega})\right\}$.
\end{theorem}
Proofs of Theorems~\ref{thm-1}–\ref{thm-3} and explicit expressions for the sandwich covariance estimator and Wald statistics are given in Appendix \ref{sup-B}. We use the resulting sandwich covariance matrix for $\hat{\boldsymbol{\Omega}}$ to compute standard errors, conduct Wald tests and construct confidence intervals.

\section{Implication of Propensity Score Misspecification}
\label{sec4}
The consistency of the causal estimators introduced above depends on the correct specification of the PS models used in the two-stage procedure. The explicit form of the possibly biased estimator obtained by PS regression adjustment based on a possibly misspecified PS model for a semi-continuous exposure under a one-stage framework is derived \citep{akkaya2021propensity}. Here we derive limiting values under misspecified PSs for the two-stage approach introduced in Section \ref{sec3} with PS regression adjustment in both stages. 

\subsection{Misspecification of the Stage I Propensity Score}
We work with expectations of the estimating functions to characterize limiting values under misspecification \citep{white1982maximum}. Since the data are i.i.d., we suppress the subscript $i$ and use $(Y,A,D,\mathbf{X})$ to denote a generic observation.

Let $\tilde{S}_1(\mathbf{X};\tilde{\boldsymbol{\alpha}}_1)$ denote the Stage I PS under misspecification parameterized by $\tilde{\boldsymbol{\alpha}}_1$. A regression adjustment based on $\tilde{S}_1(\mathbf{X};\tilde{\boldsymbol{\alpha}}_1)$ involves fitting 
\[
E(Y | A=1,D,\tilde{S}_1(\mathbf{X};\tilde{\boldsymbol{\alpha}}_1);\tilde{\boldsymbol{\gamma}}_1)
= \tilde{\gamma}_{10} + \tilde{\gamma}_{11}D 
  + \tilde{\gamma}_{12}\tilde{S}_1(\mathbf{X};\tilde{\boldsymbol{\alpha}}_1)
= \mu_{11}(\tilde{\boldsymbol{\phi}}_1),
\]
where $\tilde{\boldsymbol{\gamma}}_1=(\tilde{\gamma}_{10},\tilde{\gamma}_{11}, \tilde{\gamma}_{12})'$ and 
$\tilde{\boldsymbol{\phi}}_1=(\tilde{\boldsymbol{\gamma}}_1',\tilde{\boldsymbol{\alpha}}_1')'$.
Let $\mathcal{U}(\cdot)$ denote the corresponding estimating functions under misspecification. The limiting value of the estimator for $\boldsymbol{\gamma}_1$ is the solution to
\begin{equation}
\label{equa-4.1}
E\left\{\mathcal{U}_{11}(\mathcal{D};\tilde{\boldsymbol{\phi}}_1)\right\} = \mathbf{0},
\end{equation}
with
\[
\mathcal{U}_{11}(\mathcal{D};\tilde{\boldsymbol{\phi}}_1)
= A \, \frac{\partial \mu_{11}(\tilde{\boldsymbol{\phi}}_1)}{\partial \tilde{\boldsymbol{\gamma}}_1}
  \{ Y - \mu_{11}(\tilde{\boldsymbol{\phi}}_1)\},
\]
where the expectation is taken with respect to the true distribution of $(Y,A,D,\mathbf{X})$. Solving (\ref{equa-4.1}) yields $\boldsymbol{\gamma}_1^\ast=(\gamma_{10}^\ast,\gamma_{11}^\ast,\gamma_{12}^\ast)'$ with
\begin{align}
\label{equa-4.2}
\gamma_{11}^\ast = \theta_{12} 
+ \frac{\boldsymbol{\theta}_2'\left[\boldsymbol{\zeta}_1(\mathbf{X}| A=1) 
- \boldsymbol{\beta}_1(\mathbf{X}| A=1)\rho_1
   \sqrt{\frac{\text{var}\{E(D|\mathbf{X},A=1)\}}{\text{var}\{\tilde{S}_1(\mathbf{X};\tilde{\boldsymbol{\alpha}}_1)\}}}\right]}
 {\text{var}(D| A=1)-\text{var}\{E(D|\mathbf{X},A=1)\}\rho_1^2},
\end{align}
where $\rho_1=\text{corr}\{E(D| \mathbf{X},A=1),\tilde{S}_1(\mathbf{X};\tilde{\boldsymbol{\alpha}}_1)\}$,  
$\boldsymbol{\zeta}_1(\mathbf{X}| A=1)=\{\zeta_{11}(\mathbf{X}| A=1),\ldots,\zeta_{1k}(\mathbf{X}| A=1)\}'$,  
and $\boldsymbol{\beta}_1(\mathbf{X}| A=1)=\{\beta_{11}(\mathbf{X}| A=1),\ldots,\beta_{1k}(\mathbf{X}| A=1)\}'$, with  
$\zeta_{1j}(\mathbf{X}| A=1)=\text{cov}\{E(D|\mathbf{X},A=1),\mathbf{X}_j| A=1\}$ and  
$\beta_{1j}(\mathbf{X}| A=1)=\text{cov}\{\tilde{S}_1(\mathbf{X};\tilde{\boldsymbol{\alpha}}_1),\mathbf{X}_j| A=1\}$ for $j=1,\ldots,k$.  
Further details are provided in Appendix \ref{sup-c1}.

When the PS model for the continuous exposure is correctly specified, 
$\tilde{S}_1(\mathbf{X};\tilde{\boldsymbol{\alpha}}_1)=E(D|\mathbf{X},A=1)$. In this case 
$\rho_1=1$, $\boldsymbol{\zeta}_1(\mathbf{X}| A=1)=\boldsymbol{\beta}_1(\mathbf{X}| A=1)$, and 
$\text{var}\{E(D|\mathbf{X},A=1)\}=\text{var}\{\tilde{S}_1(\mathbf{X};\tilde{\boldsymbol{\alpha}}_1)\}$. 
Substituting these equalities into (\ref{equa-4.2}) eliminates the second term, yielding 
$\gamma_{11}^\ast=\theta_{11}$. Otherwise, when $\tilde{S}_1(\mathbf{X};\tilde{\boldsymbol{\alpha}}_1)$ is misspecified, the asymptotic bias of $\hat{\gamma}_{11}$ depends on the covariance structure among $E(D| \mathbf{X},A=1)$, $\mathbf{X}$ given $A=1$, and $\tilde{S}_1(\mathbf{X};\tilde{\boldsymbol{\alpha}}_1)$.

\subsection{Biased Estimator in Stage II}
We now derive the limiting value of the exposure effect at the reference dose compared to no exposure. Let $\tilde{S}_2(\mathbf{X};\tilde{\boldsymbol{\alpha}}_2)$ be the Stage II PS under misspecification, and then fit the regression model by treating  $\tilde{\gamma}_{11}A(D-c)$ as an offset term in
$$    E\left\{Y| D,A,\tilde{S}_2(\mathbf{X};\tilde{\boldsymbol{\alpha}}_2);\tilde{\boldsymbol{\gamma}}_1,\tilde{\boldsymbol{\gamma}}_2\right\}=\tilde{\gamma}_{20}+\tilde{\gamma}_{21}A+\tilde{\gamma}_{22}\tilde{S}_2(\mathbf{X};\tilde{\boldsymbol{\alpha}}_2)+\text{offset}\left\{\tilde{\gamma}_{11}A(D-c)\right\}$$
where $\tilde{\boldsymbol{\gamma}}_2=(\tilde{\gamma}_{21}, \tilde{\gamma}_{22},\tilde{\gamma}_{23})'$. The estimating equation for $\tilde{\boldsymbol{\gamma}}_2$ is 
\begin{equation}
\label{equa-4.3}
    E\left\{\mathcal{U}_{21}(D;\tilde{\boldsymbol{\gamma}}_1,\tilde{\boldsymbol{\phi}}_2)\right\}=\mathbf{0}
\end{equation}
where $$\mathcal{U}_{21}(D;\tilde{\boldsymbol{\gamma}}_1,\tilde{\boldsymbol{\phi}}_2)=\frac{\partial \mu_{21}(\tilde{\boldsymbol{\phi}}_2)}{\partial \tilde{\boldsymbol{\phi}}_2}\left[Y-\left\{\mu_{21}(\tilde{\boldsymbol{\phi}}_2)+\tilde{\gamma}_{11}A(D-c)\right\}\right]$$
with $\mu_{21}(\tilde{\boldsymbol{\phi}}_2)=\tilde{\gamma}_{20}+\tilde{\gamma}_{21}A+\tilde{\gamma}_{22}\tilde{S}_2(\mathbf{X};\tilde{\boldsymbol{\alpha}}_2)$ again; the expectation in (\ref{equa-4.3}) is taken with respect to the true distribution of $(Y,D,A,\mathbf{X})$.  Solving (\ref{equa-4.3}) with ${\gamma}_{11}^*$ given by (\ref{equa-4.2}) yields $\boldsymbol{\gamma}_2^*=({\gamma}_{20}^*,{\gamma}_{21}^*,{\gamma}_{22}^*)'$. Here we let $\rho_{21}=\text{corr}\bigl\{E(A|\mathbf{X}),\tilde{S}_1(\mathbf{X})\bigr\}$, $\rho_{22}=\text{corr}\bigl\{E(AD|\mathbf{X}),\tilde{S}_1(\mathbf{X})\bigr\}$,  $\boldsymbol{\zeta}_2(\mathbf{X})=\bigl\{\zeta_{21}(\mathbf{X}),...,\zeta_{2k}(\mathbf{X})\bigr\}'$, $\boldsymbol{\beta}_2(\mathbf{X}) = \left\{\beta_{21}(\mathbf{X}),...,\beta_{2k}(\mathbf{X})\right\}'$, and $\delta_1=\text{cov}\bigl\{E(A|\mathbf{X}),\tilde{S}_1(\mathbf{X})\bigr\}$ where $\zeta_{2j}(\mathbf{X})=\text{cov}\left\{E(A|\mathbf{X}),\mathbf{X}_j\right\}$ and $\beta_{2j}(\mathbf{X})=\text{cov}\bigl\{\tilde{S}_1(\mathbf{X}),\mathbf{X}_j\bigr\}$
  for $j=1,2,...,k$. Then ${\gamma}_{21}^*$ corresponding to the limiting value of the causal effect of the exposure at the reference dose $c$ compared to no exposure under misspecification is
 \begin{align}
 \label{equa-2.49}
    &\gamma_{21}^* = \theta_{11}+\frac{\boldsymbol{\theta}_2'\left[\boldsymbol{\zeta}_2(\mathbf{X})-\boldsymbol{\beta}_2(\mathbf{X})\rho_{21}\sqrt{\frac{\text{var}\left\{E(A|\mathbf{X})\right\}}{\text{var}\{\tilde{S}_2(\mathbf{X};\tilde{\boldsymbol{\alpha}}_2)\}}}\right] }{\text{var}(A)-\text{var}\left\{E(A|\mathbf{X})\right\}\rho_{21}^2}\nonumber \\
    &+\frac{(\theta_{12}-{\gamma}_{11}^*)\left[\text{cov}(AD,A)-cE\left\{\text{var}(A|\mathbf{X})\right\}-\delta_1\rho_{22}\sqrt{\text{var}\left\{E(A|\mathbf{X})\right\}\text{var}\left\{E(AD|\mathbf{X})\right\}}\right]}{\text{var}(A)-\text{var}\left\{E(A|\mathbf{X})\right\}\rho_{21}^2}.  
\end{align}
Further details are provided in Appendix \ref{sup-c2}.

Note that $\gamma_{21}^* = \theta_{12}$ when the second and third terms in (\ref{equa-2.49}) are zero. If $\tilde{S}_2(\mathbf{X};\tilde{\boldsymbol{\alpha}}_2)$ is correctly specified so that $\tilde{S}_2(\mathbf{X};\tilde{\boldsymbol{\alpha}}_2) = E(A\mid\mathbf{X})$, and the Stage~I estimator is consistent (implying ${\gamma}_{11}^* = \theta_{11}$), these terms vanish and $\gamma_{21}^* = \theta_{12}$. More generally, the asymptotic bias in the Stage~II estimator depends on the covariance and variance structure of $E(A\mid\mathbf{X})$, $\mathbf{X}$ and $\tilde{S}_2(\mathbf{X};\tilde{\boldsymbol{\alpha}}_2)$, as well as any misspecification of $\tilde{S}_1(\mathbf{X};\tilde{\boldsymbol{\alpha}}_1)$ in Stage~I. The bias decreases as $\tilde{S}_2(\mathbf{X};\tilde{\boldsymbol{\alpha}}_2)$ approaches $E(A\mid\mathbf{X})$ and as $\tilde{\gamma}_{11}$ approaches $\theta_{11}$.

 \section{Simulation Studies} 
 \label{sec5} 
\subsection{Data Generation and Simulation Design}

We consider two data generation processes.
\subsubsection{Data Generation Model 1} 
\label{sec5.1.1}
Let $\mathbf{X} = (X_1,\ldots,X_6)'$ follow a multivariate normal distribution with mean zero, unit variances and pairwise correlation $0.2$ with $\mathbf{X}_1=(X_{11},X_{12})'$, $\mathbf{X}_2=(X_{21},X_{22})'$, and $\mathbf{X}_3=(X_{31},X_{32})'$. The binary exposure $A|\mathbf{Z}_{2}$ is generated from (\ref{eq:status_model}) with $\bar{\mathbf{Z}}_2 = (1, \mathbf{X}_2', \mathbf{X}_3')'$, $\boldsymbol{\alpha}_{21}=\left(\log1.2,\log2\right)'$ and $\boldsymbol{\alpha}_{22}=\left(\log0.8,\log1.3\right)'$. We solve for $\alpha_{20}$ such that $P(A=1)=E_{\mathbf{Z}_2}\left\{P(A=1|\mathbf{Z}_2)\right\}=0.25,0.5$ or $0.75$ which give values $\alpha_{20}=-1.249$, $0.0008$ and $1.249$. For those exposed, the continuous exposure $D$ is generated from \eqref{eq:dose_model} with $\bar{\mathbf{Z}}_1 = (1, \mathbf{X}_1', \mathbf{X}_3')'$, $\alpha_{10}=0$, $\boldsymbol{\alpha}_{11}=(1,0.5)'$, $\boldsymbol{\alpha}_{12}=(0.6,0.8)'$ and a standard normal distributed error term. If $A_i=0$ for each individual $D_i$ undefined. Finally, we generate the response variable $Y$ from model (\ref{eq:outcome_mean}) with $\theta_0=0$, ${\theta}_{11}=4$, $\theta_{12}=0.5$, $\boldsymbol{\theta}_{21}=(0.6,0.8)'$, $\boldsymbol{\theta}_{22}=(0.8,0.4)'$, $\boldsymbol{\theta}_{23}=(0.3,0.7)'$, and a standard normal distributed error term.

\subsubsection{Data Generation Model 2} 
Under this data generation process, we assume a common confounder vector $\mathbf{X}$ for both the exposure status and continuous exposure (i.e., $\mathbf{Z}_1=\mathbf{Z}_2=\mathbf{X}$), and is therefore a simplified version of the data generation process described in Section \ref{sec5.1.1}. Let $\mathbf{X}=(X_1,X_2)'$ follow a bivariate normal distribution with mean zero, unit variances and correlation $0.2$. We generate $A$ based on \begin{equation} \label{eq:status2} \log \left\{\frac{P(A=1\mid \mathbf X)}{1-P(A=1\mid \mathbf X)}\right\}=\alpha_{20}+ \boldsymbol{\alpha}_{21}'\mathbf{X} \end{equation} with $\boldsymbol{\alpha}_{21}=(\log1.2,\log2)'$; we solve for $\alpha_{20}$ to give marginal probabilities of $A=1$ as $0.25,0.5$ and $0.75$, yielding $\alpha_{20}=-1.224$, $0.005$ and $1.236$. If $A=1$, the continuous exposure $D$ is generated from 
\begin{equation} \label{eq:dose2} D=\alpha_{10} + \boldsymbol{\alpha}_{11}'\mathbf{X}+W, \end{equation}
where $\alpha_{10}=0$, $\boldsymbol{\alpha}_{11}=(1,0.5)'$, $W\perp\mathbf{X}$, and $W \sim N(0,1)$. The second data generation model is the special case of (\ref{eq:outcome_mean})
\begin{equation}
\label{response2}
Y=\theta_0 +{\theta}_{11}A(D-c)+\theta_{12}A + \mathbf{X}'\boldsymbol{\theta}_{2}+ E,
\end{equation}
with $\theta_0=0$, ${\theta}_{11}=0.5$, $\theta_{12}=4$, $\boldsymbol{\theta}_2=(0.8,0.4)'$, and $E\sim N(0,1)$.



\subsection{Methods of Analysis}
\label{sec5.2}
\begin{table}[t]
\centering
\caption{Model misspecification scenarios considered in the simulation study.}
\label{tab:misspec}
\begin{threeparttable}
\small
\begin{tabular}{@{}lcc@{}}
\toprule
Scenario 
& Data Generation Model 1
& Data Generation Model 2 \\
\midrule
(i) Misspecified $S_1$
& omit $X_{12}$ from \eqref{eq:dose_model} &  omit $X_{2}$ from \eqref{eq:dose2}  \\
(ii) Misspecified $S_2$ 
& omit $X_{22}$ from \eqref{eq:status_model} & omit $X_{2}$ from \eqref{eq:status2}  \\
(iii) Misspecified $m_a$ 
& omit $X_{12}$ from \eqref{eq:outcome_mean} &  omit $X_{2}$ from \eqref{response2}  \\
(iv) Misspecified $S_1$ and $m_a$ 
&  omit $X_{12}$ from \eqref{eq:dose_model} + omit $X_{12}$ from \eqref{eq:outcome_mean} & omit $X_{2}$ from \eqref{eq:dose2} + omit $X_{2}$ from \eqref{response2}  \\
(v) Misspecified $S_2$ and $m_a$ 
& omit $X_{22}$ from \eqref{eq:status_model} + omit $X_{12}$ from \eqref{eq:outcome_mean}&  omit $X_{2}$ from \eqref{eq:status2} + omit $X_{2}$ from \eqref{response2}\\
\bottomrule
\end{tabular}
\begin{tablenotes}%
\scriptsize
  \item  $S_1$ represents the PS model for the continuous exposure $D$ conditional on $A=1$, given by $S_1= E(D|A=1, \mathbf{Z}_1)$; $S_2$ represents the PS model for the binary exposure $A$, given by $S_2 = E(A|\mathbf{Z}_2)$; $m_a$ represents the imputation models for $a=0,1$. 
\end{tablenotes}
\end{threeparttable}
\end{table}

For each data generation model, we generate $2000$ replicated datasets of size $n=1000$ and estimate the causal parameters using six methods: (i) a naive least-squares fit of the MSM (\ref{msm}); (ii) the correctly specified response model; (iii) two–part PS regression adjustment in a single stage; (iv) two-stage approach with PS regression in Stage II; (v) two-stage approach with IPW in Stage II; and (vi) two-stage approach with AIPW in Stage II.

Including covariates unrelated to the exposure but related to the outcome can reduce the variance of causal effect estimators without introducing bias \citep{brookhart2006variable}. Accordingly, we consider two propensity score specifications: (i) including only exposure-related covariates and (ii) including all outcome-related covariates. Under the second data generation model, these sets coincide, so the two specifications are identical.

To examine robustness, we consider five misspecification scenarios: (i) misspecified PS $S_1$ for the continuous exposure; (ii) misspecified PS $S_2$ for the binary exposure; (iii) misspecified imputation model $m_a$; (iv) misspecification of $S_1$ and $m_a$; and (v) misspecification of $S_2$ and $m_a$. In all cases we focus on the setting with minimal propensity scores $\mathbf{S} = (S_1(\mathbf{Z}_1),S_2(\mathbf{Z}_2))'$. Full details of how each model is misspecified are given in Table \ref{tab:misspec}.

\subsection{Finite Sample Performance} \begin{table}
\centering\small
\caption{Empirical results of estimators of causal effects with $P(A=1)=0.5$ under six methods of analysis based on correctly specified models.}
\label{table-1}
\begin{tabular}{llcccccccccccc}
\hline
&&& \multicolumn{4}{l}{ {Minimal PS: ${\mathbf{S}=\{S_1(\mathbf{Z}_1),S_2(\mathbf{Z}_2)\}'}$}} &\multicolumn{4}{l}{ {Expanded PS: ${\mathbf{S}=\{S_1(\mathbf{X}),S_2(\mathbf{X})\}'}$}} \\
\cline{4-11}
Method & Exposure & Effect & Ebias & ESE & RSE & ECP(\%) & Ebias & ESE & RSE & ECP(\%) \\
\hline
&&&\multicolumn{8}{c}{\em Data Generation Model 1}\\
 \multirow{2}{*}{Naive} & $D$ & $\psi_{11}$ & 0.760 & 0.036 & 0.035 & 0.0 & - & - & - & -\\ 
 & $A$& $\psi_{12}$ & 0.952 & 0.128 & 0.126 & 0.0 & - & - & - & -\\  [4pt]
 \multirow{2}{*}{DG model} & $D$& $\psi_{11}$ & -0.001 & 0.028 & 0.027 & 94.2 & - & - & - & -\\ 
 & $A$& $\psi_{12}$ & $<0.001$ & 0.067 & 0.068 & 95.2 & - & - & - & -\\  [4pt]
 \multirow{2}{*}{Two PS Reg} & $D$& $\psi_{11}$ & -0.002 & 0.034 & 0.034 & 94.3 & -0.002 & 0.033 &0.032 & 95.0\\ 
 & $A$& $\psi_{12}$ & $<0.001$ & 0.073 & 0.074 & 95.4 & $<0.001$& 0.067& 0.068 & 95.0\\  [4pt]
 \multirow{2}{*}{$\text{PS}_\text{I}+\text{PS}_\text{II}$} &$D$ & $\psi_{11}$ & -0.002 & 0.060 & 0.059 & 94.4 & -0.002 & 0.046 & 0.045 & 94.2\\ 
 & $A$ & $\psi_{12}$ & -0.002 &0.096  & 0.098 & 95.2 & $<0.001$&0.068 & 0.069& 95.6\\  [4pt]
 \multirow{2}{*}{$\text{PS}_\text{I}+\text{IPW}_\text{II}$} & $D$& $\psi_{11}$ & -0.002 & 0.060 & 0.059 &94.4  & -0.002 & 0.046& 0.045&94.2\\ 
 & $A$& $\psi_{12}$ &-0.002  &0.107  &0.107  &94.5  &$<0.001$  &0.082 &0.083 &95.5\\  [4pt]
 \multirow{2}{*}{$\text{PS}_\text{I}+\text{AIPW}_\text{II}$} & $D$& $\psi_{11}$ & -0.002 & 0.060 & 0.059 &94.4  & -0.002 &0.046& 0.045 &94.2\\ 
 & $A$& $\psi_{12}$ & 0.001 & 0.070 &0.071  &95.5  & $<0.001$ & 0.069 &0.070 &95.4\\ 
\hline
&&&\multicolumn{8}{c}{\em Data Generation Model 2}\\
 \multirow{2}{*}{Naive} & $D$ & $\psi_{11}$ &0.462  &0.034  &0.034  &0.0  &  -& -&  -& -\\ 
 & $A$ & $\psi_{12}$ & 0.486 &0.080  &0.080  &0.0  &  - & - & - & -\\  [4pt]
 \multirow{2}{*}{DG model} & $D$& $\psi_{11}$&0.001  &0.035  &0.034  &94.4  & - &- &-   &- \\ 
 & $A$ & $\psi_{12}$ & $<0.001$ &0.068  &0.068  &95.4  & - & - & - &- \\  [4pt]
 \multirow{2}{*}{Two PS Reg} & $D$& $\psi_{11}$ &0.001  &0.035  &0.037  & 95.4 &-& - & - &- \\ 
 &$A$ & $\psi_{12}$ & $<0.001$ &0.068  &0.068  &95.3  &-& - &-  &-\\  [4pt]
 \multirow{2}{*}{$\text{PS}_\text{I}+\text{PS}_\text{II}$} & $D$& $\psi_{11}$ & 0.002 & 0.046 &0.045  & 93.9 &- & - &  -&-\\ 
 &$A$& $\psi_{12}$ & $<0.001$ & 0.069 &0.069  & 95.1 &- &- & - &-\\  [4pt]
 \multirow{2}{*}{$\text{PS}_\text{I}+\text{IPW}_\text{II}$} & $D$& $\psi_{11}$ & 0.002 & 0.046 & 0.045 & 93.9 &-  &-  & - &- \\ 
 &$A$ & $\psi_{12}$ & $<0.001$ &0.072  &0.072  & 94.9 & - &- &- &- \\  [4pt]
 \multirow{2}{*}{$\text{PS}_\text{I}+\text{AIPW}_\text{II}$} &$D$ & $\psi_{11}$& 0.002 & 0.046 & 0.045 & 93.9&-  &-   & -  &-\\ 
 &$A$ & $\psi_{12}$ &$<0.001$  & 0.070 & 0.070 & 94.4 & - &- &- & -\\ 
\hline
\end{tabular}
\begin{tablenotes}
\scriptsize
\item DG model: Data generation model; Two PS Reg: Two-part PS regression adjustment; $\text{PS}_\text{I}+\text{PS}_\text{II}$: Using PS regression adjustment in Stage I and II; $\text{PS}_\text{I}+\text{IPW}_\text{II}$: Using PS regression adjustment in Stage I and IPW in Stage II; $\text{PS}_\text{I}+\text{AIPW}_\text{II}$: Using PS regression adjustment in Stage I and AIPW in Stage II.  
\end{tablenotes}
\end{table}

\begin{table}
\centering\small
\caption{Empirical results for estimators of causal effects with $P(A=1)=0.5$ for two-part PS regression adjustment and two-stage analysis under model misspecification.}
\label{table-2}
\begin{tabular}{llccccccccc}
\hline
  &  &   & \multicolumn{4}{c}{Misspecified $S_1$} & \multicolumn{4}{c}{Misspecified $S_2$}\\ \cline{4-11} 
 {Method} & {Exposure} &  {Effect} &Ebias & ESE & RSE & ECP(\%) & Ebias & ESE & RSE & ECP(\%)\\ 
\hline
&&&\multicolumn{8}{c}{\em Data Generation Model 1}\\
 \multirow{2}{*}{Two PS Reg} &$D$ & $\psi_{11}$  & 0.135 &0.039  &0.039  &6.9  & -0.002 &0.035  &0.036  &96.0 \\ 
 &$A$ & $\psi_{12}$ &0.034  &0.084  & 0.086 & 93.6& 0.245 & 0.074 &0.075  &9.7   \\   [4pt]
 \multirow{2}{*}{$\text{PS}_\text{I}$+$\text{PS}_\text{II}$} & $D$& $\psi_{11}$  & 0.348 & 0.062 & 0.062 &0.1 & -0.002 & 0.060 &0.059  &94.4  \\ 
 &$A$ & $\psi_{12}$& 0.089 & 0.093 & 0.094 &84.1 & 0.331 &0.095  &0.099  & 7.5   \\   [4pt]
 \multirow{2}{*}{$\text{PS}_\text{I}$+$\text{IPW}_\text{II}$} & $D$& $\psi_{11}$  & 0.348 &0.062  &0.062  & 0.1& -0.002 & 0.060 & 0.059 &94.4  \\ 
 & $A$& $\psi_{12}$  &0.089  & 0.103 &0.102  &84.3 & 0.330 & 0.097 &0.100  &7.8  \\ \hline
&&&\multicolumn{8}{c}{\em Data Generation Model 2}\\
 \multirow{2}{*}{Two PS Reg} & $D$& $\psi_{11}$  &-0.053  &0.045  &0.044  & 75.9& 0.002 & 0.046 & 0.045 &93.9  \\ 
 & $A$& $\psi_{12}$  &-0.016  & 0.073 &0.072  & 94.3& 0.208 & 0.070 & 0.070 &15.8  \\ [4pt]
 \multirow{2}{*}{$\text{PS}_\text{I}$+$\text{PS}_\text{II}$} & $D$& $\psi_{11}$ &0.144  &0.044  &0.043  &8.4 &0.002  &0.046  &0.045  &93.9   \\  
 & $A$& $\psi_{12}$ &0.044  &0.068  & 0.068 &90.5 & 0.241 & 0.069 & 0.069 &6.9   \\  [4pt]
 \multirow{2}{*}{$\text{PS}_\text{I}$+$\text{IPW}_\text{II}$} &$D$ & $\psi_{11}$  &0.144  &0.044  &0.043  &8.4& 0.002 &0.046  & 0.045 &93.9   \\ 
 & $A$& $\psi_{12}$ &0.043  &0.072  &0.071  &90.3 & 0.241 &0.070  &0.069  &6.8  \\   
\hline
\end{tabular}

\begin{tablenotes}%
\scriptsize
\item $S_1$ represents the PS model for the continuous exposure $D$ conditioned on $A=1$, given by $S_1= E(D|A=1, \mathbf{Z}_1)$; $S_2$ represents the PS model for the binary exposure $A$, given by $S_2 = E(A|\mathbf{Z}_2)$.
\end{tablenotes}

\end{table}

\begin{table}
\centering\small
\caption{Empirical results for estimators of effects for both parts with $P(A=1)=0.5$ for applying PS regression adjustment in Stage I and AIPW in Stage II under model misspecification.}
\label{table-3}
\begin{tabular}{llccccc}
\hline
{Misspecified Model} & {Exposure} & Effect & Ebias & ESE & RSE & ECP(\%)\\  
\hline
&&&\multicolumn{4}{c}{\em Data Generation Model 1}\\
 \multirow{2}{*}{$S_1$} &$D$ & $\psi_{11}$ & 0.348 & 0.063 & 0.062 & 0.1 \\ 
 &$A$ & $\psi_{12}$ & 0.091 & 0.077 & 0.077 & 79.0 \\  [4pt]
 \multirow{2}{*}{$S_2$} &$D$ & $\psi_{11}$ & -0.002 & 0.060 & 0.059 & 94.4 \\ 
 &$A$ & $\psi_{12}$ & -0.015 & 0.071 & 0.072 & 95.0 \\ 
 [4pt]
 \multirow{2}{*}{$m_a(\mathbf{X};\boldsymbol{\theta})$} & $D$& $\psi_{11}$ & -0.002 & 0.060 & 0.059 & 94.4 \\ 
 & $A$& $\psi_{12}$ & -0.003 & 0.071 & 0.074 & 95.7 \\ [4pt]
  \multirow{2}{*}{$S_1$ + $m_a(\mathbf{X};\boldsymbol{\theta})$} &$D$ & $\psi_{11}$ &0.348& 0.062 & 0.062 & 0.1\\ 
 &$A$& $\psi_{12}$ & 0.093 & 0.078 & 0.079 & 79.1 \\ [4pt]
 \multirow{2}{*}{$S_2$ + $m_a(\mathbf{X};\boldsymbol{\theta})$} &$D$ & $\psi_{11}$ & -0.002 & 0.060 & 0.059 & 94.4 \\ 
 &$A$& $\psi_{12}$ & 0.225 & 0.069 & 0.071 & 11.0 \\ 
 \hline

&&&\multicolumn{4}{c}{\em Data Generation Model 2}\\
 \multirow{2}{*}{$S_1$} &$D$ & $\psi_{11}$ & 0.144 & 0.044 & 0.043 & 8.4 \\ 
 &$A$ & $\psi_{12}$ & 0.044 & 0.070 & 0.069 & 90.8 \\ [4pt]
 \multirow{2}{*}{$S_2$} &$D$ & $\psi_{11}$ & 0.002 & 0.046 & 0.045 & 93.9 \\ 
 &$A$ & $\psi_{12}$ & $<0.001$& 0.069 & 0.069 & 95.0 \\ [4pt]
 \multirow{2}{*}{$m_a(\mathbf{X};\boldsymbol{\theta})$} &$D$ & $\psi_{11}$ & 0.002 & 0.046 & 0.045 & 93.9 \\ 
 &$A$& $\psi_{12}$ &$<0.001$& 0.070 & 0.072 & 95.2 \\ [4pt]
 \multirow{2}{*}{$S_1$ +$m_a(\mathbf{X};\boldsymbol{\theta})$} &$D$ & $\psi_{11}$ & 0.144 & 0.044 & 0.043 & 8.4 \\ 
 &$A$ & $\psi_{12}$ & 0.044 & 0.070 & 0.070 & 91.0 \\ [4pt]
  \multirow{2}{*}{$S_2$ + $m_a(\mathbf{X};\boldsymbol{\theta})$} & $D$& $\psi_{11}$ & 0.002 & 0.046 & 0.045 & 93.9 \\ 
 &$A$ & $\psi_{12}$ & 0.241 & 0.070 & 0.069 & 6.8 \\ 
 \hline
\end{tabular}
\begin{tablenotes}%
\scriptsize
  \item  $S_1$ represents the PS model for the continuous exposure $D$ conditioned on $A=1$, given by $S_1= E(D|A=1, \mathbf{Z}_1)$; $S_2$ represents the PS model for the binary exposure $A$, given by $S_2 = E(A|\mathbf{Z}_2)$; $m_a(\mathbf{X};\boldsymbol{\theta})$ represents the imputation models for $a=0,1$. 
\end{tablenotes}
\end{table}

Table \ref{table-1} presents the empirical bias (EBias), empirical standard error (ESE), average robust standard error (RSE), and empirical coverage probability (ECP) for estimators with $P(A=1)=0.5$ under correctly specified models. The naive method yields poor estimates for both parts across both data-generation models. In contrast, all other methods show small biases and ECPs close to the nominal 95\% level. Including all covariates in the PS models reduces empirical variance without inflating bias, in line with the variance reduction principle of \citep{brookhart2006variable}. The ESEs closely match the mean RSE, supporting the validity of the sandwich variance estimator in Section~\ref{sec-variance}.

Table \ref{table-2} reports empirical results under misspecified $S_1$ and $S_2$ for the two-part PS regression adjustment, PS regression in both stages, and PS regression in Stage I with IPW in Stage II with $P(A=1)=0.5$. Misspecification of $S_1$ induces bias in the Stage I estimator and consequently poor Stage II performance. Similarly, misspecification of $S_2$ yields large empirical bias in the Stage II estimator.

Table~\ref{table-3} summarizes the performance of the Stage~II AIPW estimators under model misspecification with $P(A=1)=0.5$. When either the imputation model or the $S_2$ model is misspecified, the estimator remains consistent by double robustness. In contrast, misspecified $S_1$ induces bias in both stages, and simultaneously misspecifying the imputation and $S_2$ models leads to biased Stage~II estimates and loss of nominal coverage.

Tables \ref{table-appendix1}–\ref{table-appendix3} and \ref{table-appendix4}–\ref{table-appendix6} in Appendix \ref{sup-d} present results for $P(A=1) = 0.25$ and $0.75$ respectively, and show patterns consistent with those for $P(A=1) = 0.5$.

\section{Application to the prenatal alcohol exposure study}
\label{sec6} 

\subsection{Description of the Detroit Longitudinal Cohort}

The Detroit Longitudinal Cohort is a prospective study of 480 pregnant African-American women from inner-city Detroit and their children, followed from birth to age 19 to investigate the effects of PAE \citep{jacobson2004maternal,jacobson2002validity}. Prenatal alcohol intake was assessed at each visit using a timeline follow-back interview \citep{jacobson2002validity} and summarized as average daily consumption over pregnancy; let $T$ denote the average ounces of absolute alcohol (AA) per day. We treat PAE as a semi-continuous exposure: $A = I(T>0)$ is a binary indicator of any drinking (16.2\% of mothers reported no drinking during pregnancy), and $D = \log T$ is the log daily dose among drinkers. The reference dose is set to $-2.31$, the sample mean of $D$ among exposed mothers. The outcome is the Full Scale IQ score of the Wechsler Intelligence Scales for Children, 3rd edition (WISC-III) \cite{wechsler1991wisc}, measured at age 7 years for 377 children. Among children with observed outcomes, the mean (SD) Full Scale IQ score was 84.22 (12.31).

We adjust for baseline maternal sociodemographic characteristics, smoking and other substance use during pregnancy, reproductive history, gestational age at screening, and measures of the home environment and maternal cognition (listed in Table \ref{table-appendix7} in Appendix \ref{sup-e1}). To address missingness, we generate 20 multiply imputed datasets using an imputation model including all variables; further details are given in Appendix \ref{sup-E2}. Point estimates and standard errors are combined using Rubin’s rules under a missing-at-random assumption \citep{rubin1976inference,little2019statistical}.

\subsection{Application and Findings}

\begin{table}[!t]
\centering\small
\caption{Estimated effects of drinking status and log prenatal absolute alcohol consumption per day}
\label{table-5}
\begin{tabular}{lcccc}
\hline 
  &  \multicolumn{2}{c}{$D | A=1$}
  &  \multicolumn{2}{c}{$A$} \\
  \cline{2-5}
 Method & Estimate & Standard error & Estimate & Standard error \\
 \hline
 Covariate adjustment & -0.125 & 0.515 & 0.253 & 1.655 \\
 Two–PS regression & -0.189 & 0.572 & -0.128 & 1.621 \\
 $\text{PS}_\text{I} + \text{PS}_\text{II}$ & -0.011 & 0.518 & -0.013 & 1.634 \\
 $\text{PS}_\text{I} + \text{IPW}_\text{II}$ & -- & -- & -0.268 & 2.128 \\
 $\text{PS}_\text{I} + \text{AIPW}_\text{II}$ & -- & -- & -0.176 & 1.965 \\
\hline
\end{tabular}
\begin{tablenotes}
\scriptsize
\item Two–PS regression: two-dimensional PS regression adjustment; 
$\text{PS}_\text{I} + \text{PS}_\text{II}$: PS regression adjustment in Stages~I and II; 
$\text{PS}_\text{I} + \text{IPW}_\text{II}$: PS regression in Stage~I and IPW in Stage~II;
$\text{PS}_\text{I} + \text{AIPW}_\text{II}$: PS regression in Stage~I and AIPW in Stage~II.
\end{tablenotes}
\end{table}

We estimate the causal effects of prenatal drinking status and continuous dose using five approaches: (i) conventional covariate regression adjustment; (ii) two–part PS regression adjustment; (iii) a two-stage analysis with PS regression adjustment in Stage~II; (iv) a two-stage analysis with IPW in Stage~II; and (v) a two-stage analysis with AIPW in Stage~II. Implementation details are provided in Appendix \ref{sup-e3}.

Appendix \ref{sup-e1} summarizes the fitted models for the continuous and binary exposure components, $E(D | A=1,\mathbf{X})$ and $P(A=1 | \mathbf{X})$. Both models include all candidate covariates to improve precision of causal effect estimates \citep{brookhart2006variable}. 

Table~\ref{table-5} reports the estimated causal effects under the five approaches. For all PS–based methods, the estimated effects of both $\log(\text{AA/day})$ among the exposed and drinking status are negative, with the magnitude of the drinking status effect larger than that of the dose effect. By contrast, the conventional covariate adjustment model yields a small positive estimate (0.253) for drinking status and a slightly negative estimate ($-0.125$) for log dose. Given the reliance of covariate adjustment on correct specification of the outcome model \citep{greenland1999causal,brookhart2006variable}, these estimates may be more vulnerable to bias than those from the PS–based procedures.

For the log-dose effect among drinkers, PS–based point estimates are small and similar across methods, with standard errors around 0.5. For the drinking status effect, the PS–based estimates range from approximately $-0.013$ to $-0.268$, whereas the conventional covariate adjustment estimate is slightly positive. The two-stage AIPW estimator provides a negative point estimate ($-0.176$) and benefits from its doubly robust property, but, as in our simulations, weighting-based methods yield larger standard errors than regression-based approaches (approximately 2.0 versus 1.6). Taken together, the PS–based results are compatible with modest harmful causal effects of both increased prenatal alcohol dose among drinkers and drinking status at the reference dose on WISC-III distractibility scores, although none of the estimated effects is statistically significant at the 5\% level.

\subsection{Further Considerations}

To reduce the variance of the IPW and AIPW estimators, we considered trimming weights to limit the influence of extreme values. However, weight trimming is an ad hoc procedure and it can be difficult to obtain valid variance estimates, particularly for deriving asymptotic standard errors \citep{crump2009dealing}. In addition, when applying Rubin’s Rules to combine estimates across imputations, we found that the between-imputation variance for Stage~II IPW and AIPW was substantially larger than that observed for other approaches. One potential strategy to improve stability is to average propensity scores across multiple imputed datasets, with the aim of reducing the impact of extreme scores \citep{seaman2013review}. Motivated by this idea, we explored the use of averaged propensity scores within the Stage~II IPW and AIPW procedures. Although this reduced the between-imputation variance in our analyses, concerns have been raised about the validity of such averaging. In particular, averaging propensity scores across imputations may not yield valid balancing scores and can introduce bias, especially for weighting-based estimators \citep{leyrat2019propensity}. More recent work further suggests that estimating propensity scores within each imputed dataset is more reliable for achieving adequate covariate balance and producing stable estimates \citep{nguyen2024multiple}. We therefore proceed with the conventional within-imputation strategy: propensity scores are estimated separately within each imputed dataset, point estimates are obtained within each dataset, and Rubin’s Rules are used to compute the overall variance.

The analyses reported here also differ from those in other studies of prenatal alcohol exposure (PAE) using data from the Detroit cohort, reflecting differences in the outcome definition, propensity score construction, analytic strategy, and handling of missing data. With respect to the outcome, some work has used a composite cognitive score derived from a second-order confirmatory factor analysis \citep{li2023use,jacobson2024dose} and the WISC-III Freedom from Distractibility Index measured at age 7 \citep{akkaya2021propensity}, whereas we focus on the WISC-III Full Scale IQ measured at age 7. In terms of modelling the semi-continuous exposure, some approaches treat exposure as a single construct and adjust for confounding using a single two-part propensity score \citep{akkaya2021propensity}. In contrast, our framework decomposes exposure into status and dose components and carries out a two-stage analysis that adjusts for separate propensity score models at each stage. We also use a slightly different set of baseline confounders in the propensity score models. In addition, our two-part formulation allows us to model the log dose among exposed individuals directly, avoiding the need to add an arbitrary constant prior to log transformation.


In the Detroit cohort, we find insufficient evidence to support the need for a two-part exposure representation, and the corresponding null model yields significant evidence of an effect of PAE on child cognition, consistent with previous findings \citep{akkaya2021propensity}. Finally, with respect to missing data, we use multiple imputation to address missing values in baseline confounders \citep{li2023use}, whereas complete-case analysis has also been used in this setting \citep{akkaya2021propensity}. Our primary objective here is to illustrate the proposed methodological framework using data from the Detroit cohort; for a more comprehensive investigation of PAE effects, we refer readers to \citep{jacobson2002validity,jacobson2024dose}.

\section{Discussion}
\label{sec7}

We propose a causal framework for evaluating the effects of a semi-continuous exposure using a two-stage estimation. In Stage~I, propensity score regression adjustment targets the dose-response effect among exposed individuals, while Stage~II targets the causal reference-dose effect compared to no exposure using a doubly robust AIPW estimator. We establish large-sample properties for the resulting estimators, characterize their limiting values under propensity score misspecification, and assess finite-sample performance in simulations. Taken together, these developments provide an interpretable framework that separates the consequences of any exposure from those of increasing dose, while allowing doubly robust estimation of the status effect.

A key limitation of the current work is that Stage~I does not incorporate doubly robust estimation of the continuous exposure effect. Doubly robust approaches for continuous treatments based on kernel smoothing and flexible generalized propensity score models have been proposed and could be used to strengthen Stage~I and reduce sensitivity to model misspecification \citep{kennedy2017non,zhao2020propensity,zhu2015boosting}. More broadly, adopting more flexible semiparametric modeling strategies for the exposure process, as in \citet{stringer2024semi}, may help reduce sensitivity to parametric assumptions in Stage~I. When Stage~I yields an inconsistent estimate, consistency in Stage~II generally cannot be recovered, highlighting the importance of robust modelling strategies in the first stage. Developing approaches that mitigate the impact of Stage~I misspecification on Stage~II estimation remains an important direction for future research.

In our application, children’s cognition is assessed through multidimensional neurological outcomes in several U.S.\ longitudinal cohort studies, but we focused on a single outcome (WISC-III) from the Detroit cohort. Related work has developed methods for analysing data from multiple cohort studies while accounting for correlated outcomes, including hierarchical meta-analysis and multi-domain structural equation modelling with propensity score adjustment \citep{akkaya2022hierarchical,dang2023bayesian}. Extending the proposed semi-continuous two-stage framework to accommodate multivariate neurological outcomes, and to settings that pool information across cohorts, is a natural next step.

Finally, the Detroit study contains missing data in baseline covariates. An additional extension would be to develop methods for evaluating effects of a semi-continuous exposure in the presence of missing confounders, combining flexible causal inference tools with principled missing data methodology.
 
\section*{Acknowledgements}
We thank Neil Dodge, Ph.D., for his assistance with data management support. This research was funded by grants to Sandra W. Jacobson and Joseph L. Jacobson from the National Institutes of Health/National Institute on Alcohol Abuse and Alcoholism (NIH/NIAAA; R01-AA025905) and the Lycaki-Young Fund from the State of Michigan. Richard J. Cook was supported by a grant from the Canadian Institutes for Health Research (PJT: 180551). Richard J. Cook is the Faculty of Mathematics Research Chair at the University of Waterloo. Yeying Zhu is supported by a Discovery grant from the Natural Sciences and Engineering Research Council of Canada (RGPIN-2017-04064). Louise Ryan was supported by the Australian Research Council Centre of Excellence for Mathematical and Statistical Frontiers (ACEMS) CE140100049. Data collection for the Detroit Longitudinal Study was supported by grants from NIH/NIAAA (R01-AA06966, R01-AA09524, and P50-AA07606) and NIH/ National Institute on Drug Abuse (R21- DA021034).

\section*{Conflict of interest}

The authors declare no potential conflict of interests.

\bibliographystyle{unsrtnat}
\bibliography{references}  


\appendix
\section{Proof of Balancing for Regression Adjustment with a Two-Part Propensity Score}\renewcommand{\theequation}{A.\arabic{equation}}
\setcounter{equation}{0}
\label{sup-A}

We consider the data generation model 
\begin{equation}
\label{equa-A1}
    Y = \theta_0 + \theta_{11}A(D-c) + \theta_{12}A 
    + \boldsymbol{\theta}_{21}'\mathbf{X}_{1}
    + \boldsymbol{\theta}_{22}'\mathbf{X}_{2}
    + \boldsymbol{\theta}_{23}'\mathbf{X}_{3}
    + E ,
\end{equation}
where $\boldsymbol{\theta}_2=(\boldsymbol{\theta}_{21}',\boldsymbol{\theta}_{22}',\boldsymbol{\theta}_{23}')'$, $\mathbf{X}=(\mathbf{X}_{1}',\mathbf{X}_{2}',\mathbf{X}_{3}')'$, and $E$ is an error term with mean zero and independent of $(A,D,\mathbf{X})$. 
Define the two-part propensity score vector
\[
\mathbf{S}(\mathbf{X};\boldsymbol{\alpha})
=\Bigl(S_1(\mathbf{Z}_1;\boldsymbol{\alpha}_1),\, S_2(\mathbf{Z}_2;\boldsymbol{\alpha}_2)\Bigr)',
\]
where
\[
S_1(\mathbf{Z}_1;\boldsymbol{\alpha}_1)=E(D\mid A=1,\mathbf{Z}_1;\boldsymbol{\alpha}_1),
\qquad
S_2(\mathbf{Z}_2;\boldsymbol{\alpha}_2)=P(A=1\mid \mathbf{Z}_2;\boldsymbol{\alpha}_2).
\]
Our objective is to show that regression adjustment on $\mathbf{S}(\mathbf{X};\boldsymbol{\alpha})$ suffices to identify the target causal effects.

\subsection{Dose-Response Effect among the Exposed}
Consider the effect of increasing the dose $D$ by one unit among exposed individuals ($A=1$). Under the data-generating model \eqref{equa-A1}, this contrast is
\begin{align}
&E\left\{Y \mid A=1, D=d+1,\mathbf{S}(\mathbf{X};\boldsymbol{\alpha})\right\}
- E\left\{Y \mid A=1, D=d,\mathbf{S}(\mathbf{X};\boldsymbol{\alpha})\right\} \nonumber \\
=& \left[\theta_0+\theta_{11}(d+1-c)+\theta_{12} 
   + \boldsymbol{\theta}_2' E\left\{\mathbf{X}\mid A=1,D=d+1,\mathbf{S}(\mathbf{X};\boldsymbol{\alpha})\right\}\right] \nonumber\\
&\quad - \left[\theta_0+\theta_{11}(d-c)+\theta_{12}
   + \boldsymbol{\theta}_2' E\left\{\mathbf{X}\mid A=1,D=d,\mathbf{S}(\mathbf{X};\boldsymbol{\alpha})\right\}\right] \nonumber\\
\label{equa-A2}
=& \theta_{11} 
   + \boldsymbol{\theta}_2' \left[ E\left\{\mathbf{X}\mid A=1,D=d+1,\mathbf{S}(\mathbf{X};\boldsymbol{\alpha})\right\} 
   - E\left\{\mathbf{X}\mid A=1,D=d,\mathbf{S}(\mathbf{X};\boldsymbol{\alpha})\right\} \right].
\end{align}

To recover $\theta_{11}$ the term in square brackets in \eqref{equa-A2} must equal zero, i.e.,
\begin{equation}
\label{equa-A3}
E\left\{\mathbf{X}\mid A=1,D=d+1,\mathbf{S}(\mathbf{X};\boldsymbol{\alpha})\right\}
= E\left\{\mathbf{X}\mid A=1,D=d,\mathbf{S}(\mathbf{X};\boldsymbol{\alpha})\right\}.
\end{equation}

To see why this holds, note that under the dose model for $D\mid(A=1,\mathbf{Z}_1)$ we can write
\[
D = S_1(\mathbf{Z}_1;\boldsymbol{\alpha}_1) + W,
\]
where $W\sim N(0,\sigma_W^2)$ and $W$ is independent of $\mathbf{X}$ given $(A=1,\mathbf{S}(\mathbf{X};\boldsymbol{\alpha}))$. Therefore,
\[
E\left\{\mathbf{X}\mid A=1,D,\mathbf{S}(\mathbf{X};\boldsymbol{\alpha})\right\}
=E\left\{\mathbf{X}\mid A=1,\mathbf{S}(\mathbf{X};\boldsymbol{\alpha})\right\},
\]
which is no longer a function of $D$. Hence \eqref{equa-A3} holds, and the contrast in \eqref{equa-A2} reduces to $\theta_{11}$.
This corresponds to the causal dose--response effect among exposed individuals, $\psi_{11}$, in the marginal structural model~(3) in the main text.

\subsection{Causal Effect of Reference-Dose Exposure versus No Exposure}
We next consider the effect of exposure at the reference dose $D=c$ compared with no exposure ($A=0$). Under the data-generating model \eqref{equa-A1},
\begin{align}
&E\left\{Y \mid A=1,D=c,\mathbf{S}(\mathbf{X};\boldsymbol{\alpha})\right\}  
- E\left\{Y \mid A=0,\mathbf{S}(\mathbf{X};\boldsymbol{\alpha})\right\} \nonumber\\
&\quad= \left[\theta_0+\theta_{11}(c-c)+\theta_{12}
+\boldsymbol{\theta}_2' E\left\{\mathbf{X}\mid A=1,D=c,\mathbf{S}(\mathbf{X};\boldsymbol{\alpha})\right\}\right] \nonumber\\
&\qquad - \left[\theta_0+\boldsymbol{\theta}_2' E\left\{\mathbf{X}\mid A=0,\mathbf{S}(\mathbf{X};\boldsymbol{\alpha})\right\}\right]\nonumber\\
\label{equa-A4}
&\quad= \theta_{12} + \boldsymbol{\theta}_2' \left[ E\left\{\mathbf{X}\mid A=1,D=c,\mathbf{S}(\mathbf{X};\boldsymbol{\alpha})\right\} 
- E\left\{\mathbf{X}\mid A=0,\mathbf{S}(\mathbf{X};\boldsymbol{\alpha})\right\} \right].
\end{align}

To isolate $\theta_{12}$ it suffices that
\begin{equation}
\label{equa-A5}
E\left\{\mathbf{X}\mid A=1,D=c,\mathbf{S}(\mathbf{X};\boldsymbol{\alpha})\right\}
= E\left\{\mathbf{X}\mid A=0,\mathbf{S}(\mathbf{X};\boldsymbol{\alpha})\right\}.
\end{equation}

Expanding both sides of \eqref{equa-A5} gives
\begin{align}
E\left\{\mathbf{X}\mid A=1,D=c,\mathbf{S}(\mathbf{X};\boldsymbol{\alpha})\right\}
&= E\left\{\mathbf{X}\mid D=c,\mathbf{S}(\mathbf{X};\boldsymbol{\alpha})\right\}
 =E\left\{\mathbf{X}\mid \mathbf{S}(\mathbf{X};\boldsymbol{\alpha})\right\}, \label{new.equation-A6L}\\[2mm]
E\left\{\mathbf{X}\mid A=0,\mathbf{S}(\mathbf{X};\boldsymbol{\alpha})\right\}
&=\frac{E\left\{\mathbf{X}\,I(A=0)\mid \mathbf{S}(\mathbf{X};\boldsymbol{\alpha})\right\}}{P\{A=0\mid \mathbf{S}(\mathbf{X};\boldsymbol{\alpha})\}}.  \label{new.equation-A6}
\end{align}
The first equality in \eqref{new.equation-A6L} holds since $D$ is observed only when $A=1$ and the second holds because $D=S_1(\mathbf{Z}_1;\boldsymbol{\alpha}_1)+W$ with $W\perp\mathbf{X}\mid \mathbf{S}(\mathbf{X};\boldsymbol{\alpha})$. The ratio in \eqref{new.equation-A6} follows from Bayes’ rule for conditional expectations.

To simplify \eqref{new.equation-A6}, rewrite the numerator and denominator as
\begin{align} 
E\left\{\mathbf{X}\,I(A=0)\mid \mathbf{S}(\mathbf{X};\boldsymbol{\alpha})\right\}
&= E\left[\mathbf{X}\,E\left\{I(A=0)\mid \mathbf{X},\mathbf{S}(\mathbf{X};\boldsymbol{\alpha})\right\}\mid \mathbf{S}(\mathbf{X};\boldsymbol{\alpha})\right] \nonumber\\
&= E\left\{\mathbf{X}\,P(A=0\mid \mathbf{X})\mid\mathbf{S}(\mathbf{X};\boldsymbol{\alpha})\right\}\nonumber\\
&= E\left[\mathbf{X}\,\{1-S_2(\mathbf{Z}_2;\boldsymbol{\alpha}_2)\}\mid \mathbf{S}(\mathbf{X};\boldsymbol{\alpha})\right]\nonumber\\
&= \{1-S_2(\mathbf{Z}_2;\boldsymbol{\alpha}_2)\}\,E\left\{\mathbf{X}\mid \mathbf{S}(\mathbf{X};\boldsymbol{\alpha})\right\},  \label{new.equation-A7}\\
P\{A=0\mid \mathbf{S}(\mathbf{X};\boldsymbol{\alpha})\}
&= E\left\{1-S_2(\mathbf{Z}_2;\boldsymbol{\alpha}_2)\mid \mathbf{S}(\mathbf{X};\boldsymbol{\alpha})\right\}
= 1-S_2(\mathbf{Z}_2;\boldsymbol{\alpha}_2). 
  \label{new.equation-A8}
\end{align}
The first line in \eqref{new.equation-A7} uses the law of iterated expectations and we use
$P(A=1\mid \mathbf{X})=S_2(\mathbf{Z}_2;\boldsymbol{\alpha}_2)$ so that $P(A=0\mid \mathbf{X})=1-S_2(\mathbf{Z}_2;\boldsymbol{\alpha}_2)$. Substituting \eqref{new.equation-A7} and \eqref{new.equation-A8} into \eqref{new.equation-A6} gives
\begin{equation}
E\left\{\mathbf{X}\mid A=0,\mathbf{S}(\mathbf{X};\boldsymbol{\alpha})\right\}
=E\left\{\mathbf{X}\mid \mathbf{S}(\mathbf{X};\boldsymbol{\alpha})\right\}.
\label{new.equation-A6R}
\end{equation}
Combining \eqref{new.equation-A6L} and \eqref{new.equation-A6R} yields \eqref{equa-A5}, so \eqref{equa-A4} reduces to $\theta_{12}$.
This coincides with $\psi_{12}$, the causal effect of exposure at the reference dose versus no exposure in the marginal structural model~(3) in the main text.

\section{Results on Large Sample Properties}
\label{sup-B}

\renewcommand{\theequation}{B.\arabic{equation}}
\setcounter{equation}{0}
\subsection{Proof of Theorem 1}
Under Assumptions 1 to 3, and the propensity score $S_1(\boldsymbol{Z}_1;\boldsymbol{\alpha}_1)$ is correctly specified. We aim to show 
$E_{\mathcal{D}_i}\bigl\{\mathbf{U}_{i11}(\mathcal{D}_i;\boldsymbol{\phi}_1)\bigr\}=\mathbf{0}.
$
Firstly, we take conditional expectation respect to $Y_i | D_{i},A_i, \mathbf{X}_{i1},\mathbf{X}_{i2}, \mathbf{X}_{i3}$ to obtain 
$
A_i {\partial \mu_{i 11}\left(\boldsymbol{\phi}_1\right)}/{\partial \boldsymbol{\gamma}_1}\left\{E\left(Y_i | D_i, A_i,\mathbf{X}_{i1},\mathbf{X}_{i2}, \mathbf{X}_{i3}\right)-\mu_{i 11}\left(\boldsymbol{\phi}_1\right)\right\}.
$
Then we take the expectation with respect to $\mathbf{X}_{i1} | D_i,A_i, \mathbf{X}_{i2}, \mathbf{X}_{i3}$ to obtain
$$
\int_{\mathbf{X}_{1}}A_i \frac{\partial \mu_{i 11}\left(\boldsymbol{\phi}_1\right)}{\partial \boldsymbol{\gamma}_1}\left\{E\left(Y_i | D_i, A_i,\mathbf{X}_{i1},\mathbf{X}_{i2}, \mathbf{X}_{i3}\right)-\mu_{i 11}\left(\boldsymbol{\phi}_1\right)\right\}\partial G\left(\mathbf{X}_{i1} |  D_i,A_i, \mathbf{X}_{i2}, \mathbf{X}_{i3}\right)
$$
where $ \partial G\left(\mathbf{X}_{i1} |  D_i,A_i, \mathbf{X}_{i2}, \mathbf{X}_{i3}\right)$ is given by
$$
\begin{aligned}
   \frac{\partial F(D_i|A_i,\mathbf{X}_{i1},\mathbf{X}_{i2}, \mathbf{X}_{i3})\partial G(\mathbf{X}_{i2}|A_i,\mathbf{X}_{i1}, \mathbf{X}_{i3}) }{\partial F(D_i,A_i,\mathbf{X}_{i2},\mathbf{X}_{i3})} \partial F(A_i|\mathbf{X}_{i1}, \mathbf{X}_{i3}) \partial G(\mathbf{X}_{i3}|\mathbf{X}_{i1}) \partial G(\mathbf{X}_{i1}).
\end{aligned}$$
Upon substitution
\begin{align}
& \int_{\mathbf{X}_{1}}A_i \frac{\partial \mu_{i 11}\left(\boldsymbol{\phi}_1\right)}{\partial \boldsymbol{\gamma}_1}\left\{E\left(Y_i | D_i, A_i,\mathbf{X}_{i1},\mathbf{X}_{i2}, \mathbf{X}_{i3}\right)-\mu_{i 11}\left(\boldsymbol{\phi}_1\right)\right\}\nonumber\\
\label{extra-equa-B6}
& \cdot \frac{\partial F(D_i|A_i,\mathbf{X}_{i1},\mathbf{X}_{i2}, \mathbf{X}_{i3})\partial G(\mathbf{X}_{i2}|A_i,\mathbf{X}_{i1}, \mathbf{X}_{i3})\partial F(A_i|\mathbf{X}_{i1}, \mathbf{X}_{i3}) \partial G(\mathbf{X}_{i3}|\mathbf{X}_{i1}) \partial G(\mathbf{X}_{i1})}{\partial F(D_i,A_i,\mathbf{X}_{i2},\mathbf{X}_{i3})} \\
& =\int_{\mathbf{X}_{1}}A_i \frac{\partial \mu_{i 11}\left(\boldsymbol{\phi}_1\right)}{\partial \boldsymbol{\gamma}_1}\frac{\partial F(D_i|A_i,\mathbf{X}_{i1},\mathbf{X}_{i2}, \mathbf{X}_{i3})\partial G(\mathbf{X}_{i2}|A_i,\mathbf{X}_{i1}, \mathbf{X}_{i3})\partial F(A_i|\mathbf{X}_{i1}, \mathbf{X}_{i3}) \partial G(\mathbf{X}_{i3}|\mathbf{X}_{i1})}{\partial F(D_i,A_i,\mathbf{X}_{i2},\mathbf{X}_{i3})} \nonumber\\
\label{extra-equa-B7}
& \cdot \left\{E\left(Y_i | D_i, A_i,\mathbf{X}_{i1},\mathbf{X}_{i2}, \mathbf{X}_{i3}\right)-\mu_{i 11}\left(\boldsymbol{\phi}_1\right)\right\}  \partial G(\mathbf{X}_{i1}) \\
& =\int_{\mathbf{X}_{1}}A_i \frac{\partial \mu_{i 11}\left(\boldsymbol{\phi}_1\right)}{\partial \boldsymbol{\gamma}_1}\frac{\partial F(D_i|A_i,\mathbf{X}_{i1},\mathbf{X}_{i2}, \mathbf{X}_{i3})\partial G(\mathbf{X}_{i2}|A_i,\mathbf{X}_{i1}, \mathbf{X}_{i3})\partial F(A_i|\mathbf{X}_{i1}, \mathbf{X}_{i3}) \partial G(\mathbf{X}_{i3}|\mathbf{X}_{i1})}{\partial F(D_i,A_i,\mathbf{X}_{i2},\mathbf{X}_{i3})} \nonumber\\
\label{extra-equa-B8}
& \cdot \bigl[ E\left(Y_i | D_i,\mathbf{X}_{i2}, \mathbf{X}_{i3}\right)-
E\left\{Y_i | A_i,D_i, S_1\left(\mathbf{Z}_{i 1} ; \boldsymbol{\alpha}_1\right);\boldsymbol{\gamma}_1\right\}
\bigr] \\
& = \mathbf{0}. \nonumber
\end{align}
Note (\ref{extra-equa-B7}) is an algebraic re-expression of (\ref{extra-equa-B6}). The transition from (\ref{extra-equa-B7}) to (\ref{extra-equa-B8}) follows from the equality $\mu_{i 11}\left(\boldsymbol{\phi}_1\right)=E\left\{Y_i | A_i,D_i, S_1\left(\mathbf{Z}_{i1} ; \boldsymbol{\alpha}_1\right);\boldsymbol{\gamma}_1\right\}$, together with the law of iterated expectations, under which integrating out $\mathbf{X}_{i1}$ removes its conditional dependence from the first expectation term inside the curly brackets of (\ref{extra-equa-B7}). Then, under the assumption that the generalized propensity score model $S_1\left(\mathbf{Z}_{i1} ; \boldsymbol{\alpha}_1\right)$ is correctly specified, conditioning on $S_1\left(\mathbf{Z}_{i1} ; \boldsymbol{\alpha}_1\right)$ controls for confounding by $\mathbf{X}_{i1}$ and $\mathbf{X}_{i3}$ in estimating the causal effect of $D_i$ on $Y_i$ among the exposed. As a result, two conditional expectations in the brackets in (\ref{extra-equa-B8}) are equal yielding the desired result. Therefore, the estimator $\hat{\gamma}_{11}$ is a consistent estimator of $\psi_{11}$ in the MSM (3) in the main text under Assumptions 1-3. 

\subsection{Proof of Theorem 2}
\label{sub-b2}
Under Assumptions 1 to 4, and assuming $\hat{\gamma}_{11}$ obtained from Stage I is consistent for $\psi_{11}$, we show the AIPW estimating function is an unbiased estimating function to prove consistency when at least one of the propensity score model $S_2\left(\mathbf{Z}_2; \boldsymbol{\alpha}_2\right)$ and the imputation model $m_a(\boldsymbol{\theta})$ are correctly specified. Proving consistency is equivalent to showing 
$$E_{\mathcal{D}_i}\left\{\bar{\bar{\mathbf{U}}}_{i21}(\mathcal{D}_i;\boldsymbol{\gamma}_{1},\boldsymbol{\phi}_{2})\right\}=\mathbf{0}.$$ 
We first rewrite the estimating function $\bar{\bar{\mathbf{U}}}_{i21}(\mathcal{D}_i;\boldsymbol{\gamma}_{1},\boldsymbol{\phi}_{2})$ as
\begin{align}
\label{extra-equa-newB9}
&\sum_{a=0}^1 w_i\left(a ; \boldsymbol{\alpha}_2\right) \frac{\partial \mu_{12}\left(a ; \boldsymbol{\gamma}_2\right)}{\partial \gamma_2}\left[Y_i-\left\{\mu_{12}\left(a ; \gamma_2\right)+\gamma_{11} A_i\left(D_i-c\right)\right\}\right]-\left\{w_i\left(a ; \boldsymbol{\alpha}_2\right)-1\right\} \boldsymbol{g}_i\left(a ; \boldsymbol{\theta}, \boldsymbol{\gamma}_2\right)\\
\label{extra-equa-newB10}
=&\sum_{a=0}^1 w_i\left(a ; \boldsymbol{\alpha}_2\right) \frac{\partial \mu_{12}\left(a ; \boldsymbol{\gamma}_2\right)}{\partial \gamma_2}\left\{\bar{Y}_i- \mu_{12}\left(a ; \gamma_2\right)  \right\}-\left\{w_i\left(a ; \boldsymbol{\alpha}_2\right)-1\right\} \boldsymbol{g}_i\left(a ; \boldsymbol{\theta}, \boldsymbol{\gamma}_2\right)\end{align}
where $\bar{Y}_i =\theta_0+\theta_{11} A_i+\boldsymbol{\theta}_{21}'\mathbf{X}_{1}+\boldsymbol{\theta}_{22}'\mathbf{X}_{2}+\boldsymbol{\theta}_{23}'\mathbf{X}_{3}$. Under the assumption that $\hat{\gamma}_{11}$ is a consistent estimator of $\psi_{11}$ in the MSM, where $\psi_{12}=\theta_{12}$ in the data-generating model, the term $\hat{\gamma}_{11} A_i(D_i - c)$ captures the component of $Y_i$ that is causally attributable to the continuous exposure. Subtracting this term from $Y_i$ in (\ref{extra-equa-newB9}) removes the causal dose-response effect, isolating the portion of the linear predictor that depends only on the binary exposure and the covariates. This yields (\ref{extra-equa-newB10}). The problem is therefore reduced to showing that
$
E_{\bar{Y}_{i}, A_i, \mathbf{X}_{i}}\left\{\bar{\bar{\mathbf{U}}}_{i21}(\mathcal{D}_i;\boldsymbol{\gamma}_{1},\boldsymbol{\phi}_{2})\right\} = \mathbf{0}
$
since (\ref{extra-equa-newB10}) no longer depends on the continuous exposure $D_i$, and $\bar{Y}_i$ captures only the component of the outcome explained by $A_i$ and $\mathbf{X}_i$. 

Here, we consider three scenarios separately: 1) only the imputation model is correctly specified, 2) only the propensity score model is correctly specified, 3) both are correctly specified.
First, we take the conditional expectation of $\bar{\bar{\mathbf{U}}}_{i21}(\mathcal{D}_i;\boldsymbol{\gamma}_{1},\boldsymbol{\phi}_{2})$ with respect to $\bar{Y}_{i}|  A_i, \mathbf{X}_{i1},\mathbf{X}_{i2}, \mathbf{X}_{i3}$. Then we get
\begin{align}
\label{equa-B22}
\sum_{a=0}^1 w_i\left(a ; \boldsymbol{\alpha}_2\right) \frac{\partial \mu_{12}\left(a ; \boldsymbol{\gamma}_2\right)}{\partial \boldsymbol{\gamma}_2}\left\{E(\bar{Y}_{i}| A_i, \mathbf{X}_{i})- \mu_{12}\left(a ; \boldsymbol{\gamma}_2\right)  \right\}\nonumber\\
-\left\{w_i\left(a ; \boldsymbol{\alpha}_2\right)-1\right\}\frac{\partial \mu_{12}\left(a ; \boldsymbol{\gamma}_2\right)}{\partial \boldsymbol{\gamma}_2}\left\{m_{i a}(\boldsymbol{\theta})-\mu_{12}\left(a ; \boldsymbol{\gamma}_2\right)\right\}    
\end{align}
where $w_i\left(a ; \boldsymbol{\alpha}_2\right)= {I\left(A_i=a\right)}/\left[{S_2\left(\mathbf{Z}_{i 2} ; \boldsymbol{\alpha}_2\right)^a\left\{1-S_2\left(\mathbf{Z}_{i 2} ; \boldsymbol{\alpha}_2\right)\right\}^{1-a}}\right]
.$
The first term in (\ref{equa-B22}) can be represented as 
\begin{align}
  &  \sum_{a=0}^1 \frac{I\left(A_i=a\right)}{S_2\left(\mathbf{Z}_{i 2} ; \boldsymbol{\alpha}_2\right)^a\left\{1-S_2\left(\mathbf{Z}_{i 2} ; \boldsymbol{\alpha}_2\right)\right\}^{1-a}}\frac{\partial \mu_{12}\left(a ; \boldsymbol{\gamma}_2\right)}{\partial \boldsymbol{\gamma}_2}\left\{E(\bar{Y}_{i}| A_i, \mathbf{X}_{i})- \mu_{12}\left(a ; \boldsymbol{\gamma}_2\right)  \right\}\nonumber\\
 =& \sum_{a=0}^1 \frac{I\left(A_i=a\right)}{S_2\left(\mathbf{Z}_{i 2} ; \boldsymbol{\alpha}_2\right)^a\left\{1-S_2\left(\mathbf{Z}_{i 2} ; \boldsymbol{\alpha}_2\right)\right\}^{1-a}}\left(\begin{matrix}
     1\\
     a
 \end{matrix}\right)\left\{E(\bar{Y}_{i}| A_i, \mathbf{X}_{i})- \mu_{12}\left(a ; \boldsymbol{\gamma}_2\right)  \right\}\nonumber\\
 \label{equa-B24}
=&\frac{I(A_i=0)}{1-S_2\left(\mathbf{Z}_{i 2} ; \boldsymbol{\alpha}_2\right)}\left(
\begin{matrix}
1\\
0
\end{matrix}\right)\left\{E(\bar{Y}_{i}| A_i, \mathbf{X}_{i})- \mu_{12}\left(0 ; \boldsymbol{\gamma}_2\right) \right\}+\frac{I(A_i=1)}{S_2\left(\mathbf{Z}_{i 2} ; \boldsymbol{\alpha}_2\right)}\left(
\begin{matrix}
1\\
1
\end{matrix}\right)\left\{E(\bar{Y}_{i}| A_i, \mathbf{X}_{i})- \mu_{12}\left(1 ; \boldsymbol{\gamma}_2\right) \right\}\\
\label{equa-B25}
=&
\frac{(1-A_i)}{1-S_2\left(\mathbf{Z}_{i 2} ; \boldsymbol{\alpha}_2\right)}\left(
\begin{matrix}
1\\
0
\end{matrix}\right)\left\{E(\bar{Y}_{i}| A_i, \mathbf{X}_{i})-\mu_{12}(0;\boldsymbol{\gamma}_2)\right\} +\frac{A_i}{S_2\left(\mathbf{Z}_{i 2} ; \boldsymbol{\alpha}_2\right)}\left(
\begin{matrix}
1\\
1
\end{matrix}\right)\left\{E(\bar{Y}_{i}| A_i, \mathbf{X}_{i})-\mu_{12}(1;\boldsymbol{\gamma}_2)\right\}
\end{align}
Then we take the conditional expectation to (\ref{equa-B25}) with respect to $A_i|\mathbf{X}_{i}$ to obtain
\begin{align}
   & P(A_i=1|\mathbf{X}_{i})\biggl[\frac{(1-1)}{1-S_2\left(\mathbf{Z}_{i 2} ; \boldsymbol{\alpha}_2\right)}\left(
\begin{matrix}
1\\
0
\end{matrix}\right)\left\{E(\bar{Y}_{i}|A_i=1, \mathbf{X}_{i})-\mu_{12}(0;\boldsymbol{\gamma}_2)\right\}\nonumber\\
& +\frac{1}{S_2\left(\mathbf{Z}_{i 2} ; \boldsymbol{\alpha}_2\right)}\left(
\begin{matrix}
1\\
1
\end{matrix}\right)\left\{E(\bar{Y}_{i}|A_i=1, \mathbf{X}_{i})-\mu_{12}(1;\boldsymbol{\gamma}_2)\right\} \biggl]+\nonumber\\
& P(A_i=0|\mathbf{X}_{i})\biggl[\frac{(1-0)}{1-S_2\left(\mathbf{Z}_{i 2} ; \boldsymbol{\alpha}_2\right)}\left(
\begin{matrix}
1\\
0
\end{matrix}\right)\left\{E(\bar{Y}_{i} |A_i=0 , \mathbf{X}_{i})-\mu_{12}(0;\boldsymbol{\gamma}_2)\right\}\nonumber\\
\label{equa-B26}
& +\frac{0}{S_2\left(\mathbf{Z}_{i 2} ; \boldsymbol{\alpha}_2\right)}\left(
\begin{matrix}
1\\
1
\end{matrix}\right)\left\{E(\bar{Y}_{i}|A_i=0, \mathbf{X}_{i})-\mu_{12}(1;\boldsymbol{\gamma}_2)\right\}\biggl]\\
=&P(A_i=1|\mathbf{X}_{i})\frac{1}{S_2\left(\mathbf{Z}_{i 2} ; \boldsymbol{\alpha}_2\right)}\left(
\begin{matrix}
1\\
1
\end{matrix}\right)\left\{E(\bar{Y}_{i}|A_i=1,\mathbf{X}_{i})-\mu_{12}(1;\boldsymbol{\gamma}_2)\right\} \nonumber\\
\label{equa-B27}
& +P(A_i=0|\mathbf{X}_{i})\frac{1}{1-S_2\left(\mathbf{Z}_{i 2} ; \boldsymbol{\alpha}_2\right)}\left(
\begin{matrix}
1\\
0
\end{matrix}\right)\left\{E(\bar{Y}_{i}|A_i=0,\mathbf{X}_{i})-\mu_{12}(0;\boldsymbol{\gamma}_2)\right\}\\
=&P(A_i=1|\mathbf{Z}_{i1})\frac{1}{S_2\left(\mathbf{Z}_{i 2} ; \boldsymbol{\alpha}_2\right)}\left(
\begin{matrix}
1\\
1
\end{matrix}\right)\left[E(\bar{Y}_{i}|A_i=1,\mathbf{X}_{i})-\mu_{12}(1;\boldsymbol{\gamma}_2)\right] \nonumber\\ 
& +P(A_i=0|\mathbf{Z}_{i1})\frac{1}{1-S_2\left(\mathbf{Z}_{i 2} ; \boldsymbol{\alpha}_2\right)}\left(
\begin{matrix}
1\\
0
\end{matrix}\right)\left\{E(\bar{Y}_{i}|A_i=0,\mathbf{X}_{i})-\mu_{12}(0;\boldsymbol{\gamma}_2)\right\}\\
=&\frac{S_2\left(\mathbf{Z}_{i 2} ; \boldsymbol{\alpha}_2\right)}{S_2\left(\mathbf{Z}_{i 2} ; \boldsymbol{\alpha}_2\right)}\left(
\begin{matrix}
1\\
1
\end{matrix}\right)\left\{E(\bar{Y}_{i}|A_i=1,\mathbf{X}_{i})-\mu_{12}(1;\boldsymbol{\gamma}_2)\right\} \nonumber\\ 
& +\frac{1-S_2\left(\mathbf{Z}_{i 2} ; \boldsymbol{\alpha}_2\right)}{1-S_2\left(\mathbf{Z}_{i 2} ; \boldsymbol{\alpha}_2\right)}\left(
\begin{matrix}
1\\
0
\end{matrix}\right)\left\{E(\bar{Y}_{i}|A_i=0, \mathbf{X}_{i})-\mu_{12}(0;\boldsymbol{\gamma}_2)\right\}\\
=& \boldsymbol{0} \nonumber
\end{align}
Thus we only need to show the second term in (\ref{equa-B22}) is equal to $\boldsymbol{0}$ under the following three cases.

\subsubsection*{Case 1: Only the imputation model is correctly specified. \hspace{2pt}}
Let $\tilde{S}_2\left(\mathbf{X} ; \tilde{\boldsymbol{\alpha}}_2\right)$ denote the Stage II propensity score under misspecification, and the weight corresponding to the misspecified propensity score is denoted as $\tilde{w}_i\left(a ; \boldsymbol{\alpha}_2\right)$. The second component in (\ref{equa-B22}) is $$\sum_{a=0}^1\left(\tilde{w}_i\left(a ; \boldsymbol{\alpha}_2\right)-1\right)\frac{\partial \mu_{12}\left(a ; \boldsymbol{\gamma}_2\right)}{\partial \boldsymbol{\gamma}_2}\left\{m_{i a}(\boldsymbol{\theta})-\mu_{12}\left(a ; \boldsymbol{\gamma}_2\right)\right\}.$$ Since the imputation model is correctly specified, 
\begin{equation}
\label{equa-B23}
 m_{a}(\boldsymbol{\theta})=E(\bar{Y}_{i}| A=a, \mathbf{X}_{i1},\mathbf{X}_{i2}, \mathbf{X}_{i3})=\mu_{12}(a;\boldsymbol{\gamma}_2).
\end{equation}
Under (\ref{equa-B23}), (\ref{equa-B22}) equals 0. Therefore, we conclude that $\hat{\gamma}_{21}$ is consistent for $\psi_{12}$ in the MSM (3) if the propensity score model is misspecified and the imputation model is correctly specified, under Assumptions 1–3 and based on a consistent $\hat{\gamma}_{11}$.

\subsubsection*{Case 2: Only the propensity score model is correctly specified. \hspace{2pt}}

Assume that the imputation model \( m_a(\boldsymbol{\theta}) \) is misspecified and denoted by \( \tilde{m}_a(\tilde{\boldsymbol{\theta}}) \). In this case, we assume the propensity score model \( S_2\left(\mathbf{Z}_{i 2} ; \boldsymbol{\alpha}_2\right) \) is correctly specified. The corresponding weight for the \( i \)-th subject is $w_i\left(a ; \boldsymbol{\alpha}_2\right)= {I\left(A_i=a\right)}/\left[{S_2\left(\mathbf{Z}_{i 2} ; \boldsymbol{\alpha}_2\right)^a\left\{1-S_2\left(\mathbf{Z}_{i 2} ; \boldsymbol{\alpha}_2\right)\right\}^{1-a}}\right]$. Since \( A_i \in \{0, 1\} \), the indicator functions \( I(A_i = 1) \) and \( I(A_i = 0) \) can be equivalently expressed as \( A_i \) and \( 1 - A_i \) respectively. Therefore, in the subsequent derivation, (\ref{equa-B24}) can be rewritten as (\ref{equa-B25}), and (\ref{equa-B31}) as (\ref{equa-B32}). Since the propensity score model is correctly specified, $S_2\left(\mathbf{Z}_{i 2} ; \boldsymbol{\alpha}_2\right)=E(A_i|\mathbf{Z}_{i2}; \boldsymbol{\alpha}_2)$.

The second component in (\ref{equa-B22}) can be re-expressed as
\begin{align}
   & \sum_{a=0}^1\left[ \frac{I(A_i=a)}{S_2\left(\mathbf{Z}_{i 2} ; \boldsymbol{\alpha}_2\right)^a\{1-S_2\left(\mathbf{Z}_{i 2} ; \boldsymbol{\alpha}_2\right)\}^{1-a}}-1\right]\left(
\begin{matrix}
1\\
a
\end{matrix}\right)\left\{\tilde{m}_{a}(\tilde{\boldsymbol{\theta}})-\mu_{12}(a;\boldsymbol{\gamma}_2)\right\}\nonumber\\
=& \frac{I(A_i=0)-\{1-S_2\left(\mathbf{Z}_{i 2} ; \boldsymbol{\alpha}_2\right)\}}{1-S_2\left(\mathbf{Z}_{i 2} ; \boldsymbol{\alpha}_2\right)}\left(
\begin{matrix}
1\\
0
\end{matrix}\right)\left\{\tilde{m}_{0}(\tilde{\boldsymbol{\theta}})-\mu_{12}(0;\boldsymbol{\gamma}_2)\right\}\nonumber\\
\label{equa-B31}
&+\frac{I(A_i=1)-S_2\left(\mathbf{Z}_{i 2} ; \boldsymbol{\alpha}_2\right)}{S_2\left(\mathbf{Z}_{i 2} ; \boldsymbol{\alpha}_2\right)}\left(
\begin{matrix}
1\\
1
\end{matrix}\right)\left\{\tilde{m}_{1}(\tilde{\boldsymbol{\theta}})-\mu_{12}(1;\boldsymbol{\gamma}_2)\right\}\\
=&\frac{(1-A_i)-\{1-S_2\left(\mathbf{Z}_{i 2} ; \boldsymbol{\alpha}_2\right)\}}{1-S_2\left(\mathbf{Z}_{i 2} ; \boldsymbol{\alpha}_2\right)}\left(
\begin{matrix}
1\\
0
\end{matrix}\right)\left\{\tilde{m}_{0}(\tilde{\boldsymbol{\theta}})-\mu_{12}(0;\boldsymbol{\gamma}_2)\right\}\nonumber\\
&+\frac{A_i-S_2\left(\mathbf{Z}_{i 2} ; \boldsymbol{\alpha}_2\right)}{S_2\left(\mathbf{Z}_{i 2} ; \boldsymbol{\alpha}_2\right)}\left(
\begin{matrix}
1\\
1
\end{matrix}\right)\left\{\tilde{m}_{1}(\tilde{\boldsymbol{\theta}})-\mu_{12}(1;\boldsymbol{\gamma}_2)\right\} \label{equa-B32}
\end{align} 

Now, we take the conditional expectation of (\ref{equa-B32}) with respect to $A_i|\mathbf{X}_{i}$ to obtain

\begin{align}
    & \frac{\{1-E(A_i|\mathbf{X}_{i})\}-\{1-S_2\left(\mathbf{Z}_{i 2} ; \boldsymbol{\alpha}_2\right)\}}{1-S_2\left(\mathbf{Z}_{i 2} ; \boldsymbol{\alpha}_2\right)}\left(
\begin{matrix}
1\\
0
\end{matrix}\right)\left\{\tilde{m}_{0}(\tilde{\boldsymbol{\theta}})-\mu_{12}(0;\boldsymbol{\gamma}_2)\right\}\nonumber\\
&+\frac{E(A_i|\mathbf{X}_{i})-S_2\left(\mathbf{Z}_{i 2} ; \boldsymbol{\alpha}_2\right)}{S_2\left(\mathbf{Z}_{i 2} ; \boldsymbol{\alpha}_2\right)}\left(
\begin{matrix}
1\\
1
\end{matrix}\right)\left\{\tilde{m}_{1}(\tilde{\boldsymbol{\theta}})-\mu_{12}(1;\boldsymbol{\gamma}_2)\right\}  \\
=& \frac{\{1-E(A_i|\mathbf{Z}_{i1})\}-\{1-S_2\left(\mathbf{Z}_{i 2} ; \boldsymbol{\alpha}_2\right)\}}{1-S_2\left(\mathbf{Z}_{i 2} ; \boldsymbol{\alpha}_2\right)}\left(
\begin{matrix}
1\\
0
\end{matrix}\right)\left\{\tilde{m}_{0}(\tilde{\boldsymbol{\theta}})-\mu_{12}(0;\boldsymbol{\gamma}_2)\right\}\nonumber\\
&+\frac{E(A_i|\mathbf{Z}_{i1})-S_2\left(\mathbf{Z}_{i 2} ; \boldsymbol{\alpha}_2\right)}{S_2\left(\mathbf{Z}_{i 2} ; \boldsymbol{\alpha}_2\right)}\left(
\begin{matrix}
1\\
1
\end{matrix}\right)\left\{\tilde{m}_{1}(\tilde{\boldsymbol{\theta}})-\mu_{12}(1;\boldsymbol{\gamma}_2)\right\} \\
=& \frac{\{1-S_2\left(\mathbf{Z}_{i 2} ; \boldsymbol{\alpha}_2\right)\}-\{1-S_2\left(\mathbf{Z}_{i 2} ; \boldsymbol{\alpha}_2\right)\}}{1-S_2\left(\mathbf{Z}_{i 2} ; \boldsymbol{\alpha}_2\right)}\left(
\begin{matrix}
1\\
0
\end{matrix}\right)\left\{\tilde{m}_{0}(\tilde{\boldsymbol{\theta}})-\mu_{12}(0;\boldsymbol{\gamma}_2)\right\}+\nonumber \\
&+\frac{S_2\left(\mathbf{Z}_{i 2} ; \boldsymbol{\alpha}_2\right)-S_2\left(\mathbf{Z}_{i 2} ; \boldsymbol{\alpha}_2\right)}{S_2\left(\mathbf{Z}_{i 2} ; \boldsymbol{\alpha}_2\right)}\left(
\begin{matrix}
1\\
1
\end{matrix}\right)\left\{\tilde{m}_{1}(\tilde{\boldsymbol{\theta}})-\mu_{12}(1;\boldsymbol{\gamma}_2)\right\}  \\
=&\boldsymbol{0}\nonumber
\end{align}
Therefore, (\ref{equa-B22}) equals 0 in this case. In conclusion, $\hat{\gamma}_{21}$ is consistent for $\psi_{12}$ in the MSM (3) if the imputation model is misspecified and the propensity score model is correctly specified, under Assumptions 1-3 and based on a consistent $\hat{\gamma}_{11}$.

\subsubsection*{Case 3: both the imputation model and the propensity score model are correctly specified. \hspace{2pt}}

This case represents a more strict scenario. Since Cases 1 and 2 have respectively established that the augmented estimating function remains unbiased when either model is correctly specified, it immediately follows that the estimating function remains unbiased when both are correct. 

In this setting, the bias from both sources is simultaneously removed, and the estimating function consistently recovers the causal estimand. Therefore, Case 3 is a special instance where both conditions for double robustness are satisfied, and the consistency result follows directly from the arguments in Cases 1 and 2.

\subsection{Proof of Theorem 3}
\label{sub-C3}
Here we derive the asymptotic covariance matrix of the estimator obtained from two-stage analysis focusing on applying PS regression adjustment in stage I and use of AIPW in stage II. Recall that ${\hat{\boldsymbol{\Omega}}}=(\hat{\boldsymbol{\phi}}_1',\hat{\boldsymbol{\phi}}_2')'$ is the solution of 
\begin{equation}
\bar{\bar{\mathbf{U}}}(\boldsymbol{\Omega})=\left\{\begin{array}{l}
\mathbf{U}_1\left(\mathcal{D} ; \boldsymbol{\phi}_1\right) \\
\bar{\bar{\mathbf{U}}}_2\left(\mathcal{D} ; \boldsymbol{\gamma}_1, \boldsymbol{\phi}_2\right)
\end{array}\right\}=\mathbf{0} . \nonumber
\end{equation}
where $\boldsymbol{\phi}_1=(\boldsymbol{\gamma}_1',\boldsymbol{\alpha}_1')'$ and $\boldsymbol{\phi}_2=(\boldsymbol{\gamma}_2',\boldsymbol{\alpha}_2')'$.

Since
$$\bar{\bar{\mathbf{U}}}({\hat{\boldsymbol{\Omega}}})=\bar{\bar{\mathbf{U}}}(\boldsymbol{\Omega})+\frac{\partial \bar{\bar{\mathbf{U}}}(\boldsymbol{\Omega})}{\partial \boldsymbol{\Omega}'}\left({\bar{\bar{\boldsymbol{\Omega}}}}-\boldsymbol{\Omega}\right)+o_p\left(\frac{1}{\sqrt{n}}\right)$$
then we have
$$
\sqrt{n}(\bar{\bar{\boldsymbol{\Omega}}}-\boldsymbol{\Omega})=\left\{-\frac{1}{n} \frac{\partial \bar{\bar{\mathbf{U}}}(\boldsymbol{\Omega})}{\partial \boldsymbol{\Omega}'}\right\}^{-1}\left\{\frac{1}{\sqrt{n}} \bar{\bar{\mathbf{U}}}(\boldsymbol{\Omega})\right\}+o_{p}(1)
$$
where $\frac{\partial \bar{\bar{\mathbf{U}}}(\boldsymbol{\Omega})}{\partial \boldsymbol{\Omega}'}$ is
$$
\sum_{i=1}^{n}\left\{\begin{matrix}
\frac{\partial \mathbf{U}_{i11}(\mathcal{D}_i;\boldsymbol{\phi}_1)}{ \partial \boldsymbol{\gamma}_1} &\frac{\partial \mathbf{U}_{i11}(\mathcal{D}_i;\boldsymbol{\phi}_1)}{ \partial \boldsymbol{\alpha}_1} &\boldsymbol{0}&\boldsymbol{0}&\boldsymbol{0}\\
\boldsymbol{0} &\frac{\partial \mathbf{U}_{i12}(\mathcal{D}_i;\boldsymbol{\alpha}_1)}{\partial \boldsymbol{\alpha}_1} &\boldsymbol{0}&\boldsymbol{0}&\boldsymbol{0} \\
\frac{\partial \bar{\bar{\mathbf{U}}}_{i21}(\mathcal{D}_i;\boldsymbol{\gamma}_1,\boldsymbol{\phi}_2)}{\partial \boldsymbol{\gamma}_1} &\boldsymbol{0}&\frac{\partial \bar{\bar{\mathbf{U}}}_{i21}(\mathcal{D}_i;\boldsymbol{\gamma}_1,\boldsymbol{\phi}_2)}{\partial \boldsymbol{\gamma}_2} &\frac{\partial \bar{\bar{\mathbf{U}}}_{i21}(\mathcal{D}_i;\boldsymbol{\gamma}_1,\boldsymbol{\phi}_2)}{\partial \boldsymbol{\alpha}_2} &\frac{\partial \bar{\bar{\mathbf{U}}}_{i21}(\mathcal{D}_i;\boldsymbol{\gamma}_1,\boldsymbol{\phi}_2)}{\partial \boldsymbol{\theta}_{1}} \\
\boldsymbol{0} &\boldsymbol{0}&\boldsymbol{0} &\frac{\partial \mathbf{U}_{i22}\left(\mathcal{D}_i;\boldsymbol{\alpha}_2\right)}{\partial \boldsymbol{\alpha}_2}&\boldsymbol{0} \\
\frac{\partial \mathbf{U}_{i23}(\mathcal{D}_i;\boldsymbol{\gamma}_{1},\boldsymbol{\theta})}{\partial \boldsymbol{\gamma}_1}&\boldsymbol{0}&\boldsymbol{0}&\boldsymbol{0}&\frac{\partial \mathbf{U}_{i23}(\mathcal{D}_i;\boldsymbol{\gamma}_{1},\boldsymbol{\theta})}{\partial \boldsymbol{\theta}}
\end{matrix}\right\}
$$
As $n\to \infty$, $\mathcal{A}$ converges to 
$$
E\left\{-\frac{\partial {\bar{\bar{\mathbf{U}}}_{i}}(\boldsymbol{\Omega})}{\partial \boldsymbol{\Omega}'}\right\}=\left\{\begin{matrix}
\mathcal{A}_{11} &\mathcal{A}_{12} &\boldsymbol{0} &\boldsymbol{0}&\boldsymbol{0}\\
\boldsymbol{0} &\mathcal{A}_{22} &\boldsymbol{0} &\boldsymbol{0}&\boldsymbol{0}\\
\mathcal{A}_{31}&\boldsymbol{0}&\mathcal{A}_{33} &\mathcal{A}_{34} &\mathcal{A}_{35} \\
\boldsymbol{0} &\boldsymbol{0}&\boldsymbol{0} &\mathcal{A}_{44} &\boldsymbol{0}\\
\mathcal{A}_{51}&\boldsymbol{0}&\boldsymbol{0}&\boldsymbol{0} &\mathcal{A}_{55}  
\end{matrix}\right\}
$$
where $$\mathcal{A}_{11}=E\left\{-\frac{\partial \mathbf{U}_{i11}(\mathcal{D}_i;\boldsymbol{\phi}_1)}{ \partial \boldsymbol{\gamma}_1}\right\} \text{, } \mathcal{A}_{12}=E\left\{-\frac{\partial \mathbf{U}_{i11}(\mathcal{D}_i;\boldsymbol{\phi}_1)}{ \partial \boldsymbol{\alpha}_1}\right\}, $$

$$\mathcal{A}_{22}=E\left\{-\frac{\partial \mathbf{U}_{i12}(\boldsymbol{\boldsymbol{\alpha}_1)}}{ \partial \boldsymbol{\alpha}_1}\right\} \text{, } \mathcal{A}_{31}=E\left\{\frac{\partial \bar{\bar{\mathbf{U}}}_{i21}(\mathcal{D}_i;\boldsymbol{\gamma}_1,\boldsymbol{\phi}_2)}{\partial \boldsymbol{\gamma}_1}\right\},$$

$$\mathcal{A}_{33}=E\left\{\frac{\partial \bar{\bar{\mathbf{U}}}_{i21}(\mathcal{D}_i;\boldsymbol{\gamma}_1,\boldsymbol{\phi}_2)}{\partial \boldsymbol{\gamma}_2}  \right\}\text{, } \mathcal{A}_{34}=E\left\{\frac{\partial \bar{\bar{\mathbf{U}}}_{i21}(\mathcal{D}_i;\boldsymbol{\gamma}_1,\boldsymbol{\phi}_2)}{\partial \boldsymbol{\alpha}_2}  \right\},$$
$$\mathcal{A}_{35}=E\left\{ \frac{\partial \bar{\bar{\mathbf{U}}}_{i21}(\mathcal{D}_i;\boldsymbol{\gamma}_1,\boldsymbol{\phi}_2)}{\partial \boldsymbol{\theta}_{1}}\right\}\text{, }\mathcal{A}_{44}=E\left\{ \frac{\partial \mathbf{U}_{i22}\left(\mathcal{D}_i;\boldsymbol{\alpha}_2\right)}{\partial \boldsymbol{\alpha}_2}\right\},$$
$$\mathcal{A}_{51}=E\left\{ \frac{\partial \mathbf{U}_{i231}(\mathcal{D}_i;\boldsymbol{\gamma}_{1},\boldsymbol{\theta}_1)}{\partial \boldsymbol{\gamma}_1}\right\}\text{ and }\mathcal{A}_{55}=E\left\{\frac{\partial \mathbf{U}_{i231}(\mathcal{D}_i;\boldsymbol{\gamma}_{1},\boldsymbol{\theta}_1)}{\partial \boldsymbol{\theta}_1} \right\}.$$
As $n\to \infty$, 
$$
\frac{1}{n} \sum_{i=1}^{n}{\bar{\bar{\mathbf{U}}}_{i}}(\boldsymbol{\Omega}) {\bar{\bar{\mathbf{U}}}_{i}}(\boldsymbol{\Omega})'
$$
converges in probability to 
$$
E\left\{{\bar{\bar{\mathbf{U}}}_{i}}(\boldsymbol{\Omega}) {\bar{\bar{\mathbf{U}}}_{i}}(\boldsymbol{\Omega})'\right\}=\mathcal{B}(\boldsymbol{\Omega})
$$
Due to the consistency property of estimators for $\boldsymbol{\gamma}$, we have
$$\sqrt{n}\left({\hat{{\boldsymbol{\Omega}}}}-\boldsymbol{\Omega}\right)\stackrel{D}{\rightarrow}MVN \left\{\mathbf{0},\boldsymbol{\Sigma}(\boldsymbol{\Omega})\right\}$$
where $\boldsymbol{\Sigma}(\boldsymbol{\Omega})=\mathcal{A}^{-1}(\boldsymbol{\Omega}) \mathcal{B}(\boldsymbol{\Omega})\left\{\mathcal{A}^{-1}(\boldsymbol{\Omega})\right\}'$ and $$
\mathcal{A}(\boldsymbol{\Omega})=\left\{\begin{matrix}
\mathcal{A}_{11} &\mathcal{A}_{12} &\boldsymbol{0} &\boldsymbol{0}&\boldsymbol{0}\\
\boldsymbol{0} &\mathcal{A}_{22} &\boldsymbol{0} &\boldsymbol{0}&\boldsymbol{0}\\
\mathcal{A}_{31}&\boldsymbol{0}&\mathcal{A}_{33} &\mathcal{A}_{34} &\mathcal{A}_{35} \\
\boldsymbol{0} &\boldsymbol{0}&\boldsymbol{0} &\mathcal{A}_{44} &\boldsymbol{0}\\
\mathcal{A}_{51}&\boldsymbol{0}&\boldsymbol{0}&\boldsymbol{0} &\mathcal{A}_{55}  
\end{matrix}\right\}
.$$
\subsection{Sandwich Variance Estimator and Wald-Based Inference}

For inference, we replace the population expectations in $\mathcal{A}(\cdot)$ and $\mathcal{B}(\cdot)$ with their empirical counterparts, yielding the sandwich covariance estimator
\[
\hat{\boldsymbol{\Sigma}}(\hat{\boldsymbol{\Omega}}) 
= \hat{\mathcal{A}}^{-1}(\hat{\boldsymbol{\Omega}})\,
  \hat{\mathcal{B}}(\hat{\boldsymbol{\Omega}})\,
  \{\hat{\mathcal{A}}^{-1}(\hat{\boldsymbol{\Omega}})\}' ,
\]
where
\[
\hat{\mathcal{A}}(\hat{\boldsymbol{\Omega}}) 
= -\frac{1}{n}\sum_{i=1}^n 
   \frac{\partial \mathbf{U}_i(\boldsymbol{\Omega})}{\partial \boldsymbol{\Omega}'} 
   \Big|_{\boldsymbol{\Omega}=\hat{\boldsymbol{\Omega}}},
\quad
\hat{\mathcal{B}}(\hat{\boldsymbol{\Omega}}) 
= \frac{1}{n}\sum_{i=1}^n 
   \mathbf{U}_i(\hat{\boldsymbol{\Omega}})
   \mathbf{U}_i(\hat{\boldsymbol{\Omega}})' .
\]

The estimated covariance matrix $\hat{\boldsymbol{\Sigma}}(\hat{\boldsymbol{\Omega}})$ provides standard errors for all components of $\hat{\boldsymbol{\Omega}}$, from which Wald tests and confidence intervals can be constructed in the usual way. For instance, to test $H_0:\gamma_{11}=0$, the Wald statistic is
\[
W = \frac{\hat{\gamma}_{11}}{\text{s.e.}(\hat{\gamma}_{11})}, 
\quad
\text{s.e.}(\hat{\gamma}_{11}) = \sqrt{\hat{\boldsymbol{\Sigma}}(\hat{\boldsymbol{\Omega}})_{[2,2]}},
\]
with an analogous procedure for $\gamma_{21}$ and other parameters of interest.

\section{Proof of the Bias Induced by Misspecified Propensity Scores in Regression Adjustment} 
\renewcommand{\theequation}{C.\arabic{equation}}
\setcounter{equation}{0}
\label{sup-C}
Here, we aim to show the derivations to evaluate the form of the limiting values of estimators of causal effects under the two-stage approach, especially apply propensity score regression adjustment in both stages. 

The response model is in the form of 
\begin{equation}
    Y=\theta_0 +{\theta}_{11}A + {\theta}_{12}A(D-c) + \mathbf{X}'\boldsymbol{\theta}_{2}+E
    \nonumber
\end{equation}
where $E\sim N(0, \tau^2)$ and $(A,D,\mathbf{X})\perp E$. 
\subsection{Bias Caused by Misspecified Propensity Scores in Stage I}
\label{sup-c1}
In the first stage, we let $\tilde{S}_1(\mathbf{X};\tilde{\boldsymbol{\alpha}}_1)$ denote the Stage I propensity score under misspecification, and consider the regression adjustment 
\[
E(Y | A=1,D,\tilde{S}_1(\mathbf{X};\tilde{\boldsymbol{\alpha}}_1);\tilde{\boldsymbol{\gamma}}_1)
= \tilde{\gamma}_{10} + \tilde{\gamma}_{11}D 
  + \tilde{\gamma}_{12}\tilde{S}_1(\mathbf{X};\tilde{\boldsymbol{\alpha}}_1)
= \mu_{11}(\tilde{\boldsymbol{\phi}}_1),
\]
where $\tilde{\boldsymbol{\gamma}}_1=(\tilde{\gamma}_{10},\tilde{\gamma}_{11}, \tilde{\gamma}_{12})'$ and 
$\tilde{\boldsymbol{\phi}}_1=(\tilde{\boldsymbol{\gamma}}_1',\tilde{\boldsymbol{\alpha}}_1')'$.
Here $\tilde{\boldsymbol{\gamma}}_1$ is the solution to
\begin{equation}
\label{equa-sup4.1}
E\left\{\mathcal{U}_{11}(\mathcal{D};\tilde{\boldsymbol{\phi}}_1)\right\} = \mathbf{0}
\end{equation}
where
\[
\mathcal{U}_{11}(\mathcal{D};\tilde{\boldsymbol{\phi}}_1)
= A \, \frac{\partial \mu_{11}(\tilde{\boldsymbol{\phi}}_1)}{\partial \tilde{\boldsymbol{\gamma}}_1}
  \{ Y - \mu_{11}(\tilde{\boldsymbol{\phi}}_1)\}.
\]
By applying the law of iterated expectations with respect to $A$, we obtain  
$$E\bigl\{\mathcal U_{11}(\mathcal D;\tilde{\boldsymbol\phi}_1)\bigr\}
= E\left[E\left[A \frac{\partial \mu_{11}(\tilde{\boldsymbol\phi}_1)}{\partial \tilde{\boldsymbol\gamma}_1}\{Y-\mu_{11}(\tilde{\boldsymbol\phi}_1)\}\middle|A\right]\right].$$
Since $A$ is binary,  
$$E\left[A \frac{\partial \mu_{11}(\tilde{\boldsymbol\phi}_1)}{\partial \tilde{\boldsymbol\gamma}_1}\{Y-\mu_{11}(\tilde{\boldsymbol\phi}_1)\}\,\middle|\,A\right]
= A E\left[\frac{\partial \mu_{11}(\tilde{\boldsymbol\phi}_1)}{\partial \tilde{\boldsymbol\gamma}_1}\{Y-\mu_{11}(\tilde{\boldsymbol\phi}_1)\}\,\middle|A\right].$$
Therefore,  
$$E\bigl\{\mathcal U_{11}(\mathcal D;\tilde{\boldsymbol\phi}_1)\bigr\}
= P(A=1)\,
E\left[\frac{\partial \mu_{11}(\tilde{\boldsymbol\phi}_1)}{\partial \tilde{\boldsymbol\gamma}_1}\{Y-\mu_{11}(\tilde{\boldsymbol\phi}_1)\}\,\middle|\,A=1\right].$$
Since $P(A=1)>0$, solving the estimating equation (\ref{equa-sup4.1})
is equivalent to solving  
$$E\left\{\mathcal{U}_{11}(\mathcal{D};\tilde{\boldsymbol{\phi}}_1)\middle|A=1\right\}=0.$$

Here, we let $Y^*$ represent $Y|A=1$, $D^*$ represent $D|A=1$, $\mathbf{X}^*$ represent $\mathbf{X}|A=1$, and $\mathcal{D}^*=(Y^*,S^*,\boldsymbol{X}^*)$. Now, we aim to evaluate this expectation. Firstly, we take the expectation with respect to $Y^*$ given $D^*$ and $\mathbf{X}^*$, then we have
 \begin{align}
     E_{Y^*}\bigl\{\mathcal{U}_{11}(\mathcal{D};\tilde{\boldsymbol{\phi}}_1)|D^*,\mathbf{X}^*\bigr\}&=E_{Y^*|D^*,\mathbf{X}^*}\bigl\{\mathcal{U}_{11}(\mathcal{D};\tilde{\boldsymbol{\phi}}_1)\bigr\}\nonumber \\
     &= \bar{\mathbf{V}}_1^*\bigl\{E(Y^*|D^*,\mathbf{X}^*;\boldsymbol{\theta})-\bar{\mathbf{V}}_1^{*'}\boldsymbol{\gamma}_{1}\bigr\}
 \end{align}
where $\bar{\mathbf{V}}_1^*=\{1,D^*, \tilde{S}_1(\mathbf{X}^*;\tilde{\boldsymbol{\alpha}}_1)\}'$.

Then we take the expectation with respect to $D^*|\mathbf{X}^*$. This gives a system of equations as follows:
\begin{align}
\label{equa-B3}
&\theta_0^*+\theta_{12}E(D^*|\mathbf{X}^*)+\boldsymbol{\theta}_2'\mathbf{X}^*-\bigl\{\gamma_{10}+\gamma_{11}E(D^*|\mathbf{X}^*)+\gamma_{12}\tilde{S}_1(\mathbf{X}^*;\tilde{\boldsymbol{\alpha}}_1)\bigr\}=0  \\
\label{equa-B4}
&\theta_0^*E(D^*|\mathbf{X}^*)+\theta_{12}E(D^{*2}|\mathbf{X}^*)+\boldsymbol{\theta}_2'E\bigl\{(D^*|\mathbf{X}^*)\mathbf{X}^*\bigr\}- \nonumber\\
&\left[\gamma_{10}E(D^*|\mathbf{X}^*)+\gamma_{11}E(D^{*2}|\mathbf{X}^*)+\gamma_{12}E\bigl\{\log (D^*|\mathbf{X}^*)\tilde{S}_1(\mathbf{X}^*;\tilde{\boldsymbol{\alpha}}_1)\bigr\}\right]  =0  \\
\label{equa-B5}
&\theta_0^*\tilde{S}_1(\mathbf{X}^*;\tilde{\boldsymbol{\alpha}}_1)+\theta_{12}E(D^*|\mathbf{X}^*)\tilde{S}_1(\mathbf{X}^*;\tilde{\boldsymbol{\alpha}}_1)+\boldsymbol{\theta}_2'\mathbf{X}^*\tilde{S}_1(\mathbf{X}^*;\tilde{\boldsymbol{\alpha}}_1)-\nonumber \\
&\bigl\{\gamma_{10}\tilde{S}_1(\mathbf{X}^*;\tilde{\boldsymbol{\alpha}}_1)+\gamma_{11}E(D^*|\mathbf{X}^*)\tilde{S}_1(\mathbf{X}^*;\tilde{\boldsymbol{\alpha}}_1)+\gamma_{12}\tilde{S}_1(\mathbf{X}^*;\tilde{\boldsymbol{\alpha}}_1)^2\bigr\}=0  
\end{align}
where $\theta_0^*=\theta_{0}+\theta_{11}-\theta_{12}c$. 
Next, we take expectation with respect to $\mathbf{X}^*$. From (\ref{equa-B3}), we get 
\begin{equation}
\label{equa-B6}
    \theta_0^*+\theta_{12}E(D^* )+\boldsymbol{\theta}_2'E(\mathbf{X}^*)-\left[\gamma_{10}+\gamma_{11}E(D^* )+\gamma_{12}E\bigl\{\tilde{S}_1(\mathbf{X}^*;\tilde{\boldsymbol{\alpha}}_1)\bigr\}\right]=0
\end{equation}
From (\ref{equa-B4}), we get 
\begin{align}
\label{equa-B7}
    &\theta_0^*E(D^*)+\theta_{12}E(D^{*2})+\boldsymbol{\theta}_2'\{\boldsymbol{\zeta}_1(\mathbf{X}^*)+E(D^*)E(\mathbf{X}^*)\}- \nonumber\\
&\biggl[\gamma_{10}E(D^*)+\gamma_{11}E(D^{*2})+\gamma_{12}\left[\text{cov}\bigl\{E(D^*|\mathbf{X}^*),\tilde{S}_1(\mathbf{X}^*;\tilde{\boldsymbol{\alpha}}_1)\bigr\}+E(D^*)E\bigl\{\tilde{S}_1(\mathbf{X}^*;\tilde{\boldsymbol{\alpha}}_1)\bigr\}\right]\biggr]=0  
\end{align}
where $\boldsymbol{\zeta}_1(\mathbf{X}^*)=\bigl\{\zeta_{11}(\mathbf{X}^*),...,\zeta_{1k}(\mathbf{X}^*)\bigr\}'$ where $\zeta_{1j}(\mathbf{X}^*)=\text{cov}\bigl\{E(D^*|\mathbf{X}^*_j),\mathbf{X}^*\bigr\}$ for $j=1,2,...,k$.

From (\ref{equa-B5}), we get 
\begin{align}
\label{equa-B8}
    &\theta_0^*E\bigl\{\tilde{S}_1(\mathbf{X}^*;\tilde{\boldsymbol{\alpha}}_1)\bigr\}+\theta_{12}\left[\text{cov}\bigl\{E(D^*|\mathbf{X}^*)\tilde{S}_1(\mathbf{X}^*;\tilde{\boldsymbol{\alpha}}_1)\bigr\}+E(D^*)E\bigl\{\tilde{S}_1(\mathbf{X}^*;\tilde{\boldsymbol{\alpha}}_1)\bigr\}\right]\nonumber\\
    &+\boldsymbol{\theta}_2'\bigl[\boldsymbol{\phi}_1(\mathbf{X}^*)+E\{\tilde{S}_1(\mathbf{X}^*;\tilde{\boldsymbol{\alpha}}_1)\}E(\mathbf{X}^*)\bigr]\nonumber \\
&-\biggl[\gamma_{10}E\bigl\{\tilde{S}_1(\mathbf{X}^*;\tilde{\boldsymbol{\alpha}}_1)\bigr\}+\gamma_{11}\left[\text{cov}\bigl\{E(D^*|\mathbf{X}^*)\tilde{S}_1(\mathbf{X}^*;\tilde{\boldsymbol{\alpha}}_1)\bigr\}+E(D^*)E\bigl\{\tilde{S}_1(\mathbf{X}^*;\tilde{\boldsymbol{\alpha}}_1)\bigr\}\right] \nonumber\\
&+\gamma_{12}E\bigl\{\tilde{S}_1(\mathbf{X}^*;\tilde{\boldsymbol{\alpha}}_1)^2\bigr\}\biggr]=0
\end{align}
where $\boldsymbol{\phi}_1(\mathbf{X}^*) = \bigl\{\phi_{11}(\mathbf{X}^*),...,\phi_{1k}(\mathbf{X}^*)\bigr\}'$ where $\phi_{1j}(\mathbf{X}^*)=\text{cov}\bigl\{\tilde{S}_1(\mathbf{X}^*;\tilde{\boldsymbol{\alpha}}_1),\mathbf{X}_j^*\bigr\} $ for $j=1,2,...,k$. Then from (\ref{equa-B6}), we can obtain the form of $\gamma_{10}$:
\begin{align}
    \gamma_{10} = \theta_0^*+\theta_{12}E(D^* )+\boldsymbol{\theta}_2'E(\mathbf{X}^*)-\gamma_{11}E(D^* )-\gamma_{12}E\bigl\{\tilde{S}_1(\mathbf{X}^*;\tilde{\boldsymbol{\alpha}}_1)\bigr\}
\nonumber
\end{align}
Then, we replace $\gamma_{10}$ in (\ref{equa-B7})
\begin{align}
        &\theta_0^*E(D^*)+\theta_{12}E(D^{*2})+\boldsymbol{\theta}_2'\bigl\{\boldsymbol{\zeta}_1(\mathbf{X}^*)+E(D^*)E(\mathbf{X}^*)\bigr\}- \nonumber\\
&\biggl[\left[\theta_0^*+\theta_{12}E(D^* )+\boldsymbol{\theta}_2'E(\mathbf{X}^*)-\gamma_{11}E(D^* )-\gamma_{12}E\bigl\{\tilde{S}_1(\mathbf{X}^*;\tilde{\boldsymbol{\alpha}}_1)\bigr\}\right]E(D^*)\nonumber\\
&+\gamma_{11}E(D^{*2})+\gamma_{12}\left[\text{cov}\bigl\{E(D^*|\mathbf{X}^*),\tilde{S}_1(\mathbf{X}^*;\tilde{\boldsymbol{\alpha}}_1)\bigr\}+E(D^*)E\bigl\{\tilde{S}_1(\mathbf{X}^*;\tilde{\boldsymbol{\alpha}}_1)\bigr\}\right]\biggr]  =0  \nonumber
\end{align}
\begin{align}
        &\theta_0^*E(D^*)+\theta_{12}E(D^{*2})+\boldsymbol{\theta}_2'\bigl\{\boldsymbol{\zeta}_1(\mathbf{X}^*)+E(D^*)E(\mathbf{X}^*)\bigr\}- \nonumber\\
&\biggl[\left[\theta_0^*+\theta_{12}E(D^* )+\boldsymbol{\theta}_2'E(\mathbf{X}^*)-\gamma_{11}E(D^* )-\gamma_{12}E\bigl\{\tilde{S}_1(\mathbf{X}^*;\tilde{\boldsymbol{\alpha}}_1)\bigr\}\right]E(D^*)\nonumber\\
&+\gamma_{11}E(D^{*2})+\gamma_{12}\left[\text{cov}\bigl\{E(D^*|\mathbf{X}^*),\tilde{S}_1(\mathbf{X}^*;\tilde{\boldsymbol{\alpha}}_1)\bigr\}+E(D^*)E\bigl\{\tilde{S}_1(\mathbf{X}^*;\tilde{\boldsymbol{\alpha}}_1)\bigr\}\right]\biggr]   =0  \nonumber
\end{align}
After some algebra, we get 
\begin{align}
\label{equa-B9}
   & \theta_{12}[E(D^{*2})-E(D^*)^2]+\boldsymbol{\theta}_2'\boldsymbol{\zeta}_1(\mathbf{X}^*)-\gamma_{11}[E(D^{*2})-E(D^*)^2] \nonumber\\
   &-\gamma_{12}\text{cov}\bigl[E(D^*|\mathbf{X}^*),\tilde{S}_1(\mathbf{X}^*;\tilde{\boldsymbol{\alpha}}_1)\bigr]=0.
\end{align}
After replacing $\gamma_{10}$ in (\ref{equa-B8}), we get 
\begin{align}
\label{equa-B10}
&\theta_0^*E\bigl\{\tilde{S}_1(\mathbf{X}^*;\tilde{\boldsymbol{\alpha}}_1)\bigr\}+\theta_{12}\left[\text{cov}\bigl\{E(D^*|\mathbf{X}^*),\tilde{S}_1(\mathbf{X}^*;\tilde{\boldsymbol{\alpha}}_1)\bigr\}+E(D^*)E\bigl\{\tilde{S}_1(\mathbf{X}^*;\tilde{\boldsymbol{\alpha}}_1)\bigr\}\right]\nonumber\\
&+\boldsymbol{\theta}_2'\left[\boldsymbol{\zeta}_1(\mathbf{X}^*)+E\bigl\{\tilde{S}_1(\mathbf{X}^*;\tilde{\boldsymbol{\alpha}}_1)\bigr\}E(\mathbf{X}^*)\right]\nonumber \\
&-\biggl[\left[\theta_0^*+\theta_{12}E(D^* )+\boldsymbol{\theta}_2'E(\mathbf{X}^*)-\gamma_{11}E(D^* )-\gamma_{12}E\bigl\{\tilde{S}_1(\mathbf{X}^*;\tilde{\boldsymbol{\alpha}}_1)\bigr\}\right]E\bigl\{\tilde{S}_1(\mathbf{X}^*;\tilde{\boldsymbol{\alpha}}_1)\bigr\}\nonumber \\
&+\gamma_{11}\left[\text{cov}\bigl\{E(D^*|\mathbf{X}^*),\tilde{S}_1(\mathbf{X}^*;\tilde{\boldsymbol{\alpha}}_1)\bigr\}+E(D^*)E\bigl\{\tilde{S}_1(\mathbf{X}^*;\tilde{\boldsymbol{\alpha}}_1)\bigr\}\right]+\gamma_{12}E\bigl[\tilde{S}_1(\mathbf{X}^*;\tilde{\boldsymbol{\alpha}}_1)^2\bigr]\biggr]=0.
\end{align}
Similarly, after some algebra, from (\ref{equa-B10}) we finally get 
\begin{align}
 &\theta_{12}\text{cov}\bigl\{E(D^*|\mathbf{X}^*),\tilde{S}_1(\mathbf{X}^*;\tilde{\boldsymbol{\alpha}}_1)\bigr\}+\boldsymbol{\theta}_2'\boldsymbol{\zeta}_1(\mathbf{X}^*)-\gamma_{11}\text{cov}\bigl\{E(D^*|\mathbf{X}^*),\tilde{S}_1(\mathbf{X}^*;\tilde{\boldsymbol{\alpha}}_1)\bigr\}\nonumber\\
 &-\gamma_{12}\left\{E\bigl\{\tilde{S}_1(\mathbf{X}^*;\tilde{\boldsymbol{\alpha}}_1)^2\bigr\}-E\bigl\{\tilde{S}_1(\mathbf{X}^*;\tilde{\boldsymbol{\alpha}}_1)\bigr\}^2\right\}
=0 \nonumber\end{align}
Then we can obtain $\gamma_{12}$ expressed as 
\begin{align}
    \gamma_{12} = \frac{\left[(\theta_{12}-\gamma_{11})\text{cov}\bigl\{E(D^*|\mathbf{X}^*),\tilde{S}_1(\mathbf{X}^*;\tilde{\boldsymbol{\alpha}}_1)\bigr\}+\boldsymbol{\theta}_2'\boldsymbol{\zeta}_1(\mathbf{X}^*)\right] }{\text{var}\bigl\{\tilde{S}_1(\mathbf{X}^*;\tilde{\boldsymbol{\alpha}}_1)\bigr\}}.
\end{align}
By replacing $\gamma_{12}$ in (\ref{equa-B9}), we get 
\begin{align}
       & \theta_{12}\{E(D^{*2})-E(D^*)^2]+\boldsymbol{\theta}_2'\boldsymbol{\zeta}_1(\mathbf{X}^*)-\gamma_{11}\{E(D^{*2})-E(D^*)^2\} \nonumber\\
   &-\frac{\left[(\theta_{12}-\gamma_{11})\text{cov}\bigl\{E(D^*|\mathbf{X}^*),\tilde{S}_1(\mathbf{X}^*;\tilde{\boldsymbol{\alpha}}_1)\bigr\}+\boldsymbol{\theta}_2'\boldsymbol{\zeta}_1(\mathbf{X}^*)\right]}{\text{var}\bigl\{\tilde{S}_1(\mathbf{X}^*;\tilde{\boldsymbol{\alpha}}_1)\bigr\}}\text{cov}\bigl\{E(D^*|\mathbf{X}^*),\tilde{S}_1(\mathbf{X}^*;\tilde{\boldsymbol{\alpha}}_1)\bigr\}=0. \nonumber
\end{align}
After simplifying, we obtain
\begin{align}
    \gamma_{11} &= \theta_{12} +\frac{\boldsymbol{\theta}_2'\left[\boldsymbol{\zeta}_1(\mathbf{X}^*)-\boldsymbol{\zeta}_1(\mathbf{X}^*)\text{corr}\bigl\{E(D^*|\mathbf{X}^*),\tilde{S}_1(\mathbf{X}^*;\tilde{\boldsymbol{\alpha}}_1)\bigr\}\sqrt{\frac{\text{var}\{E(D^*|\mathbf{X}^*)\}}{\text{var}\{\tilde{S}_1(\mathbf{X}^*;\tilde{\boldsymbol{\alpha}}_1)\}}}\right]}{\text{var}(D^*)-\text{var}\bigl\{E(D^*|\mathbf{X}^*)\bigr\}\text{corr}^2\bigl\{E(D^*|\mathbf{X}^*),\tilde{S}_1(\mathbf{X}^*;\tilde{\boldsymbol{\alpha}}_1)\bigr\}}\nonumber \\
    &=\theta_{12} +\frac{\boldsymbol{\theta}_2'\left[\boldsymbol{\zeta}_1(\mathbf{X}|A=1)-\boldsymbol{\zeta}_1(\mathbf{X}|A=1)\text{corr}\bigl[E(D|\mathbf{X},A=1),\tilde{S}_1(\mathbf{X}^*;\tilde{\boldsymbol{\alpha}}_1)\bigr]\sqrt{\frac{\text{var}\{E(D|\mathbf{X},A=1)\}}{\text{var}\{\tilde{S}_1(\mathbf{X}^*;\tilde{\boldsymbol{\alpha}}_1)\}}}\right]}{\text{var}(D|A=1)-\text{var}\bigl\{E(D|\mathbf{X},A=1)\bigr\}\text{corr}^2\bigl\{E(D|\mathbf{X},A=1),\tilde{S}_1(\mathbf{X}^*;\tilde{\boldsymbol{\alpha}}_1)\bigr\}} \nonumber
\end{align}
To make the form of the limiting value more clear, we rewrite it in the form of 
\begin{align}
    \gamma_{11}=\theta_{12} +\frac{\boldsymbol{\theta}_2'\left[\boldsymbol{\zeta}_1(\mathbf{X}|A=1)-\boldsymbol{\zeta}_1(\mathbf{X}|A=1)\rho\sqrt{\frac{\text{var}\{E(D|\mathbf{X},A=1)\}}{\text{var}\{\tilde{S}_1(\mathbf{X};\tilde{\boldsymbol{\alpha}}_1)\}}}\right]}{\text{var}(D|A=1)-\text{var}\bigl\{E(D|\mathbf{X},A=1)\bigr\}\rho^2} \nonumber
\end{align}
where $\rho = \text{corr}\bigl\{E(D|\mathbf{X},A=1),\tilde{S}_1(\mathbf{X}^*;\tilde{\boldsymbol{\alpha}}_1)\bigr\}$.

\subsection{Bias Caused by Misspecified Propensity Scores in Stage II}
\label{sup-c2}
Let $\tilde{S}_2(\mathbf{X};\tilde{\boldsymbol{\alpha}}_2)$ be the Stage II propensity score under misspecification. We fit the propensity score regression adjustment model
$$    E\left\{Y| D,A,\tilde{S}_2(\mathbf{X};\tilde{\boldsymbol{\alpha}}_2);\tilde{\boldsymbol{\gamma}}_1,\tilde{\boldsymbol{\gamma}}_2\right\}=\tilde{\gamma}_{20}+\tilde{\gamma}_{21}A+\tilde{\gamma}_{22}\tilde{S}_2(\mathbf{X};\tilde{\boldsymbol{\alpha}}_2)+\text{offset}\left\{\tilde{\gamma}_{11}A(D-c)\right\}$$
where $\tilde{\boldsymbol{\gamma}}_2=(\tilde{\gamma}_{21}, \tilde{\gamma}_{22},\tilde{\gamma}_{23})'$. The estimating equation for $\tilde{\boldsymbol{\gamma}}_2$ is 
\begin{equation}
\label{equa-sup4.3}
    E\left\{\mathcal{U}_{21}(D;\tilde{\boldsymbol{\gamma}}_1,\tilde{\boldsymbol{\phi}}_2)\right\}=\mathbf{0}
\end{equation}
where $$\mathcal{U}_{21}(D;\tilde{\boldsymbol{\gamma}}_1,\tilde{\boldsymbol{\phi}}_2)=\frac{\partial \mu_{21}(\tilde{\boldsymbol{\phi}}_2)}{\partial \tilde{\boldsymbol{\phi}}_2}\left[Y-\left\{\mu_{21}(\tilde{\boldsymbol{\phi}}_2)+\tilde{\gamma}_{11}A(D-c)\right\}\right]$$
with $\mu_{21}(\tilde{\boldsymbol{\phi}}_2)=\tilde{\gamma}_{20}+\tilde{\gamma}_{21}A+\tilde{\gamma}_{22}\tilde{S}_2(\mathbf{X};\tilde{\boldsymbol{\alpha}}_2)$. 

We let $\bar{\mathbf{V}}_2=\{1,A,\tilde{S}_2(\mathbf{X};\tilde{\boldsymbol{\alpha}}_2)\}'$, the estimating function for $\boldsymbol{\gamma}_2$ is expressed as 
\begin{equation}
\mathcal{U}_{21}(D;\tilde{\boldsymbol{\gamma}}_1,\tilde{\boldsymbol{\phi}}_2)=\bar{\mathbf{V}}_2\bigl\{Y -\tilde{\gamma}_{11}A(D-c)- \bar{\mathbf{V}}_2'\boldsymbol{\gamma}_2\bigr\}.
\nonumber
\end{equation} 
We obtain $\hat{\boldsymbol{\gamma}}_{2}$ by solving the equation 
\begin{equation}
E\{\mathcal{U}_{21}(D;\tilde{\boldsymbol{\gamma}}_1,\tilde{\boldsymbol{\phi}}_2)\}=\mathbf{0} \nonumber
\end{equation}
 Firstly, we take expectation with respect to $Y|D,A,\mathbf{X}$ and we get
 \begin{equation}
 E_{Y|D, A, \mathbf{X}}\bigl\{\mathcal{U}_{21}(D;\tilde{\boldsymbol{\gamma}}_1,\tilde{\boldsymbol{\phi}}_2)\bigr\}=\bar{\mathbf{V}}_2\bigl\{E(Y|D, A, \mathbf{X};\boldsymbol{\theta}) -\tilde{\gamma}_{11}A(D-c)- \bar{\mathbf{V}}_2'\boldsymbol{\gamma}_2\bigr\}. 
     \end{equation}
Then we take expectation with respect to $D,A|\mathbf{X}$ and we can get a $3\times 1$ vector expressed as 
\begin{align}
\label{equa-B13}
   & \theta_0+\theta_{11}E(A|\mathbf{X})+(\theta_{12}-\tilde{\gamma}_{11})\bigl\{E(AD |\mathbf{X})-cE(A|\mathbf{X})\bigr\}+\boldsymbol{\theta}_{2}'\mathbf{X} \nonumber \\
   &- \{\gamma_{20} +\gamma_{21}E(A|\mathbf{X})+\gamma_{22}\tilde{S}_2(\mathbf{X};\tilde{\boldsymbol{\alpha}}_2)\}=0 \\
   \label{equa-B14}
& \theta_0E(A|\mathbf{X})+\theta_{11}E(A^2|\mathbf{X})+(\theta_{12}-\tilde{\gamma}_{11})\bigl\{E(A^2D|\mathbf{X})-cE(A^2|\mathbf{X})\bigr\}+\boldsymbol{\theta}_{2}'E\bigl\{(A|\mathbf{X})\mathbf{X}\bigr\} \nonumber\\
& - \left[\gamma_{20}E(A|\mathbf{X}) +\gamma_{21}E(A^2|\mathbf{X})+\gamma_{22}E\bigl\{(A|\mathbf{X})\tilde{S}_2(\mathbf{X};\tilde{\boldsymbol{\alpha}}_2)\bigr\}\right]=0 \\
   \label{equa-B15}
& \theta_0{S}_1(\mathbf{X})+\theta_{11}E\bigl\{(A|\mathbf{X})\tilde{S}_2(\mathbf{X};\tilde{\boldsymbol{\alpha}}_2)\bigr\}+(\theta_{12}-\tilde{\gamma}_{11})\bigl\{E(AD |\mathbf{X})\tilde{S}_2(\mathbf{X};\tilde{\boldsymbol{\alpha}}_2)\bigr\}-cE\bigl\{(A|\mathbf{X})\tilde{S}_2(\mathbf{X};\tilde{\boldsymbol{\alpha}}_2)\bigr\}\nonumber \\ 
&+\boldsymbol{\theta}_{2}'\tilde{S}_2(\mathbf{X};\tilde{\boldsymbol{\alpha}}_2)\mathbf{X} -\left[\tilde{S}_2(\mathbf{X};\tilde{\boldsymbol{\alpha}}_2)\gamma_{20} +\gamma_{21}E\bigl\{(A|\mathbf{X})\tilde{S}_2(\mathbf{X};\tilde{\boldsymbol{\alpha}}_2)\bigr\}+\gamma_{22}\tilde{S}_2(\mathbf{X};\tilde{\boldsymbol{\alpha}}_2)^2\right]=0.
\end{align}
Our next step is to take expectation with respect to $\mathbf{X}$. From 
(\ref{equa-B13}) we get 
\begin{align}
\label{equa-B16}
   & \theta_0+\theta_{11}E(A )+(\theta_{12}-\tilde{\gamma}_{11})\bigl\{E(AD  )-cE(A )\bigr\}+\boldsymbol{\theta}_{2}'E(\mathbf{X}) - \bigl\{\gamma_{20} +\gamma_{21}E(A )+\gamma_{22}\tilde{S}_2(\mathbf{X};\tilde{\boldsymbol{\alpha}}_2)\bigr\}=0 
\end{align}
From (\ref{equa-B14}) we get 
\begin{align}
\label{equa-B17}
& \theta_0E(A)+\theta_{11}E(A^2)+(\theta_{12}-\tilde{\gamma}_{11})\bigl\{E(A^2D)-cE(A^2)\bigr\}+\boldsymbol{\theta}_{2}'\bigl\{\boldsymbol{\zeta}_2(\mathbf{X})+E(A)E(\mathbf{X})\bigr\}\nonumber\\
& - \left[\gamma_{20}E(A) +\gamma_{21}E(A^2)+\gamma_{22}\bigl[\text{cov}\bigl\{E(A|\mathbf{X}),\tilde{S}_2(\mathbf{X};\tilde{\boldsymbol{\alpha}}_2)\bigr\}+E(A)E\bigl\{\tilde{S}_2(\mathbf{X};\tilde{\boldsymbol{\alpha}}_2)\bigr\}\bigr]\right]=0
\end{align}
where $\boldsymbol{\zeta}_2(\mathbf{X})=\bigl\{\zeta_{21}(\mathbf{X}),\ldots,\zeta_{2k}(\mathbf{X})\bigr\}'$ where $\zeta_{2j}(\mathbf{X})=\text{cov}\bigl\{E(A|\mathbf{X}),\mathbf{X}_j\bigr\}$ for $j=1,2,...,k$.
 
From (\ref{equa-B15}) we get 
\begin{align}
\label{equa-B18}
    & \theta_0 E\bigl\{\tilde{S}_2(\mathbf{X};\tilde{\boldsymbol{\alpha}}_2)\bigr\}+\theta_{11}\left[\text{cov}\bigl\{E(A|\mathbf{X}),\tilde{S}_2(\mathbf{X};\tilde{\boldsymbol{\alpha}}_2)\bigr\}+E(A)E\bigl\{\tilde{S}_2(\mathbf{X};\tilde{\boldsymbol{\alpha}}_2)\bigr\}\right]\nonumber\\
    &+(\theta_{12}-\tilde{\gamma}_{11})\biggl[\text{cov}\bigl\{AD|\mathbf{X},\tilde{S}_2(\mathbf{X};\tilde{\boldsymbol{\alpha}}_2)\bigr\}+E(AD)E\bigl\{\tilde{S}_2(\mathbf{X};\tilde{\boldsymbol{\alpha}}_2)\bigr\}\nonumber \\
    &-c\Bigl[\text{cov}\bigl\{E(A|\mathbf{X}),\tilde{S}_2(\mathbf{X};\tilde{\boldsymbol{\alpha}}_2)\bigr\}+E(A)E\bigl\{\tilde{S}_2(\mathbf{X};\tilde{\boldsymbol{\alpha}}_2)\bigr\}\Bigr]\biggr] \nonumber\\
    &+\boldsymbol{\theta}_{2}'\left[\text{cov}\bigl\{\tilde{S}_2(\mathbf{X};\tilde{\boldsymbol{\alpha}}_2),\mathbf{X}\bigr\}+E\bigl\{\tilde{S}_2(\mathbf{X};\tilde{\boldsymbol{\alpha}}_2)\bigr\}E(\mathbf{X})\right]\nonumber\\
    & -\left[\gamma_{20}\tilde{S}_2(\mathbf{X};\tilde{\boldsymbol{\alpha}}_2) +\gamma_{21}\bigl[\text{cov}\bigl\{E(A|\mathbf{X}),\tilde{S}_2(\mathbf{X};\tilde{\boldsymbol{\alpha}}_2)\bigr\}+E(A)E\bigl\{\tilde{S}_2(\mathbf{X};\tilde{\boldsymbol{\alpha}}_2)\bigr\}\bigr]+\gamma_{22}E\bigl\{\tilde{S}_2(\mathbf{X};\tilde{\boldsymbol{\alpha}}_2)^2\bigr\}\right]=0.
\end{align}
Then from (\ref{equa-B16}), we obtain $\gamma_{20}$:
$$\gamma_{20} =\theta_0+\theta_{11}E(A)+(\theta_{12}-\tilde{\gamma}_{11})\bigl\{E(AD  )-cE(A )\bigr\}+\boldsymbol{\theta}_{2}'E(\mathbf{X}) -\gamma_{21}E(A )-\gamma_{22}\tilde{S}_2(\mathbf{X};\tilde{\boldsymbol{\alpha}}_2) $$
We replace $\gamma_{20}$ in (\ref{equa-B17}) and get:
\begin{align}
    & \theta_0E(A)+\theta_{11}E(A^2)+(\theta_{12}-\tilde{\gamma}_{11})\bigl\{E(A^2D)-cE(A^2)\bigr\}+\boldsymbol{\theta}_{2}'\bigl\{\boldsymbol{\zeta}_2(\mathbf{X})+E(A)E(\mathbf{X})\bigr\}\nonumber\\
& - \left[\theta_0+\theta_{11}E(A)+(\theta_{12}-\tilde{\gamma}_{11})\bigl\{E(AD  )-cE(A )\bigr\}+\boldsymbol{\theta}_{2}'E(\mathbf{X}) -\gamma_{21}E(A )-\gamma_{22}\tilde{S}_2(\mathbf{X};\tilde{\boldsymbol{\alpha}}_2)\right]E(A) \nonumber\\
&-\gamma_{21}E(A^2) -\gamma_{22}\left[\text{cov}\bigl\{E(A|\mathbf{X}),\tilde{S}_2(\mathbf{X};\tilde{\boldsymbol{\alpha}}_2)\bigr\}+E(\mathbf{X})E\bigl\{\tilde{S}_2(\mathbf{X};\tilde{\boldsymbol{\alpha}}_2)\bigr\}\right]=0. \nonumber
\end{align}
After algebra, we get 
\begin{align}
\label{equa-B19}
    &\theta_{11}\bigl\{E(A^2)-E(A)^2\bigr\}+\boldsymbol{\theta}_2'\boldsymbol{\zeta}_2(\mathbf{X})-\gamma_{21}\bigl\{E(A^2)-E(A)^2\bigr\}-\gamma_{22}\text{cov}\bigl\{E(A|\mathbf{X}),\tilde{S}_2(\mathbf{X};\tilde{\boldsymbol{\alpha}}_2)\bigr\} \nonumber \\
    &+(\theta_{12}-\tilde{\gamma}_{11})\bigl\{E(A^2D)-E(AD)E(A)+cE(A)^2-cE(A^2)\bigr\}=0.
\end{align}
To solve for $\gamma_{22}$, we replace $\gamma_{20}$ in (\ref{equa-B18}) to get 
\begin{align}
        & \theta_0E\bigl\{\tilde{S}_2(\mathbf{X};\tilde{\boldsymbol{\alpha}}_2)\bigr\}+\theta_{11}\left[\text{cov}\bigl\{E(A|\mathbf{X}),\tilde{S}_2(\mathbf{X};\tilde{\boldsymbol{\alpha}}_2)\bigr\}+E(A)E\bigl\{\tilde{S}_2(\mathbf{X};\tilde{\boldsymbol{\alpha}}_2)\bigr\}\right]\nonumber\\
    &+(\theta_{12}-\tilde{\gamma}_{11})\Bigl[\text{cov}\bigl\{E(AD|\mathbf{X}),\tilde{S}_2(\mathbf{X};\tilde{\boldsymbol{\alpha}}_2)\bigr\}+E(AD)E\bigl\{\tilde{S}_2(\mathbf{X};\tilde{\boldsymbol{\alpha}}_2)\bigr\}\nonumber \\
    &-c\bigl[\text{cov}\big\{E(A|\mathbf{X}),\tilde{S}_2(\mathbf{X};\tilde{\boldsymbol{\alpha}}_2)\bigr\}+E(A)E\bigl\{\tilde{S}_2(\mathbf{X};\tilde{\boldsymbol{\alpha}}_2)\bigr\}\bigr]\Bigr]\nonumber\\
    &+\boldsymbol{\theta}_{2}'\left[\text{cov}\bigl\{\tilde{S}_2(\mathbf{X};\tilde{\boldsymbol{\alpha}}_2),\mathbf{X}\bigr\}+E\bigl\{\tilde{S}_2(\mathbf{X};\tilde{\boldsymbol{\alpha}}_2)\bigr\}E(\mathbf{X})\right]\nonumber\\
    & -\tilde{S}_2(\mathbf{X};\tilde{\boldsymbol{\alpha}}_2)\left[\theta_0+\theta_{11}E(A)+(\theta_{12}-\tilde{\gamma}_{11})\bigl\{E(AD  )-cE(A )\bigr\}+\boldsymbol{\theta}_{2}'E(\mathbf{X}) -\gamma_{21}E(A )-\gamma_{22}\tilde{S}_2(\mathbf{X};\tilde{\boldsymbol{\alpha}}_2)\right]\nonumber \\
    &-\gamma_{21}\left[\text{cov}\bigl\{E(A|\mathbf{X}),\tilde{S}_2(\mathbf{X};\tilde{\boldsymbol{\alpha}}_2)\bigr\}\right]-\gamma_{22}E\bigl\{\tilde{S}_2(\mathbf{X};\tilde{\boldsymbol{\alpha}}_2)^2\bigr\}=0. \nonumber
\end{align}
We simplify it to get
\begin{align}
    &\theta_{11}\text{cov}\bigl\{E(A|\mathbf{X}),\tilde{S}_2(\mathbf{X};\tilde{\boldsymbol{\alpha}}_2)\bigr\}+\boldsymbol{\theta}_2'\text{cov}\bigl\{E(\tilde{S}_2(\mathbf{X};\tilde{\boldsymbol{\alpha}}_2)|\mathbf{X}),\mathbf{X}\bigr\}-\gamma_{21}\text{cov}\bigl\{E(A|\mathbf{X}),\tilde{S}_2(\mathbf{X};\tilde{\boldsymbol{\alpha}}_2)\bigr\}\nonumber\\
    &-\gamma_{22}\left[E\bigl\{\tilde{S}_2(\mathbf{X};\tilde{\boldsymbol{\alpha}}_2)^2\bigr\}-E\bigl\{\tilde{S}_2(\mathbf{X};\tilde{\boldsymbol{\alpha}}_2)\bigr\}^2\right] \nonumber\\
    &+ (\theta_{12}-\tilde{\gamma}_{11})\left[\text{cov}\bigl\{E(AD|\mathbf{X}),\tilde{S}_2(\mathbf{X};\tilde{\boldsymbol{\alpha}}_2)\bigr\}-c\cdot\text{cov}\bigl\{E(A|\mathbf{X}),\tilde{S}_2(\mathbf{X};\tilde{\boldsymbol{\alpha}}_2)\bigr\}\right]=0. \nonumber
\end{align}
Then we can obtain $\gamma_{22}$ expressed by 
\begin{align}
    \gamma_{22} = \frac{(\theta_{11}-\gamma_{21})\delta_1+\boldsymbol{\theta}_2'\text{cov}\bigl\{\tilde{S}_2(\mathbf{X};\tilde{\boldsymbol{\alpha}}_2),\mathbf{X}\bigr\}+ (\theta_{12}-\tilde{\gamma}_{11})(\delta_2-c\delta_1)}{\text{var}\bigl\{\tilde{S}_2(\mathbf{X};\tilde{\boldsymbol{\alpha}}_2)\bigr\}}
\end{align}
where $\delta_1=\text{cov}\bigl\{E(A|\mathbf{X}),\tilde{S}_2(\mathbf{X};\tilde{\boldsymbol{\alpha}}_2)\bigr\}$ and $\delta_2=\text{cov}\bigl\{E(AD|\mathbf{X}),\tilde{S}_2(\mathbf{X};\tilde{\boldsymbol{\alpha}}_2)\bigr\}$.
Next, we substitute $\gamma_{22}$ in (\ref{equa-B19}):
\begin{align}
    &\theta_{11}\bigl\{E(A^2)-E(A)^2\bigr\}+\boldsymbol{\theta}_2'\boldsymbol{\zeta}_2(\mathbf{X})-\gamma_{21}\bigl\{E(A^2)-E(A)^2\bigr\}\nonumber \\
    &-\frac{(\theta_{11}-\gamma_{21})\delta_1+\boldsymbol{\theta}_2'\text{cov}\bigl\{\tilde{S}_2(\mathbf{X};\tilde{\boldsymbol{\alpha}}_2),\mathbf{X}\bigr\}+ (\theta_{12}-\tilde{\gamma}_{11})(\delta_2-c\delta_1)}{\text{var}\bigl\{\tilde{S}_2(\mathbf{X};\tilde{\boldsymbol{\alpha}}_2)\bigr\}}\delta_1 \nonumber \\
    &+(\theta_{12}-\tilde{\gamma}_{11})\bigl\{E(A^2D)-E(AD)E(A)+cE(A)^2-cE(A^2)\bigr\}=0. \nonumber 
\end{align}
By simplifying, we get 
\begin{align}
    &\theta_{11}\text{var}(A)+\boldsymbol{\theta}_2'\boldsymbol{\zeta}_2(\mathbf{X})-\gamma_{21}\text{var}(A)\nonumber \\
    &-\frac{(\theta_{11}-\gamma_{21})\delta_1+\boldsymbol{\theta}_2'\text{cov}\bigl\{\tilde{S}_2(\mathbf{X};\tilde{\boldsymbol{\alpha}}_2),\mathbf{X}\bigr\}+ (\theta_{12}-\tilde{\gamma}_{11})(\delta_2-c\delta_1)}{\text{var}\bigl\{\tilde{S}_2(\mathbf{X};\tilde{\boldsymbol{\alpha}}_2)\bigr\}}\delta_1 \nonumber \\
    &+(\theta_{12}-\tilde{\gamma}_{11})\bigl\{\text{cov}(AD,A)-c\text{var}(A)\bigr\}=0. \nonumber 
\end{align}
Then we re-organize it and rewrite it as
\begin{align}
\label{equa-B21}
    &\theta_{11}\left[\text{var}(A)-\frac{\delta_1^2}{\text{var}\bigl\{\tilde{S}_2(\mathbf{X};\tilde{\boldsymbol{\alpha}}_2)\bigr\}}\right]+\boldsymbol{\theta}_2'\left[\boldsymbol{\zeta}_2(\mathbf{X})-\frac{\text{cov}\bigl\{\tilde{S}_2(\mathbf{X};\tilde{\boldsymbol{\alpha}}_2),\mathbf{X}\bigr\}\delta_1}{\text{var}\bigl\{\tilde{S}_2(\mathbf{X};\tilde{\boldsymbol{\alpha}}_2)\bigr\}}\right] \nonumber\\
    &+(\theta_{12}-\tilde{\gamma}_{11})\left[\text{cov}(AD,A)-c\text{var}(A)-\frac{\delta_1\delta_2-c\delta_1^2}{\text{var}\bigl\{\tilde{S}_2(\mathbf{X};\tilde{\boldsymbol{\alpha}}_2)\bigr\}}\right]=\gamma_{21}\left[\text{var}(A)-\frac{\delta_1^2}{\text{var}\bigl\{\tilde{S}_2(\mathbf{X};\tilde{\boldsymbol{\alpha}}_2)\bigr\}}\right].
\end{align}
Here we simplify some terms in (\ref{equa-B21}),
\begin{align}
    &\theta_{11}\left[\text{var}(A)-\frac{\delta_1^2}{\text{var}\bigl\{\tilde{S}_2(\mathbf{X};\tilde{\boldsymbol{\alpha}}_2)\bigr\}}\right]=\theta_{11}\left[\text{var}(A)-\text{var}\bigl\{E(A|\mathbf{X})\bigr\}\text{corr}^2\bigl\{E(A|\mathbf{X})\tilde{S}_2(\mathbf{X};\tilde{\boldsymbol{\alpha}}_2)\bigr\}\right],\nonumber \\
        &(\theta_{12}-\tilde{\gamma}_{11})\left[\text{cov}(AD,A)-c\text{var}(A)-\frac{\delta_1\delta_2-c\delta_1^2}{\text{var}\bigl\{\tilde{S}_2(\mathbf{X};\tilde{\boldsymbol{\alpha}}_2)\bigr\}}\right]\nonumber \\
        &=(\theta_{12}-\tilde{\gamma}_{11})\biggl[\text{cov}(AD,A)-cE\bigl\{\text{var}(A|\mathbf{X})\bigr\}\biggr.\nonumber\\
        &\left.-\delta_1\Bigl[\sqrt{\text{var}\bigl\{E(A|\mathbf{X})\bigr\}\text{var}\{E(AD|\mathbf{X})\}}\text{corr}\bigl\{E(AD|\mathbf{X}),\tilde{S}_2(\mathbf{X};\tilde{\boldsymbol{\alpha}}_2)\bigr\}\Bigr]\right] \nonumber \\
    & \gamma_{21}\left[\text{var}(A)-\frac{\delta_1^2}{\text{var}\bigl\{\tilde{S}_2(\mathbf{X};\tilde{\boldsymbol{\alpha}}_2)\bigr\}}\right]=\gamma_{21}\left[\text{var}(A)-\text{var}\bigl\{E(A|\mathbf{X})\bigr\}\text{corr}^2\bigl\{E(A|\mathbf{X})\tilde{S}_2(\mathbf{X};\tilde{\boldsymbol{\alpha}}_2)\bigr\}\right]. \nonumber
\end{align}
 Here we let $$\rho_{21}=\text{corr}\bigl\{E(A|\mathbf{X}),\tilde{S}_2(\mathbf{X};\tilde{\boldsymbol{\alpha}}_2)\bigr\}; \rho_{22}=\text{corr}\bigl\{E(AD|\mathbf{X}),\tilde{S}_2(\mathbf{X};\tilde{\boldsymbol{\alpha}}_2)\bigr\};$$ $$\boldsymbol{\zeta}_2(\mathbf{X})=\bigl\{\zeta_{21}(\mathbf{X}),...,\zeta_{2k}(\mathbf{X})\bigr\}'$$ where $\zeta_{2j}(\mathbf{X})=\text{cov}\bigl\{E(A|\mathbf{X}),\mathbf{X}_j\bigr\}.$
 Let 
 $$\boldsymbol{\phi}_2(\mathbf{X}) = \bigl\{\phi_{21}(\mathbf{X}),...,\phi_{2k}(\mathbf{X})\bigr\}'$$ where $\phi_{2j}(\mathbf{X})=\text{cov}\bigl\{\tilde{S}_2(\mathbf{X};\tilde{\boldsymbol{\alpha}}_2),\mathbf{X}_j\bigr\}$ for $j=1,2,...,k$.
Then (\ref{equa-B21}) can be expressed as 
\begin{align}
    &\gamma_{21} = \theta_{11}+\frac{\boldsymbol{\theta}_2'\Bigl[\boldsymbol{\zeta}_2(\mathbf{X})-\frac{\boldsymbol{\phi}_2(\mathbf{X})\delta_1}{\text{var}\{\tilde{S}_2(\mathbf{X};\tilde{\boldsymbol{\alpha}}_2)\}}\Bigr] }{\text{var}(A)-\text{var}\{E(A|\mathbf{X})\}\rho_{21}^2}\nonumber \\
    &+\frac{(\theta_{12}-\tilde{\gamma}_{11})\biggl[\text{cov}(AD,A)-cE\bigl\{\text{var}(A|\mathbf{X})\bigr\}-\delta_1\Bigl[\sqrt{\text{var}\{E(A|\mathbf{X})\}\text{var}\{E(AD|\mathbf{X})\}}\rho_{22}\Bigr]\biggr]}{\text{var}(A)-\text{var}\bigl\{E(A|\mathbf{X})\bigr\}\rho_{21}^2} \nonumber
\end{align}
and we can obtain $\gamma_{21}$ expressed as 
\begin{align}
    &\gamma_{21} = \theta_{11}+\frac{\boldsymbol{\theta}_2'\left[\boldsymbol{\zeta}_2(\mathbf{X})-\boldsymbol{\phi}_2(\mathbf{X})\rho_{21}\sqrt{\frac{\text{var}\{E(A|\mathbf{X})\}}{\text{var}\{\tilde{S}_2(\mathbf{X};\tilde{\boldsymbol{\alpha}}_2)\}}}\right] }{\text{var}(A)-\text{var}\bigl\{E(A|\mathbf{X})\bigr\}\rho_{21}^2}\nonumber \\
    &+\frac{(\theta_{12}-\tilde{\gamma}_{11})\left[\text{cov}(AD,A)-cE\bigl\{\text{var}(A|\mathbf{X})\bigr\}-\delta_1\rho_{22}\sqrt{\text{var}\{E(A|\mathbf{X})\}\text{var}\{E(AD|\mathbf{X})\}}\right]}{\text{var}(A)-\text{var}\{E(A|\mathbf{X})\}\rho_{21}^2} . \nonumber 
\end{align}

\section{Additional Simulation Results}
\label{sup-d}
\renewcommand{\thetable}{A.\arabic{table}}
\setcounter{table}{0}

\begin{table}
\centering\small
\caption{Empirical results for estimators of causal effects with $P(A=1)=0.25$ under six methods of analysis based on correctly specified models.}
\label{table-appendix1}
\begin{tabular}{llcccccccccccc}
\hline
&&& \multicolumn{4}{l}{ {Minimal PS: ${\mathbf{S}=\{S_1(\mathbf{Z}_1),S_2(\mathbf{Z}_2)\}'}$}} &\multicolumn{4}{l}{ {Expanded PS: ${\mathbf{S}=\{S_1(\mathbf{X}),S_2(\mathbf{X})\}'}$}} \\
\cline{4-11}
Method & Exposure & Effect & Ebias & ESE & RSE & ECP(\%) & Ebias & ESE & RSE & ECP(\%) \\
\hline 
&&&\multicolumn{8}{c}{\em Data Generation Model 1}\\
 \multirow{2}{*}{Naive} & $D$ & $\psi_{11}$ & 0.762 & 0.050 & 0.049 & 0.0 & - & - & - & -\\ 
 & $A$& $\psi_{12}$ & 0.971 & 0.134 & 0.134 & 0.0 & - & - & - & -\\  [4pt]
 \multirow{2}{*}{DG model} & $D$& $\psi_{11}$ & -0.001 & 0.034 & 0.033 & 94.4 & - & - & - & -\\ 
 & $A$& $\psi_{12}$ & -0.004 & 0.076 & 0.077 & 94.8 & - & - & - & -\\  [4pt]
 \multirow{2}{*}{Two PS Reg} & $D$& $\psi_{11}$ & -0.023 & 0.042 & 0.042 & 92.0 & -0.023 & 0.041 & 0.041 & 91.7\\ 
 & $A$& $\psi_{12}$ & -0.008 & 0.084 & 0.085 & 94.8 & -0.007 & 0.078 & 0.078 & 94.7\\  [4pt]
 \multirow{2}{*}{$\text{PS}_\text{I}+\text{PS}_\text{II}$} & $D$ & $\psi_{11}$ & -0.002 & 0.085 & 0.083 & 94.1 & -0.003 & 0.064 & 0.063 & 94.4\\ 
 & $A$ & $\psi_{12}$ & -0.004 & 0.112 & 0.112 & 95.1 & -0.005 & 0.080 & 0.081 & 95.2\\  [4pt]
 \multirow{2}{*}{$\text{PS}_\text{I}+\text{IPW}_\text{II}$} & $D$& $\psi_{11}$ & -0.002 & 0.085 & 0.083 & 94.1 & -0.003 & 0.064 & 0.063 & 94.4\\ 
 & $A$& $\psi_{12}$ & -0.004 & 0.160 & 0.152 & 94.0 & -0.006 & 0.138 & 0.130 & 93.1\\  [4pt]
 \multirow{2}{*}{$\text{PS}_\text{I}+\text{AIPW}_\text{II}$} & $D$& $\psi_{11}$ & -0.002 & 0.085 & 0.083 & 94.1 & -0.003 & 0.064 & 0.063 & 94.4\\ 
 & $A$& $\psi_{12}$ & -0.004 & 0.092 & 0.092 & 94.9 & -0.005 & 0.089 & 0.089 & 95.0\\ 
 \hline
&&&\multicolumn{8}{c}{\em Data Generation Model 2}\\
 \multirow{2}{*}{Naive} & $D$ & $\psi_{11}$ & 0.463 & 0.049 & 0.048 & 0.0 & - & - & - & -\\ 
 & $A$ & $\psi_{12}$ & 0.496 & 0.090 & 0.089 & 0.0 & - & - & - & -\\  [4pt]
 \multirow{2}{*}{DG model} & $D$& $\psi_{11}$ & 0.001 & 0.045 & 0.044 & 93.7 & - & - & - & -\\ 
 & $A$ & $\psi_{12}$ & 0.001 & 0.079 & 0.077 & 94.3 & - & - & - & -\\  [4pt]
 \multirow{2}{*}{Two PS Reg} & $D$& $\psi_{11}$ & 0.001 & 0.045 & 0.049 & 96.9 & - & - & - & -\\ 
 & $A$& $\psi_{12}$ & 0.001 & 0.080 & 0.080 & 94.8 & - & - & - & -\\  [4pt]
 \multirow{2}{*}{$\text{PS}_\text{I}+\text{PS}_\text{II}$} & $D$& $\psi_{11}$ & 0.001 & 0.065 & 0.063 & 93.8 & - & - & - & -\\ 
 & $A$& $\psi_{12}$ & 0.002 & 0.081 & 0.078 & 94.3 & - & - & - & -\\  [4pt]
 \multirow{2}{*}{$\text{PS}_\text{I}+\text{IPW}_\text{II}$} & $D$& $\psi_{11}$ & 0.001 & 0.065 & 0.063 & 93.8 & - & - & - & -\\ 
 & $A$& $\psi_{12}$ & 0.003 & 0.101 & 0.096 & 93.7 & - & - & - & -\\  [4pt]
 \multirow{2}{*}{$\text{PS}_\text{I}+\text{AIPW}_\text{II}$} & $D$& $\psi_{11}$ & 0.001 & 0.065 & 0.063 & 93.8 & - & - & - & -\\ 
 & $A$& $\psi_{12}$ & 0.002 & 0.091 & 0.088 & 94.5 & - & - & - & -\\  
\hline  
\end{tabular}
\begin{tablenotes}
\scriptsize
\item DG model: Data generation model; Two PS Reg: Two-part propensity score regression adjustment; $\text{PS}_\text{I}+\text{PS}_\text{II}$: Using PS regression adjustment in Stage I and II; $\text{PS}_\text{I}+\text{IPW}_\text{II}$: Using PS regression adjustment in Stage I and IPW in Stage II; $\text{PS}_\text{I}+\text{AIPW}_\text{II}$: Using PS regression adjustment in Stage I and AIPW in Stage II. \\
\end{tablenotes}
\end{table}

\begin{table}
\centering
\caption{Empirical results for estimators of causal effects with $P(A=1)=0.25$ for two-part PS regression adjustment and two-stage analysis under model misspecification.}
\label{table-appendix2}
\begin{tabular}{llccccccccc}
\hline
 &   &   & \multicolumn{4}{c}{Misspecified $S_1$} & \multicolumn{4}{c}{Misspecified $S_2$} \\ 
\cline{4-11} 
 {Method} &  {Exposure} &  {Effect} & Ebias & ESE & RSE & ECP(\%) & Ebias & ESE & RSE & ECP(\%)\\
\hline 
&&&\multicolumn{8}{c}{\em Data Generation Model 1}\\
 \multirow{2}{*}{Two PS Reg} & $D$ & $\psi_{11}$ & 0.082 & 0.047 & 0.051 & 64.2 & -0.014 & 0.042 & 0.047 & 96.2 \\ 
 & $A$ & $\psi_{12}$ & 0.008 & 0.097 & 0.098 & 95.0 & 0.243 & 0.086 & 0.087 & 18.4 \\  [4pt]
 \multirow{2}{*}{$\text{PS}_\text{I}+\text{PS}_\text{II}$} & $D$ & $\psi_{11}$ & 0.349 & 0.090 & 0.088 & 2.7 & -0.002 & 0.085 & 0.083 & 94.1 \\ 
 & $A$ & $\psi_{12}$ & 0.050 & 0.105 & 0.106 & 92.4 & 0.335 & 0.113 & 0.114 & 15.7 \\  [4pt]
 \multirow{2}{*}{$\text{PS}_\text{I}+\text{IPW}_\text{II}$} & $D$ & $\psi_{11}$ & 0.349 & 0.090 & 0.090 & 2.7 & -0.002 & 0.085 & 0.083 & 94.1 \\ 
 & $A$ & $\psi_{12}$ & 0.135 & 0.140 & 0.134 & 77.3 & 0.338 & 0.126 & 0.126 & 24.2 \\  \hline
&&&\multicolumn{8}{c}{\em Data Generation Model 2}\\
 \multirow{2}{*}{Two PS Reg} & $D$ & $\psi_{11}$ & -0.098 & 0.065 & 0.062 & 65.3 & 0.001 & 0.065 & 0.063 & 93.7 \\ 
 & $A$ & $\psi_{12}$ & -0.104 & 0.087 & 0.084 & 75.2 & 0.192 & 0.090 & 0.087 & 39.8 \\ [4pt]
 \multirow{2}{*}{$\text{PS}_\text{I}+\text{PS}_\text{II}$} & $D$ & $\psi_{11}$ & 0.145 & 0.061 & 0.060 & 30.8 & 0.001 & 0.065 & 0.063 & 93.8 \\  
 & $A$ & $\psi_{12}$ & 0.027 & 0.081 & 0.078 & 92.7 & 0.247 & 0.081 & 0.079 & 13.1 \\  [4pt]
 \multirow{2}{*}{$\text{PS}_\text{I}+\text{IPW}_\text{II}$} & $D$ & $\psi_{11}$ & 0.145 & 0.061 & 0.060 & 30.8 & 0.001 & 0.065 & 0.063 & 93.8 \\ 
 & $A$ & $\psi_{12}$ & 0.070 & 0.095 & 0.090 & 85.4 & 0.248 & 0.084 & 0.082 & 16.3 \\   
\hline
\end{tabular}
\begin{tablenotes}
\scriptsize
\item$S_1$ represents the propensity score model for the continuous exposure $D$ conditioned on $A=1$, given by $S_1 = E(D|A=1, \mathbf{Z}_1)$; $S_2$ represents the propensity score model for the binary exposure $A$, given by $S_2 = E(A|\mathbf{Z}_2)$.
\end{tablenotes}
\end{table}

\begin{table}
\centering
\caption{Empirical results for estimators of effects for both parts with $P(A=1)=0.25$ for applying PS regression adjustment in Stage I and AIPW in Stage II under model misspecification.}
\label{table-appendix3}
\begin{tabular}{llccccc}
\hline
{Misspecified Model} & {Exposure} & Effect & Ebias & ESE & RSE & ECP(\%)\\
\hline
&& &\multicolumn{4}{c}{\em Data Generation Model 1}\\
 \multirow{2}{*}{$S_1$} & $D$ & $\theta_{12}$ & 0.349 & 0.090 & 0.088 & 2.7 \\ 
 & $A$ & $\theta_{11}$ & 0.135 & 0.101 & 0.100 & 73.3 \\ [4pt]
 \multirow{2}{*}{$S_2$} & $D$ & $\theta_{12}$ & -0.002 & 0.085 & 0.083 & 94.1 \\ 
 & $A$ & $\theta_{11}$ & -0.011 & 0.090 & 0.090 & 94.8 \\ [4pt]
 \multirow{2}{*}{$m_a(\mathbf{X};\boldsymbol{\theta})$} & $D$ & $\theta_{12}$ & -0.002 & 0.085 & 0.083 & 94.1 \\ 
 & $A$ & $\theta_{11}$ & -0.001 & 0.096 & 0.100 & 95.9 \\ [4pt]
 \multirow{2}{*}{$S_1+m_a(\mathbf{X};\boldsymbol{\theta})$} & $D$ & $\theta_{12}$ & 0.349 & 0.090 & 0.088 & 2.7 \\ 
 & $A$ & $\theta_{11}$ & 0.138 & 0.104 & 0.107 & 74.4 \\ [4pt]
 \multirow{2}{*}{$S_2+m_a(\mathbf{X};\boldsymbol{\theta})$} & $D$ & $\theta_{12}$ & -0.002 & 0.085 & 0.083 & 94.1 \\ 
 & $A$ & $\theta_{11}$ & 0.226 & 0.086 & 0.088 & 26.0 \\ 
\hline
&& &\multicolumn{4}{c}{\em Data Generation Model 2}\\
 \multirow{2}{*}{$S_1$} & $D$ & $\theta_{12}$ & 0.145 & 0.061 & 0.060 & 30.8 \\ 
 & $A$ & $\theta_{11}$ & 0.068 & 0.088 & 0.085 & 86.3 \\ [4pt]
 \multirow{2}{*}{$S_2$} & $D$ & $\theta_{12}$ & 0.001 & 0.065 & 0.063 & 93.8 \\ 
 & $A$ & $\theta_{11}$ & 0.002 & 0.089 & 0.086 & 94.3 \\ [4pt]
 \multirow{2}{*}{$m_a(\mathbf{X};\boldsymbol{\theta})$} & $D$ & $\theta_{12}$ & 0.001 & 0.065 & 0.063 & 93.8 \\ 
 & $A$ & $\theta_{11}$ & 0.003 & 0.094 & 0.096 & 95.5 \\ [4pt]
 \multirow{2}{*}{$S_1+m_a(\mathbf{X};\boldsymbol{\theta})$} & $D$ & $\theta_{12}$ & 0.145 & 0.061 & 0.060 & 30.8 \\ 
 & $A$ & $\theta_{11}$ & 0.069 & 0.090 & 0.090 & 87.0 \\ [4pt]
 \multirow{2}{*}{$S_2+m_a(\mathbf{X};\boldsymbol{\theta})$} & $D$ & $\theta_{12}$ & 0.001 & 0.065 & 0.063 & 93.8 \\ 
 & $A$ & $\theta_{11}$ & 0.247 & 0.083 & 0.081 & 15.0 \\  
\hline
\end{tabular}
\begin{tablenotes}
\scriptsize
\item $S_1$ represents the propensity score model for the continuous exposure $D$ conditioned on $A=1$, given by $S_1 = E(D|A=1, \mathbf{Z}_1)$; $S_2$ represents the propensity score model for the binary exposure $A$, given by $S_2 = E(A|\mathbf{Z}_2)$; $m_a(\mathbf{X};\boldsymbol{\theta})$ represents the imputation models for $a=0,1$. 
\end{tablenotes}
\end{table}

\begin{table}
\centering
\caption{Empirical results for estimators of causal effects with $P(A=1)=0.75$ under six methods of analysis based on correctly specified models.}
\label{table-appendix4}
\begin{tabular}{llcccccccccccc}
\hline
& && \multicolumn{4}{l}{{Minimal PS: ${\mathbf{S}=\{S_1(\mathbf{Z}_1),S_2(\mathbf{Z}_2)\}'}$}} & \multicolumn{4}{l}{{Expanded PS: ${\mathbf{S}=\{S_1(\mathbf{X}),S_2(\mathbf{X})\}'}$}} \\
\cline{4-11}
Method & Exposure & Effect & Ebias & ESE & RSE & ECP(\%) & Ebias & ESE & RSE & ECP(\%) \\
\hline 
&&&\multicolumn{8}{c}{\em Data Generation Model 1}\\
\multirow{2}{*}{Naive} & $D$ & $\psi_{11}$ & 0.765 & 0.029 & 0.029 & 0.0 & - & - & - & -\\
 & $A$ & $\psi_{12}$ & 0.976 & 0.160 & 0.157 & 0.0 & - & - & - & -\\[4pt]
\multirow{2}{*}{DG model} & $D$ & $\psi_{11}$ & -0.001 & 0.026 & 0.027 & 95.9 & - & - & - & -\\
 & $A$ & $\psi_{12}$ & 0.002 & 0.077 & 0.078 & 94.9 & - & - & - & -\\[4pt]
\multirow{2}{*}{Two PS Reg} & $D$ & $\psi_{11}$ & 0.014 & 0.033 & 0.033 & 92.3 & 0.013 & 0.031 & 0.031 & 92.4\\
 & $A$ & $\psi_{12}$ & 0.007 & 0.084 & 0.085 & 95.1 & 0.007 & 0.078 & 0.079 & 94.8\\[4pt]
\multirow{2}{*}{$\text{PS}_\text{I}+\text{PS}_\text{II}$} & $D$ & $\psi_{11}$ & -0.001 & 0.048 & 0.048 & 95.5 & -0.001 & 0.036 & 0.036 & 95.6\\
 & $A$ & $\psi_{12}$ & 0.004 & 0.109 & 0.112 & 95.4 & 0.002 & 0.079 & 0.081 & 95.2\\[4pt]
\multirow{2}{*}{$\text{PS}_\text{I}+\text{IPW}_\text{II}$} & $D$ & $\psi_{11}$ & -0.001 & 0.048 & 0.048 & 95.5 & -0.001 & 0.036 & 0.036 & 95.6\\
 & $A$ & $\psi_{12}$ & 0.007 & 0.150 & 0.147 & 93.7 & 0.004 & 0.132 & 0.126 & 94.3\\[4pt]
\multirow{2}{*}{$\text{PS}_\text{I}+\text{AIPW}_\text{II}$} & $D$ & $\psi_{11}$ & -0.001 & 0.048 & 0.048 & 95.5 & -0.001 & 0.036 & 0.036 & 95.6\\
 & $A$ & $\psi_{12}$ & 0.001 & 0.083 & 0.085 & 95.3 & 0.001 & 0.084 & 0.085 & 95.1\\
\hline
&&&\multicolumn{8}{c}{\em Data Generation Model 2}\\
\multirow{2}{*}{Naive} & $D$ & $\psi_{11}$ & 0.465 & 0.028 & 0.028 & 0.0 & - & - & - & -\\
 & $A$ & $\psi_{12}$ & 0.498 & 0.099 & 0.097 & 0.15 & - & - & - & -\\[4pt]
\multirow{2}{*}{DG model} & $D$ & $\psi_{11}$ & 0.001 & 0.050 & 0.044 & 93.7 & - & - & - & -\\
 & $A$ & $\psi_{12}$ & 0.001 & 0.080 & 0.077 & 94.0 & - & - & - & -\\[4pt]
\multirow{2}{*}{Two PS Reg} & $D$ & $\psi_{11}$ & 0.0002 & 0.032 & 0.032 & 95.1 & - & - & - & -\\
 & $A$ & $\psi_{12}$ & 0.001 & 0.081 & 0.078 & 94.5 & - & - & - & -\\[4pt]
\multirow{2}{*}{$\text{PS}_\text{I}+\text{PS}_\text{II}$} & $D$ & $\psi_{11}$ & 0.001 & 0.037 & 0.037 & 95.0 & - & - & - & -\\
 & $A$ & $\psi_{12}$ & 0.002 & 0.082 & 0.079 & 94.1 & - & - & - & -\\[4pt]
\multirow{2}{*}{$\text{PS}_\text{I}+\text{IPW}_\text{II}$} & $D$ & $\psi_{11}$ & 0.001 & 0.037 & 0.037 & 95.0 & - & - & - & -\\
 & $A$ & $\psi_{12}$ & 0.002 & 0.082 & 0.079 & 94.1 & - & - & - & -\\[4pt]
\multirow{2}{*}{$\text{PS}_\text{I}+\text{AIPW}_\text{II}$} & $D$ & $\psi_{11}$ & 0.001 & 0.037 & 0.037 & 95.0 & - & - & - & -\\
 & $A$ & $\psi_{12}$ & 0.001 & 0.087 & 0.084 & 94.4 & - & - & - & -\\
\hline
\end{tabular}
\begin{tablenotes}
\scriptsize
\item DG model: Data generation model; Two PS Reg: Two-part propensity score regression adjustment; $\text{PS}_\text{I}+\text{PS}_\text{II}$: Using PS regression adjustment in Stage I and II; $\text{PS}_\text{I}+\text{IPW}_\text{II}$: Using PS regression adjustment in Stage I and IPW in Stage II; $\text{PS}_\text{I}+\text{AIPW}_\text{II}$: Using PS regression adjustment in Stage I and AIPW in Stage II. \\
\end{tablenotes}
\end{table}

\begin{table}
\centering
\small
\caption{Empirical results for estimators of causal effects with $P(A=1)=0.75$ for two-propensity-score regression adjustment and two-stage analysis under model misspecification.}
\label{table-appendix5}
\begin{tabular}{llccccccccc}
\hline
 &   &   & \multicolumn{4}{c}{Misspecified $S_1$} & \multicolumn{4}{c}{Misspecified $S_2$} \\ 
\cline{4-11} 
 {Method} &  {Exposure} &  {Effect} & Ebias & ESE & RSE & ECP(\%) & Ebias & ESE & RSE & ECP(\%)\\ 
\hline 
&&&\multicolumn{8}{c}{\em Data Generation Model 1}\\
 \multirow{2}{*}{Two PS Reg} & $D$ & $\psi_{11}$ & 0.206 & 0.026 & 0.027 & 0.0 & 0.007 & 0.032 & 0.034 & 95.3 \\ 
 & $A$ & $\psi_{12}$ & 0.077 & 0.070 & 0.070 & 80.7 & 0.255 & 0.085 & 0.087 & 15.2 \\  
[4pt]
 \multirow{2}{*}{$\text{PS}_\text{I}+\text{PS}_\text{II}$} & $D$ & $\psi_{11}$ & 0.350 & 0.050 & 0.051 & 0.0 & -0.001 & 0.048 & 0.048 & 95.5 \\ 
 & $A$ & $\psi_{12}$ & 0.134 & 0.109 & 0.112 & 78.5 & 0.344 & 0.120 & 0.114 & 13.2 \\  
[4pt]
 \multirow{2}{*}{$\text{PS}_\text{I}+\text{IPW}_\text{II}$} & $D$ & $\psi_{11}$ & 0.350 & 0.050 & 0.051 & 0.0 & -0.001 & 0.048 & 0.048 & 95.5 \\ 
 & $A$ & $\psi_{12}$ & 0.053 & 0.150 & 0.157 & 90.7 & 0.347 & 0.118 & 0.123 & 19.6 \\  
\hline
&&&\multicolumn{8}{c}{\em Data Generation Model 2}\\
 \multirow{2}{*}{Two PS Reg} & $D$ & $\psi_{11}$ & -0.011 & 0.036 & 0.036 & 93.4 & 0.001 & 0.037 & 0.037 & 95.1 \\ 
 & $A$ & $\psi_{12}$ & 0.089 & 0.086 & 0.083 & 81.1 & 0.276 & 0.084 & 0.081 & 8.6 \\ 
[4pt]
 \multirow{2}{*}{$\text{PS}_\text{I}+\text{PS}_\text{II}$} & $D$ & $\psi_{11}$ & 0.146 & 0.036 & 0.035 & 1.6 & 0.001 & 0.037 & 0.036 & 95.0 \\  
 & $A$ & $\psi_{12}$ & 0.066 & 0.081 & 0.079 & 86.0 & 0.246 & 0.082 & 0.080 & 14.4 \\  
[4pt]
 \multirow{2}{*}{$\text{PS}_\text{I}+\text{IPW}_\text{II}$} & $D$ & $\psi_{11}$ & 0.146 & 0.036 & 0.035 & 1.6 & 0.001 & 0.037 & 0.036 & 95.0 \\ 
 & $A$ & $\psi_{12}$ & 0.024 & 0.099 & 0.093 & 92.0 & 0.247 & 0.084 & 0.081 & 16.0 \\   
\hline
\end{tabular}
\begin{tablenotes}
\scriptsize
\item $S_1$ represents the propensity score model for the continuous exposure $D$ conditioned on $A=1$, given by $S_1 = E(D|A=1, \mathbf{Z}_1)$; $S_2$ represents the propensity score model for the binary exposure $A$, given by $S_2 = E(A|\mathbf{Z}_2)$.
\end{tablenotes}
\end{table}

\begin{table}
\centering
\caption{Empirical results for estimators of effects for both parts with $P(A=1)=0.75$ for applying PS regression adjustment in Stage I and AIPW in Stage II under model misspecification.}
\label{table-appendix6}
\begin{tabular}{llccccc}
\hline
{Misspecified Model} & {Exposure} & Effect & Ebias & ESE & RSE & ECP(\%)\\
\hline 
&& &\multicolumn{4}{c}{\em Data Generation Model 1}\\
 \multirow{2}{*}{$S_1$} & $D$ & $\psi_{11}$ & 0.350 & 0.050 & 0.051 & 0.0 \\ 
 & $A$ & $\psi_{12}$ & 0.048 & 0.087 & 0.089 & 92.0 \\ [4pt]
 \multirow{2}{*}{$S_2$} & $D$ & $\psi_{11}$ & -0.001 & 0.048 & 0.048 & 95.5 \\ 
 & $A$ & $\psi_{12}$ & -0.020 & 0.087 & 0.089 & 95.1 \\ [4pt]
 \multirow{2}{*}{$m_a(\mathbf{X};\boldsymbol{\theta})$} & $D$ & $\psi_{11}$ & -0.001 & 0.048 & 0.048 & 95.5 \\ 
 & $A$ & $\psi_{12}$ & 0.005 & 0.087 & 0.088 & 95.0 \\ [4pt]
 \multirow{2}{*}{$S_1 + m_a(\mathbf{X};\boldsymbol{\theta})$} & $D$ & $\psi_{11}$ & 0.350 & 0.050 & 0.051 & 0.0 \\ 
 & $A$ & $\psi_{12}$ & 0.051 & 0.091 & 0.091 & 90.7 \\ [4pt]
 \multirow{2}{*}{$S_2 + m_a(\mathbf{X};\boldsymbol{\theta})$} & $D$ & $\psi_{11}$ & -0.001 & 0.048 & 0.048 & 95.5 \\ 
 & $A$ & $\psi_{12}$ & 0.231 & 0.079 & 0.082 & 18.5 \\ 
\hline
&&& \multicolumn{4}{c}{\em Data Generation Model 2}\\
 \multirow{2}{*}{$S_1$} & $D$ & $\psi_{11}$ & 0.146 & 0.036 & 0.034 & 1.6 \\ 
 & $A$ & $\psi_{12}$ & 0.023 & 0.087 & 0.084 & 92.9 \\ [4pt]
 \multirow{2}{*}{$S_2$} & $D$ & $\psi_{11}$ & 0.001 & 0.037 & 0.036 & 95.0 \\ 
 & $A$ & $\psi_{12}$ & 0.001 & 0.085 & 0.082 & 94.4 \\ [4pt]
 \multirow{2}{*}{$m_a(\mathbf{X};\boldsymbol{\theta})$} & $D$ & $\psi_{11}$ & 0.001 & 0.037 & 0.036 & 95.0 \\ 
 & $A$ & $\psi_{12}$ & 0.001 & 0.090 & 0.086 & 93.5 \\ [4pt]
 \multirow{2}{*}{$S_1 + m_a(\mathbf{X};\boldsymbol{\theta})$} & $D$ & $\psi_{11}$ & 0.146 & 0.036 & 0.035 & 1.6 \\ 
 & $A$ & $\psi_{12}$ & 0.234 & 0.090 & 0.087 & 92.8 \\ [4pt]
 \multirow{2}{*}{$S_2 + m_a(\mathbf{X};\boldsymbol{\theta})$} & $D$ & $\psi_{11}$ & 0.001 & 0.037 & 0.036 & 95.0 \\ 
 & $A$ & $\psi_{12}$ & 0.246 & 0.083 & 0.080 & 15.1 \\  
\hline
\end{tabular}
\begin{tablenotes}
\scriptsize
\item $S_1$ represents the propensity score model for the binary exposure $A$, given by $S_1 = E(A|\mathbf{Z}_1)$; $S_2$ represents the propensity score model for the continuous exposure $D$ conditioned on $A=1$, given by $S_2 = E(D|A=1, \mathbf{Z}_2)$; $m_a(\mathbf{X};\boldsymbol{\theta})$ represents the imputation models for $a=0,1$.
\end{tablenotes}
\end{table}

We consider two scenarios where the probability of exposure is set to \( P(A=1)  = 0.25\) and \( 0.75 \). Tables \ref{table-appendix1}, \ref{table-appendix2}, and \ref{table-appendix3} provide empirical results for estimators of causal effects under $P(A=1) = 0.25$. These results are consistent with those when $P(A=1) = 0.5$. The empirical biases for the effect in Stage II are large when either $S_1$ or $S_2$ is misspecified, leading to reduced performance. The double robustness property of the AIPW approach ensures consistency when only one of the models in Stage II is correctly specified. Tables \ref{table-appendix4}, \ref{table-appendix5}, and \ref{table-appendix6} display the simulation results for $P(A=1) = 0.75$. The findings again reflect similar patterns observed under $P(A=1) = 0.5$ and $P(A=1) = 0.25$. When $S_1$ is misspecified in the first stage, the estimator performance suffers in both stages, indicating a significant sensitivity to model misspecification in the continuous part of the exposure.

\section{Additional Information about the Application}
\label{sup-e}

\begin{table}
\caption{Estimated effects of prenatal covariates on the continuous and binary exposure components}
\label{table-appendix7}
\centering
\begin{tabular}{lcccc}
\hline
  &  \multicolumn{2}{c}{$E(D | A=1,\mathbf{X})$}
  &  \multicolumn{2}{c}{$P(A=1 | \mathbf{X})$} \\
  \cline{2-5}
  & Estimate & Standard error & Estimate & Standard error \\
 \hline
 Intercept & -4.183 & 1.003 & -0.965 & 1.926 \\
 Smoking during pregnancy (cigarettes/day) & 0.016 & 0.007 & 0.088 & 0.025 \\
 Socioeconomic status at birth & -0.007 & 0.003 & -0.003 & 0.005 \\
 Child's gender (female) & -0.165 & 0.165 & -0.608 & 0.314 \\
 Mother's age at delivery & 0.057 & 0.018 & 0.056 & 0.036 \\
 Prenatal Beck score & 0.026 & 0.011 & 0.026 & 0.023 \\
 HOME score at infancy & -0.004 & 0.019 & 0.039 & 0.036 \\
 Biological mother's PPVT & -0.010 & 0.007 & 0.001 & 0.014 \\
 Biological mother's marital status & -0.042 & 0.276 & -0.103 & 0.487 \\
 Prenatal cocaine exposure (days/month) & -0.061 & 0.038 & -0.102 & 0.070 \\
 Prenatal marijuana exposure (days/month) & -0.012 & 0.032 & 0.220 & 0.198 \\
 Prenatal opiate exposure (days/month) & 0.193 & 0.079 & 0.057 & 0.253 \\
 Number of prenatal visits & -0.121 & 0.032 & -0.021 & 0.061 \\
 Biological mother's education (years) & 0.098 & 0.059 & -0.041 & 0.115 \\
 Parity & -0.003 & 0.095 & -0.471 & 0.220 \\
 Gravidity & 0.036 & 0.057 & 0.271 & 0.145 \\
 Gestational age at screening & 0.018 & 0.013 & -0.010 & 0.025 \\
\hline
\end{tabular}
\end{table}
\subsection{Effects of Prenatal Covariates on PAE}
\label{sup-e1}
The estimated effects of baseline covariates on the semi-continuous exposure are given in Table \ref{table-appendix7}.
\subsection{Details of Multiple Imputation Procedure}
\label{sup-E2}
In our data application, the Detroit dataset contains missing values in the outcome and covariates. 181 women have at least one missing value (41 with missing WISC-III and 159 with at least one missing covariate, with overlap between these groups). We address this by generating 20 multiply imputed datasets using an imputation model including all variables; multiple imputation relies on the assumption that data are missing at random (MAR), which means the probability of missingness may depend on observed data but not on the unobserved values themselves. Following this assumption, we use the \texttt{mice} package in R to perform multiple imputation by chained equations (MICE) and generate $20$ imputed datasets \citep{van2011mice}. The imputation model included all variables used in the outcome model, the two propensity score models, and several auxiliary variables associated with the outcome or with missingness \citep{white2011multiple,van2012flexible}.

Specifically, the imputation model included the outcome variable: the WISC-III Freedom from Distractibility Index measured at child age 7; the exposure variables: drinking status and average daily absolute alcohol consumption (AA/day) during pregnancy; and all covariates used in the propensity score models. In addition, auxiliary variables, which are the proportion of drinking days and the amount of alcohol consumed per drinking occasion across pregnancy are included. As shown in Li et al.\citep{li2023use}, these measures of prenatal alcohol exposure reflect distinct patterns of drinking behaviour and have differing impacts on children’s cognition; they also exhibit different relationships with covariates when constructing propensity score models. Therefore, including them as auxiliary variables enhances the plausibility of the MAR assumption and improves the accuracy of imputed values \citep{white2011multiple}.

Two binary variables, child’s gender and mother’s marital status, are imputed using logistic regression. All other variables are continuous and imputed using predictive mean matching (PMM). The default chained equations algorithm in \texttt{mice} is run for 5 iterations per imputation. 

After imputation, the analysis under the proposed methods is described in Section 6 of the main text. The resulting point estimates and standard errors are then combined using Rubin’s Rules to obtain valid overall estimates and associated uncertainty \citep{rubin1976inference}.

\subsection{Details of Method Implementation in Application}
\label{sup-e3}
\subsubsection{Conventional Covariate Regression Adjustment} 
\label{sec-f.2.1}
Let $A$ be a binary exposure representing whether the mother drinks during pregnancy, $T$ denote the absolute alcohol consumption per day (in ounces), and $D=\log T$ represent the logarithm of daily alcohol consumption if $A=1$. The vector of all confounders is denoted by $\mathbf{X}$. The regression model we consider is: 
\begin{equation}
    E(Y_i \mid A_i = 1, D_i, \mathbf{X}_i) = \theta_0 + \theta_{11} A_i(D_i-c) + \theta_{12} A_i + \mathbf{X}'_i \boldsymbol{\theta}_2, \nonumber
\end{equation}
where $c$ is a fixed constant representing the mean logarithm of daily alcohol consumption (in ounces). In this analysis, we set $c=-2.31$ which corresponds to the empirical mean value of $D$ among mothers in this study who reported drinking.

\subsubsection{Two-Part Propensity Score Regression Adjustment}
Firstly, we define two propensity score models: 
$$S_{i1} = E(D|A=1,\boldsymbol{X}_i)=\alpha_{10}+\boldsymbol{\alpha}_{11}\boldsymbol{X}_i,$$
$$S_{i2} = E(A|\boldsymbol{X}_i)=\text{expit}(\alpha_{20}+\boldsymbol{\alpha}_{21}\boldsymbol{X}_i).$$
Then the two–part PS regression adjustment is specified by the linear model
\begin{equation}
        E(Y_i| A_i=1,D_i,\mathbf{S}_i) = \eta_0 +{\eta}_{11}A_i(D_i-c) + {\eta}_{12}A_i + {\eta}_{21}S_{i1} + \eta_{22}S_{i2} \nonumber
\end{equation}
where $\mathbf{S}_i = (S_{i1},S_{i2})'$. 
\subsubsection{The Two-Stage Approach}
We assess the causal effects with a semi-continuous exposure by a two-stage approach. In the first stage, we aim to obtain the effect of the logarithm of the absolute alcohol consumption among exposed by propensity score regression adjustment. Then we treat $\hat{\gamma}_{11}A(D-c)$ as an offset term where $\hat{\gamma}_{11}$ is the estimated causal effect of alcohol consumption among exposed. In stage II, we apply propensity score regression adjustment, IPW, and AIPW approaches to find the effect of prenatal drinking status.

\end{document}